\documentclass[a4paper,11pt]{article}
\pdfoutput=1 % if your are submitting a pdflatex (i.e. if you have
\usepackage{jcappub} % for details on the use of the package, please
\usepackage[T1]{fontenc} % if needed
\usepackage{aas_macros}
\usepackage{multirow}
\usepackage{subcaption}
\usepackage{array}
\usepackage{float}
\usepackage{amsmath}
\usepackage{amssymb}
\usepackage{physics}
\newcommand\msun{M_\odot}

\title{\fontsize{22}{18}\selectfont DEMNUni: the Sunyaev-Zel'dovich effect in the presence of massive neutrinos and dynamical dark energy}

\author[a,b,c.d]{Davide Luchina,}
\author[e]{Mauro Roncarelli,}
\author[f,b]{Matteo Calabrese,}
\author[g]{Giulio Fabbian}
\author[b]{and Carmelita Carbone}
\affiliation[a]{Dipartimento di Fisica ``Aldo Pontremoli'', Universit\`{a} degli Studi di Milano, via Celoria 16, I-20133 Milano, Italy}
\affiliation[b]{INAF -- Istituto di Astrofisica Spaziale e Fisica cosmica di Milano (IASF-MI), via Alfonso Corti 12, I-20133 Milano, Italy}
\affiliation[c]{Dipartimento di Fisica,  Universit\`{a} di Trento, via Sommarive 14, I-38123 Povo (TN), Italy }
\affiliation[d]{Scuola Internazionale Superiore di Studi Avanzati (SISSA), via Bonomea 265, I-34136, Trieste, Italy}
\affiliation[e]{INAF -- Osservatorio di Astrofisica e Scienza dello Spazio di Bologna (OAS-BO), via Piero Gobetti 93/3, I-40129 Bologna, Italy}
\affiliation[f]{Astronomical Observatory of the Autonomous Region of the Aosta Valley (OAVdA), Loc. Lignan 39, I-11020, Nus (Aosta Valley), Italy}
\affiliation[g]{Université Paris-Saclay, CNRS, Institut d’Astrophysique Spatiale, F-91405, Orsay, France}
\emailAdd{dluchina@sissa.it}
\emailAdd{mauro.roncarelli@inaf.it}
\emailAdd{calabrese@oavda.it}
\emailAdd{giulio.fabbian@universite-paris-saclay.fr}
\emailAdd{carmelita.carbone@inaf.it}

\abstract{In recent years, the study of secondary anisotropies in the Cosmic Microwave Background has become a fundamental instrument to test our understanding of the Universe. Using a set of lightcones produced with the ``Dark Energy and Massive Neutrino Universe'' $N$-body simulations, we study how different dark energy equations of state and neutrino masses impact the properties of the thermal Sunyaev-Zel'dovich (tSZ) effect, focusing on the signal arising from galaxy clusters and groups. We analyse the distribution of values for the Compton-$y$ parameter and study its angular power spectrum. We find that the distribution of the logarithmic Compton parameter can be fitted with a skewed Gaussian, with a mean that, at fixed dark energy model, decreases linearly with an approximate slope of $10 f_\nu$. Regarding the power spectrum of the thermal SZ effect, we find that an increase in $\sum {m_\nu}$ is observed as a power-law scaling with respect to $\sigma_8^{\mathrm{cb}}$, with exponents ranging from 7.3 to 8.1. We also find that models with massless neutrinos typically overestimate Compton-$y$ data extracted from Planck measurements; a better agreement with the simulations is obtained for $\sum m_\nu = 0.16$ or $\sum m_\nu=0.32$ eV. For all the \texttt{DEMNUni} models we forecast the cumulative signal-to-noise ratio for thermal SZ observations with the LAT instrument of the Simons Observatory; furthermore, we compute a tailored $\chi_\mathrm{SNR}^2$ estimator to infer if such models can be distinguished from the reference $\Lambda$CDM.}


\begin{document}
\maketitle
\flushbottom

\section{Introduction}\label{introduction}
Since the earliest years of this century, measurements of the CMB anisotropies such as those from WMAP~\cite{wmap1,wmap7}, Planck~\cite{planck2013Cosmo,planck2015cosmo,planck2018} or ACT~\cite{act2025ps,act2025extended} provided key estimates of different cosmological parameters. While for many probes the standard $\Lambda$CDM works excellently, in some cases tensions arise, leading to the necessity of extending the common cosmological framework (see e.g. the recent results from DESI~\cite{desiBao2025}). In this work we investigate the extended models which make up the \texttt{DEMNUni} N-body simulation set~\cite{Carbone_2016, parimbelli_2022}, characterised by the presence of massive neutrinos and a dynamical dark energy, through the Sunyaev-Zel'dovich effect. This effect has reached a role of key importance in astrophysics and cosmology, being nowadays routinely implemented in many diverse studies~\cite{flamingoClusters,Xray+tSZ,wl+kSZBigwood,tSZACT+DESI,pairwise_kSZ_cmbS4,pairwisekSZ_soergel,chiang2020}.

 Key discoveries made at the end of the 90's with detection experiments~\cite{superK} led to the necessity of introducing a non-vanishing mass for neutrinos in the Standard Model of Particle Physics. For what concerns cosmology, while massless neutrinos have been considered a fundamental ingredient in studies for many years, only in the last 10--15 years models with massive neutrinos became common in the analyses of different phenomena~\cite{Roncarelli_2015,mauro2017,sdssBaoClustering2017,planck2018,desY3ext,kSZtomography}. In such scenarios, aside from variations in the neutrino background itself (for a treatment see~\cite{lesgourgues2006} or, more extensively,~\cite{lesgNeutrinoBook}), there are changes in both the CMB and in the distribution of matter, depending at first order solely on the total mass of all neutrinos $M_{\nu} \equiv \sum m_\nu $, which has a lower bound of $0.06$ eV imposed by flavour oscillation measurements~\cite{pdg2022}. As neutrinos interact only through gravity they constitute a collisionless fluid: one where there is no bulk behaviour and no pressure propagation. In such fluids particles on average travel long distances in between interactions, and for neutrinos at low $z$ this \textit{mean free path} typically exceeds the scales of galaxy clusters, meaning that they can't be confined inside of them. This has a great impact on the Halo Mass Function (HMF), favouring smaller sized structures and almost erasing the very high-mass end, composed of the rarest clusters. This behaviour is known as free streaming, and significantly affects the observed (mostly thermal) SZ effect signal.

 Regarding the nature of dark energy, the Chevallier-Polarski-Linder (CPL)~\cite{chevPolanki,linder} is a widely used model, which describes a dynamical dark energy fluid with equation of state (EoS) parametrised as $w (z) = w_0 + w_a \, z/(1+z)$. The case of the cosmological constant, $\Lambda$, can then be retrieved in the limit where $w_0 = -1$ and $w_a = 0$.
This parametrisation is of simple understanding and also solves problems that other choices might create, such as divergences at high redshift for the case of a linear $w(z)$. Despite its popularity, the CPL parametrisation has the downside of not being linked to any physical model and as such can exhibit phantom crossing when $w(z) < -1$ ; this violates the null energy condition, meaning that the energy density of the dark energy component increases with the universe size. Nonetheless, the recent DESI-BAO analysis~\cite{desiBao2025} has favoured values of $w_0$ and $w_a$ which are indeed characterised by this behaviour at $z>0.5$, even with the inclusion of CMB data.

 This paper is structured as follows. In Section~\ref{demnUni} we outline the simulation set and halo catalogues used for the analysis. In Section~\ref{models} we describe the Sunyaev-Zel'dovich effect, together with the pressure profile utilised to characterise the electrons in the intracluster medium (ICM) and to create synthetic maps of the thermal component. In Section~\ref{sec::tszResults} we present the results of the analysis of the maps in all the different simulations, then summarised in Section~\ref{sec::Conclusions}. In Appendix~\ref{sec::kszResults} we also provide a simple modelling of the kinetic counterpart of the SZ anisotropy, limited to the same halo sample used in the main analysis.

\section{The synthetic catalogues}\label{demnUni}

\subsection{The \texttt{DEMNUni} simulations}
The simulation set used in this work is the ``Dark Energy and Massive Neutrino universe''~\cite{Carbone_2016} (\texttt{DEMNUni}): a total of 15 N-body simulations that implement the presence of massive neutrinos in a flat background with a dynamical dark energy. The \texttt{DEMNUni} simulations have been produced with the aim of investigating the large-scale structure of the Universe in the presence of massive neutrinos and dynamical dark energy, and they were conceived for the nonlinear analysis and modelling of different probes, including dark matter, halo~\cite{Hernandez_2024a, Hernandez_2024b} and galaxy clustering~\cite{Castorina_2015, Moresco_2017, Zennaro_2017, ruggeri_2018, bel_2019, parimbelli_2021, parimbelli_2022, Guidi_2022, Baratta_2022, Gouyou_Beauchamps_2023, Verdiani_2025, SHAM-Carella_in_prep}, weak lensing~\cite{Ingoglia_2024}, CMB lensing, Sunyaev-Zel'dovich and integrated Sachs-Wolfe effects~\cite{Carbone_2016, Roncarelli_2015, Fabbian_2018}, cosmic void statistics~\cite{kreisch_2019, schuster_2019, verza_2019, verza_2022a, verza_2022b}, as well as cross-correlations among these probes~\cite{Vielzeuf_2022, Cuozzo2022}.

\begin{figure}[!ht]
    \centering
    \includegraphics[width=0.6\textwidth]{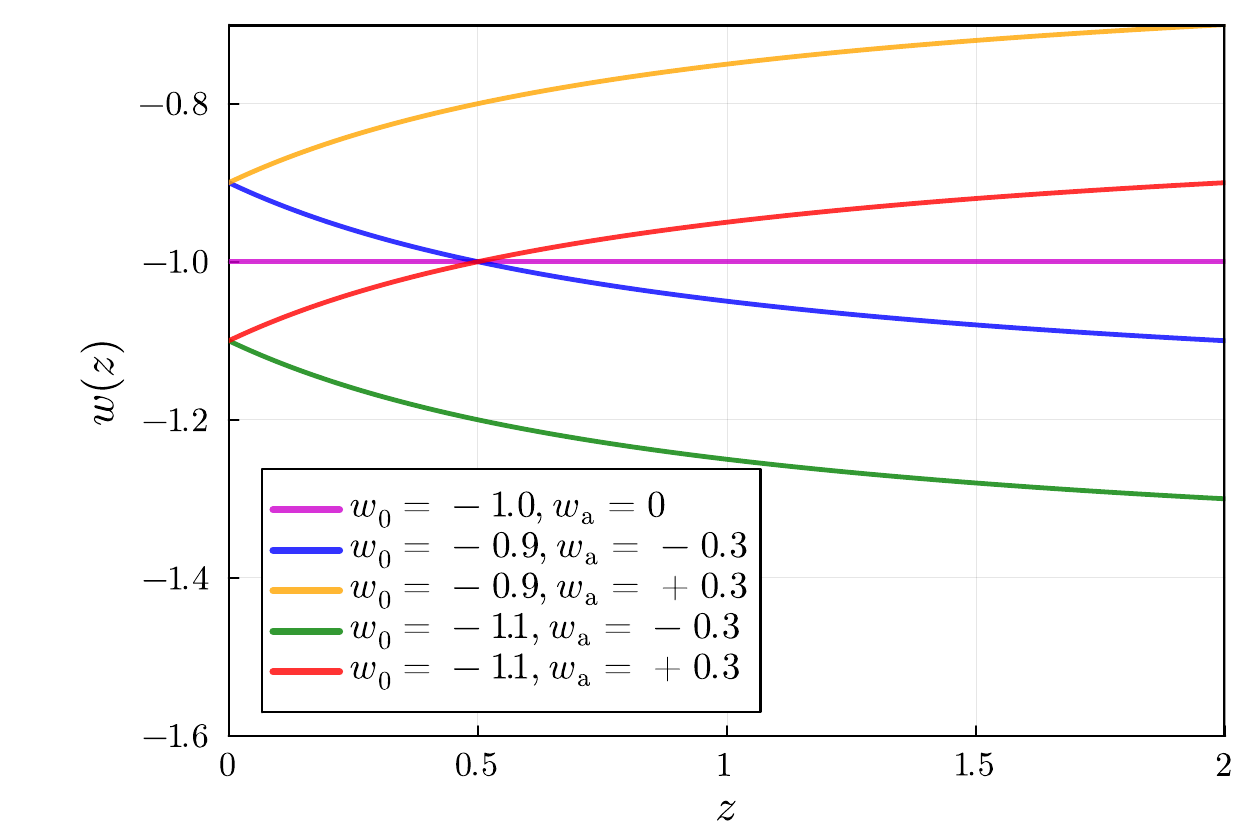}
    \caption{Redshift evolution of the EoS of dark energy in the different \texttt{DEMNUni} cosmologies, parametrised à la CPL. Notice the phantom crossing exhibited by the blue curve with $w(z) < -1$ at $z > 0.5$.}
    \label{fig::EoS}
\end{figure}

The \texttt{DEMNUni} simulations have a large $2 h^{-1}$Gpc comoving size side box, filled with $2048^3$ dark-matter particles (and $2048^3$ neutrino particles, when present) evolving from $z=99$ to $z=0$. The simulations were run using the tree particle mesh-smoothed particle hydrodynamics code \texttt{GADGET-3}, an upgraded version of the one presented in~\cite{Springel_2005}, and are characterised by a softening length of $20 h^{-1}$ kpc. The simulations are initialised at $z_\mathrm{ini}=99$ with Zel'dovich initial conditions.
The initial power spectrum is rescaled to the initial redshift via the rescaling method developed in~\cite{Zennaro_2017}. Initial conditions are then generated with a modified version of the \texttt{N-GenIC} software, assuming Rayleigh random amplitudes and uniform random phases. The common cosmological parameters of the simulations are analogous to those resulting from the analysis of Planck 2013~\cite{planck2013Cosmo} and shown in Table~\ref{tab:demnuni_sets_table}, together with the varied parameters.
The presence of a dynamical dark energy EoS was carried out using the CPL parametrisation and the four possible combinations of $w_0 =-0.9,\, -1.1$ and $w_a=-0.3,\, +0.3$, as shown in Figure~\ref{fig::EoS}. Concerning massive neutrinos, they were implemented in the simulations by using the modifications to \texttt{GADGET-3} described in~\cite{Viel_2010} and a tabulated Hubble parameter, $H(z)$, as computed in~\cite{Zennaro_2017}. This version of the code follows the evolution of CDM and neutrino particles, treating them as two separated collisionless species. Given the relatively high velocity dispersion, neutrinos have a characteristic clustering scale larger than the CDM one, allowing to save computational time by neglecting the calculation of the short-range gravitational force. This results in a different spatial resolution for the two components, which for neutrinos is fixed by the PM grid (that we have chosen to be eight times larger than the particle number), while for CDM particles is about one order of magnitude higher. Furthermore, neutrinos are assumed to be in a degenerate mass scenario\footnote{Implying that all three mass eigenstates $m_i$ have the same mass value.} with values of the total mass of $M_\nu=0,\,0.16,\,0.32$ eV, and the budget for $\Omega_\nu$ is taken from $\Omega_{\mathrm{c}}$. For each simulation, 63 snapshots, logarithmically equispaced in the scale factor $a$, are saved. Moreover, about 300 TB of data in particle comoving snapshots, halo and galaxy catalogues, projected density maps, and power spectra of different particle species are stored and available upon request.

\begin{table}[!ht]
 \centering
 \setlength{\tabcolsep}{1pt}
\begin{tabular}{|c|c|c|c|c|c|c|c|}
\hline
    \textbf{No.} & Type & $M_\nu\equiv 93.14\, h^2\,\Omega_\nu \, [\mathrm{eV}]$  &  $\Omega_{\rm c}$ & $(w_0,w_a)$& $m_{\mathrm{c}}^{p} \, [\msun h^{-1}]$ & $m_{\nu}^{p} \, [\msun h^{-1}]$ & $\sigma_8 (z=0)$      \\ 
\hline
    \textbf{1} & $\Lambda$CDM & 0 & 0.2700 & $(-1,0)$ & $8.27\times10^{10}$ & - & 0.830\\
\hline
    \textbf{2} & \multirow{2}{*}{$\nu\Lambda$CDM} & 0.16 & 0.2662 &  \multirow{2}{*}{$(-1,0)$} & $8.17\times10^{10}$ & $9.97\times10^{8}$ & 0.793\\
    \textbf{3} & & 0.32 & 0.2623 & & $8.07\times10^{10}$ & $1.99\times10^{9}$ & 0.752\\
\hline
    \textbf{4} & \multirow{4}{*}{$w_0 w_a$CDM}& \multirow{4}{*}{0} & \multirow{4}{*}{0.2700} & $(-0.9,-0.3)$ & \multirow{4}{*}{ $8.27\times10^{10}$} & \multirow{4}{*}{-} & 0.828\\
    \textbf{5} & &  &  & $(-0.9,+0.3)$ & &  & 0.777\\
    \textbf{6} & &  &  & $(-1.1,-0.3)$ & &  & 0.861\\
    \textbf{7} & &  &  & $(-1.1,+0.3)$ & &  & 0.831\\
\hline
    \textbf{8}  & \multirow{8}{*}{$w_0 w_a \nu$CDM} & \multirow{4}{*}{0.16} & \multirow{4}{*}{0.2662} & $(-0.9,-0.3)$ & \multirow{4}{*}{$8.17\times10^{10}$}& \multirow{4}{*}{$9.97\times10^{8}$} & 0.791\\
    \textbf{9}  &                             &  &  & $(-0.9,+0.3)$ & &  & 0.742\\
    \textbf{10}  &                             &  &  & $(-1.1,-0.3)$ & &  & 0.822\\
    \textbf{11} &                             &  &  & $(-1.1,+0.3)$ & &  & 0.794\\\cline{3-4}\cline{6-7}
    \textbf{12} &                             &  \multirow{4}{*}{0.32} & \multirow{4}{*}{0.2623} & $(-0.9,-0.3)$ & \multirow{4}{*}{$8.07\times10^{10}$} & \multirow{4}{*}{$1.99\times10^{9}$} & 0.750\\
    \textbf{13} &                             & & & $(-0.9,+0.3)$ & &  & 0.705\\
    \textbf{14} &                             & & & $(-1.1,-0.3)$ & &  & 0.780\\
    \textbf{15} &                             & & & $(-1.1,+0.3)$ & &  & 0.753\\
    \hline
\end{tabular}
 \caption{Cosmological parameter values of the \texttt{DEMNUni} simulation suite. The reference cosmology for all the models has $\{\Omega_{\rm b}, \Omega_{\rm m}\equiv \Omega_{\rm c}+\Omega_{\rm b}+\Omega_{\rm \nu}, h, n_{\rm s}, A_{\rm s} \} = \{0.05, 0.32, 0.67, 0.96, 2.1265 \times 10^{-9} \}.$ Above we show only the values of the varied parameters for the different cosmological scenarios considered in this work.}
\label{tab:demnuni_sets_table}
\end{table}

\subsection{The halo catalogues}
\label{halocat}
Each of the 63 snapshots was processed first with a \texttt{Friends of Friends} (\texttt{FoF}) algorithm with a linking length of $0.2$, and subsequently with the \texttt{Subfind} algorithm, both already included in \texttt{GADGET-3}~\citep{springel01,dolag09}. These allow the correct identification of gravitationally bound structures, and produce catalogues of haloes each with its mass, position and (peculiar) velocity. We apply a mass cut to the halo catalogue, selecting only those structures with $M_{200} > 3.4 \times 10^{13} h^{-1} \msun$, that we can associate with galaxy clusters and groups; this was done to ensure the applicability of a suitable pressure profile (see Section~\ref{subsec::tsz}).
We verified that this selection has a negligible impact on our main findings, related to the power spectrum of the thermal SZ effect, if not for very small scales, which are however dominated by smaller structures for which the pressure modelling is highly uncertain.

\subsection{The creation of the lightcones}
To obtain \textit{mock} universes from the \texttt{DEMNUni} halo catalogues and use them to study the tSZ effect, it is necessary to create full-sky backward lightcones.
The construction of the halo lightcone follows a procedure similar in its geometrical approach to the one employed for CMB lensing (CMBL) maps, as described in~\cite{melitaLensing,calabrese2015} and validated in~\cite{Fabbian_2018,Hilbert_2020}. In this case, instead of dark matter particles, the halo catalogue is extracted from the \texttt{DEMNUni} simulations using the \texttt{Subfind} algorithm and placed within a full-sky, 3D framework around a central observer. The methodology involves replicating the finite simulation volume to cover the entire past lightcone up to a chosen redshift, ensuring a continuous structure along the line of sight. Specifically, the volume is divided into concentric spherical shells of fixed comoving thickness, within which all simulation outputs are subjected to coherent translations and rotations, rather than independent random transformations. This approach, originally developed for weak lensing applications, preserves the continuity of the gravitational potential across transverse directions. The resulting halo lightcones extend to a maximum redshift of $z_{\rm max} \approx 2.3$, aligning with the redshift range relevant for the Euclid mission~\cite{euclidOverview}. Each halo within this framework is characterised by ten parameters, namely its mass $M_{200}$, angular coordinates, comoving distance, Cartesian position, and peculiar velocity. This process is applied to all simulations, yielding full-sky, 3D halo catalogues across the multiple cosmological scenarios shown in Table~\ref{tab:demnuni_sets_table}.

\section{The Sunyaev-Zel'dovich effect}\label{models}

The SZ effect is a source of CMB secondary anisotropies, emerging due to the Compton scattering of the radiation off ionised electrons~\cite{kompaneets,sz,szKinematic2,sunyaevKinematic}.
The thermal random motion of the electrons leads to a specific temperature variation in the spectrum of the CMB, given by
    \begin{equation}\label{deltaToverT}
    \frac{\Delta T_{\mathrm{CMB}}}{T_{\mathrm{CMB}}} = y \left( \frac{x}{\tanh{x/2} } - 4\right) \equiv y \, g(x)\ ,
\end{equation}
where
\begin{align}
    x &\equiv \frac{h \nu}{k_{\mathrm{B}} T_{\mathrm{CMB}}} \ ,\label{dimensionlessFreq} \\
    y &\equiv \frac{\sigma_{\mathrm{T}}}{m_{\mathrm{e}} c^2} \int_0^{\infty} \dd \lambda \, n_{\mathrm{e}} k_{\mathrm{B}} T_{\mathrm{e}} = \frac{\sigma_{\mathrm{T}}}{m_{\mathrm{e}} c^2} \int_0^{\infty} \dd \lambda \, P_{\mathrm{e}} (\lambda) \ . \label{y}
\end{align}
$m_{\mathrm{e}}$, $\sigma_{\mathrm{T}}$ and $k_{\mathrm{B}}$ are respectively the electron mass, the Thomson cross-section and the Boltzmann constant, while $c$ is the light speed in vacuum. The quantities $n_{\mathrm{e}}$, $T_{\mathrm{e}}$ and $P_{\mathrm{e}}$ are number density, temperature and pressure of the electron gas.
The variable $x$ is the \textit{dimensionless frequency}, while $y$, better known as Compton parameter, is a measure of the integrated electron pressure along the physical line of sight $\lambda$. This is the thermal SZ effect.

 On the other hand, ionised electrons also possess a non-zero velocity that leads to an additional, frequency independent, variation in the CMB temperature:
\begin{equation}\label{kszeq}
    \frac{\Delta T_{\mathrm{CMB}}}{T_{\mathrm{CMB}}} = - \int_0^\infty \dd \lambda \, n_{\mathrm{e}} \sigma_{\mathrm{T}} \, \mathrm{e}^{-\tau} \frac{v_{\mathrm{los}}}{c} \ .
\end{equation}
In the above $\tau$ is the optical depth for Compton scattering, while $v_{\mathrm{los}}$ is the proper velocity of the electron plasma along the line of sight.
This is the kinematic, or kinetic, SZ (kSZ) effect. Our analysis, which focuses solely on the component of galaxy clusters and groups, makes for an excellent approximation of the tSZ effect; this happens since it is sourced by the energy density of ionised electrons, which is most prominent inside large bound structures due to their deep gravitational potential wells. In turn, since the kSZ effect does not depend directly on the electron temperature, it receives a significant contribution from the gas in non-virialised structures at temperature between $10^5$ and $10^7$ K and would require a more sophisticated treatment with hydrodynamical simulations (see~\cite{mauro2017,mauro2018} and references therein). A precise estimate of the kSZ effect and its dependence with cosmology is therefore beyond the scope of this work. Nonetheless, with the \texttt{DEMNUni} simulations we can define a modelling of the halo component of the kSZ signal that we describe in Appendix~\ref{sec::kszResults}: 
its prediction has to be regarded as an underestimate of the true one.

\subsection{The tSZ model}\label{subsec::tsz}
\begin{figure}
\centering
\subfloat{\includegraphics[width = 0.45\textwidth]{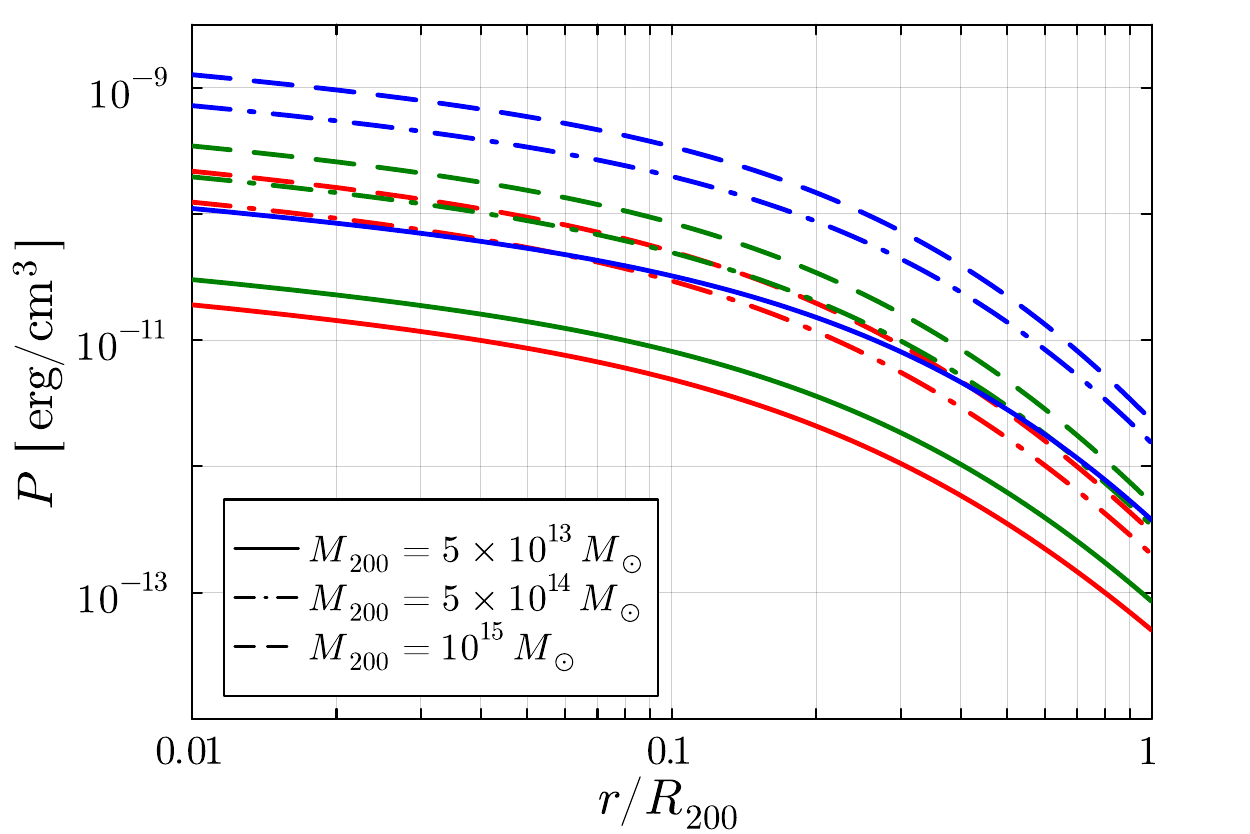}} 
\subfloat{\includegraphics[width = 0.45\textwidth]{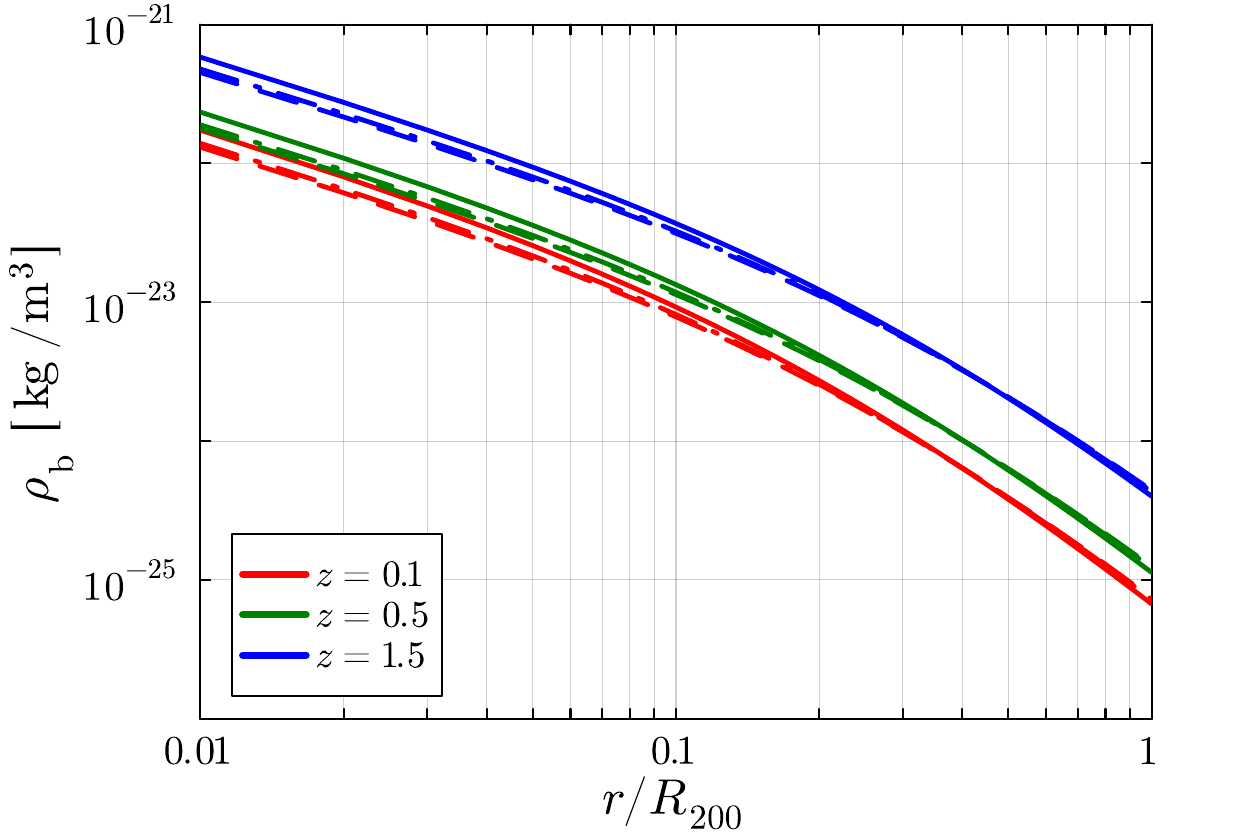}}
\caption{LEFT: Battaglia pressure profile, as in Equation~\eqref{battagliaProf1}, as a function of $r/ R_{200}$, at different redshifts and $M_{200}$ values. RIGHT: NFW profile for the baryon density $\rho_{\mathrm{b}}$, as in Equation~\eqref{nfwBaryons}, as a function of $r/R_{200}$, for the same $z$, $M_{200}$ combinations. For both we assume the $\Lambda$CDM baseline model of the \texttt{DEMNUni} suite.}

\label{multiFigProfiles}
\end{figure}

We model the gas pressure in the selected haloes using the Battaglia profile
\begin{equation}\label{battagliaProf1}
    P(X) = P_{200}\, P_0 \, (X/x_{\mathrm{c}})^{\gamma} \left[ 1 + (X/x_{\mathrm{c}})^{\alpha} \right]^{-\beta} \  ,
\end{equation}
resulting from the hydrodynamical simulations described in~\cite{battaglia1}.
Here $X \equiv r/ R_{200}$, while $\alpha$ and $\gamma$ are fixed respectively to the values $1.0$ and $-0.3$. The other 3 parameters $P_0$, $x_{\mathrm{c}}$ and $\beta$ vary with respect to $M_{200}$ and $z$ (see~\cite{battaglia1} for an accurate description). We stress here that the overdensity threshold, set equal to 200, has to be intended with respect to the \textit{critical} density $\rho_{\mathrm{cr}}$.
The scale value $P_{200}$ is the self-similar pressure given by
\begin{equation} \label{battagliaprof2}
    P_{200} = G M_{200} \  200\rho_{\mathrm{cr}}(z) \, \frac{1}{2R_{200}} \,f_{\mathrm{b}}\,  \ ,
\end{equation}
where $f_{\mathrm{b}}$ is the ratio of baryonic matter with respect to total matter in the halo. We compute this ratio as 
\begin{equation}\label{fbMassive}
    f_{\mathrm{b}} = \frac{\Omega_{\mathrm{b}}}{\Omega_{\mathrm{m}} - \Omega_\nu} \ 
\end{equation}
to account for the free-streaming of neutrinos. In Eq.~\eqref{battagliaprof2} the values of $M_{200}$ and $R_{200}$ are taken individually for each galaxy cluster/group from our halo catalogues (Section~\ref{halocat}), while the value of $f_{\mathrm{b}}$ is fixed for every cosmological model: with the 3 different $M_\nu$ in the simulations (see Table~\ref{tab:demnuni_sets_table}) we obtain the values $0.1563,0.1581,0.1601$.
This follows from the fact that, in a scenario with fixed $\Omega_{\mathrm{m}}$ and $\Omega_{\mathrm{b}}$ as in the \texttt{DEMNUni} case, haloes with the same mass have increasingly higher baryon content with greater $M_\nu$.
We show examples of the resulting pressure profiles in Figure~\ref{multiFigProfiles}.
Once the total gas pressure is defined, the electron pressure is obtained assuming a pristine cosmological gas in full ionization, through
\begin{equation}\label{electronPress}
    P_\mathrm{e} = \frac{2(f_\mathrm{H}+1)}{(5f_\mathrm{H}+3)} \, P \simeq 0.52 \, P \ ,
\end{equation}
where $f_\mathrm{H}=0.76$ is the primordial hydrogen mass fraction.

 In Section~\ref{sec::tszResults} we compute the power spectrum of the tSZ effect, calibrating the process by comparing our results to the theoretical expectations of the halo model in the $\Lambda$CDM scenario.
In the small scale, $\ell\gg1$, limit  this power spectrum has the following 1-halo and 2-halo terms~\cite{cooray2005,hillPaj2013}:
\begin{align}
    C_\ell^{1\mathrm{h}} &= \int_{z_{\mathrm{min}}}^{z_{\mathrm{max}}} \dd z \, \frac{\dd^2V}{\dd \Omega \dd z} \int_{M_{\mathrm{min}}}^{M_{\mathrm{max}}} \dd M \, n(M,z) \abs{\tilde{y}_\ell (M,z)}^2 \ , \label{tsz1halo}\\
    C_\ell^{2\mathrm{h}} &= \int_{z_{\mathrm{min}}}^{z_{\mathrm{max}}} \dd z \, \frac{\dd^2V}{\dd \Omega \dd z} \left [\int_{M_{\mathrm{min}}}^{M_{\mathrm{max}}} \dd M \, n(M,z) \tilde{y}_\ell (M,z) \, B(M,z) \right ]^2 P_{\mathrm{m}} \left(k,z \right) \ . \label{tsz2halo}
\end{align}
Here $V$ is the comoving volume, $n(M,z)$ is the HMF, $P_{\mathrm{m}}(k = (\ell+1/2)/d_{\mathrm{C}},z)$ is the matter linear power spectrum and $B(M,z)$ the linear halo bias.
The profile $\tilde{y}_\ell (M,z)$  is the 2-dimensional Fourier transform of the Compton parameter
\begin{equation}\label{yFourier}
    \tilde{y}_\ell (M,z) = \frac{4\pi r_{\mathrm{c}}}{\ell_{\mathrm{c}}^2} \frac{\sigma_{\mathrm{T}}}{m_{\mathrm{e}}c^2} \int_0^{\infty} \dd s \, s^2 P_{\mathrm{e}} (s,M,z) \frac{\sin \left( \ell s / \ell_{\mathrm{c}} \right)}{\left( \ell s / \ell_{\mathrm{c}} \right)} \ ,
\end{equation}
in which $r_{\mathrm{c}}$ is the scale radius of the pressure profile $P_{\mathrm{e}}$, $s = r / r_{\mathrm{c}}$ and $\ell_{\mathrm{c}} = d_{\mathrm{A}}/r_{\mathrm{c}}$.

\begin{figure}[!ht]
\centering
\begin{tabular}{ll}
\subfloat{\includegraphics[width=0.37\textwidth]{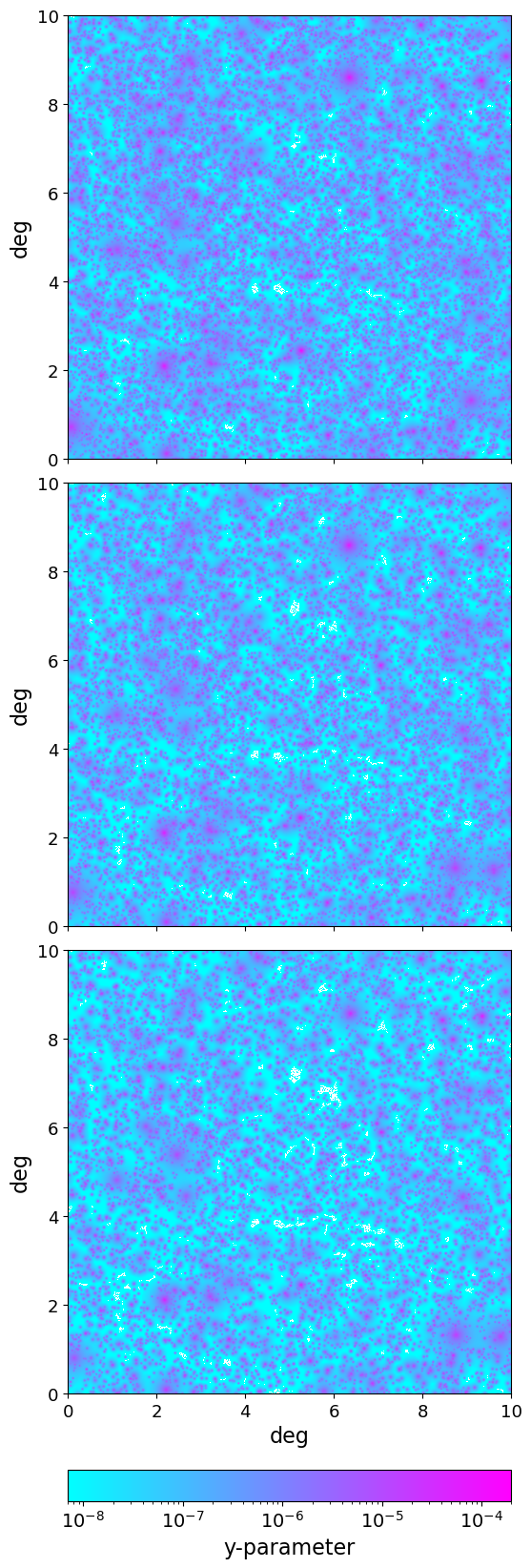}} 
\subfloat{\includegraphics[width=0.37\textwidth]{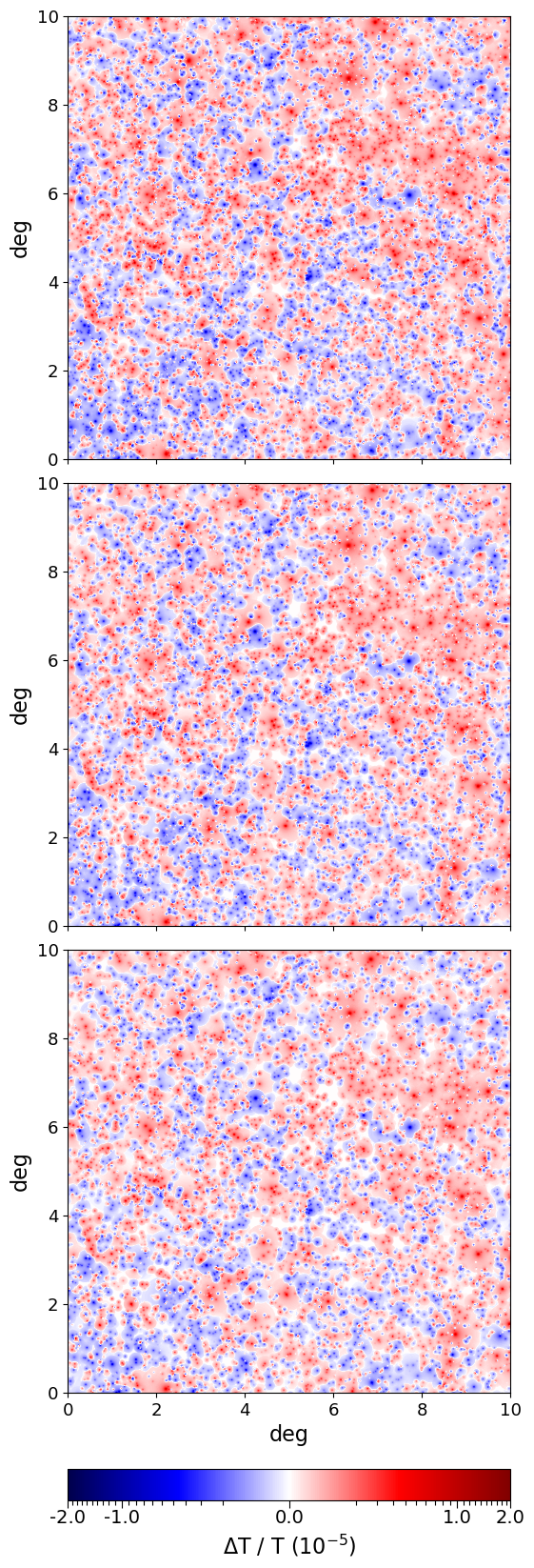}}
\end{tabular}
\caption{Example of three synthetic maps of the tSZ Compton-$y$ parameter (left column) and of the relative kSZ temperature difference (right column, described in~\ref{ksz}) for the \texttt{DEMNUni} $\nu\Lambda$CDM models, with $M_\nu=0,\,0.16,\,0.32$ eV increasing from top to bottom (see Table~\ref{tab:demnuni_sets_table}). In the maps it is possible to observe many massive haloes, with $y$ values up to $10^{-4}$ and $\Delta T^{\mathrm{kSZ}} / T$ reaching approximately $\pm 10^{-5}$. 
While the analyses described in the paper are performed using full-sky maps, here we only show a patch of $10^{\circ} \times 10 ^{\circ}$ selected near the equator, to avoid the visual deformation at the poles induced by the CAR pixelisation scheme.}

\label{multiFigMaps}
\end{figure}

\subsection{The map-making process}\label{mapmaking}
Once defined the model describing the pressure of the electrons, we apply it to the 15 halo lightcones deriving from the \texttt{DEMNUni} set (see Section~\ref{demnUni}). 
For each halo we integrate the relevant quantity along the lines of sight within a radius of $4 R_{200}$ from its centre, to obtain estimates for the Compton-$y$. Summing then the contributions from all haloes in each sky direction allows us to create synthetic tSZ maps.
For this purpose we use the code \texttt{XGPaint}\footnote{\url{https://websky-cita.github.io/XGPaint.jl/stable/}}, which was specifically developed to paint emission of extragalactic foregrounds on dark matter haloes. As our baseline choice we produce our maps in the CAR pixelisation scheme~\cite{carWCS}.
We choose to generate, for each of the two effects and each of the simulations, two types of map:
\begin{itemize}
    \item a full-sky, lower resolution (with a pixel size of $0.5 ^{\circ}$) one, for the one-point statistics;
    \item a full-sky, higher resolution (with a pixel size of $0.5^\prime$ ) one, for the computation of the power spectrum.
\end{itemize}
We show examples of smaller maps in Figure~\ref{multiFigMaps}, of both the SZ effects (see Appendix~\ref{sec::kszResults} for a description of our modelling of the kSZ limited to DM haloes). It is possible to see how the presence of massive neutrinos lowers the signal for the two effects, but more markedly for the tSZ. Nonetheless, the formed structures appearing in the different maps resemble each other, highlighting the specific imprint of free streaming.

\section{Analysis of the tSZ effect}\label{sec::tszResults}
\subsection{One-point statistics}
\begin{figure}[!ht]
    \centering
    \includegraphics[width=0.6\textwidth]{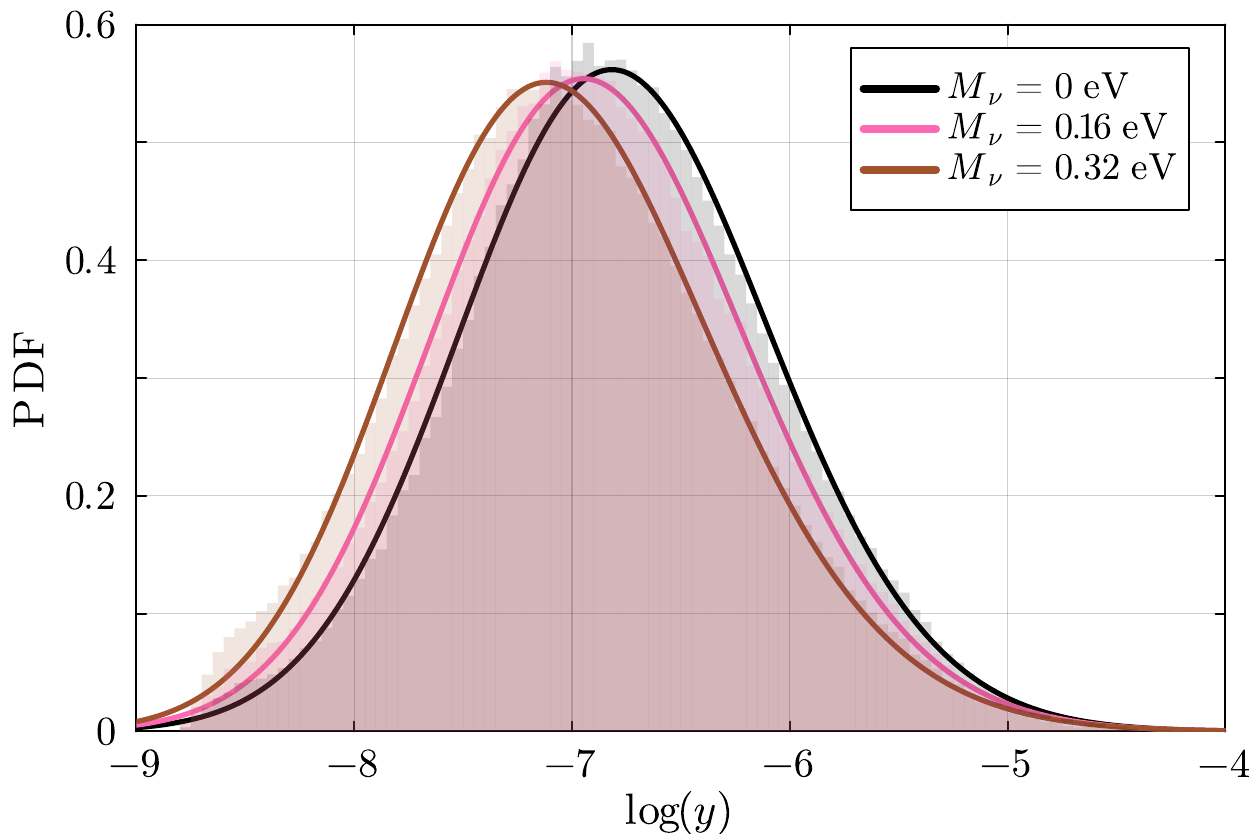}
    \caption{Normalised distribution of the Compton parameter for the three values of $M_\nu=0,\,0.16,\,0.32$ eV in the \texttt{DEMNUni} $\nu\Lambda$CDM models (see Table~\ref{tab:demnuni_sets_table}). The best-fitting skewed Gaussian distributions are also shown as solid lines.}
    \label{pdfFit}
\end{figure}

For the analysis of the one-dimensional pixel distribution of the tSZ effect we choose to work with $\log (y)$ rather than $y$ itself, as the ensemble of values spans different orders of magnitude, approximately from $10^{-9}$ to $10^{-4}$. Before working in logarithmic scale every null pixel in the map was removed: these are associated to a lack of contribution in the region by haloes, implying that the line of sight associated to the pixel is not within a distance of $4 R_{200}$ from the centre of any structure, and are present in our mock maps due to the absence of a modelling for the diffuse gas. These $y=0$ pixels amount to around $3\%$ to $10 \%$ of the map, depending on the chosen simulation. The resulting $\log (y)$ distributions can be fitted with a skewed Gaussian~\cite{azzaliniSkew}:
\begin{equation}\label{skewNorm1}
    P (x) = \frac{2}{\sigma \sqrt{2\pi}} \mathrm{e}^{-\frac{(x-\mu)^2}{2\sigma^2}} \int_{-\infty}^{\alpha \, \frac{x-\mu}{\sigma}} \dd t \, \frac{1}{\sqrt{2\pi}} \mathrm{e}^{-\frac{t^2}{2}} \ ,
\end{equation}
where $\mu$ and $\sigma$ represent the standard Gaussian parameters, while $\alpha$ quantifies the skewness.
The results of the fitting for $\alpha$ and $\sigma$ are listed in Table~\ref{pdfTable}, while in Figure~\ref{pdfFit} we compare the actual distributions to the best-fitting curves, in the $\Lambda$ cosmological models.
In all cases an increase in neutrino mass, with fixed dark energy EoS, leads to larger values both for $\sigma$ and $\alpha$. The latter, in particular, is always positive, meaning that the distribution favours values larger than the one corresponding to the peak. We ascribe the behaviour of $\alpha$ with respect to $M_\nu$ to the fact that the bulk of the distribution, which results from an integral over the whole mass range considered, is highly affected by the HMF damping. On the other hand, the highest values of $y$ depend only on a small number of massive systems whose temperature and density are only slightly reduced by the increase of $M_\nu$, thus resulting in an increased skewness of the distribution.

\begin{table}[!t]
\begin{center}
\begin{tabular}{ | c l||   r |  r| }
\hline
    \multicolumn{2}{|c||}{Simulation} & $\alpha \ \, \,$ & $\sigma \ \, \,$ \\ 
\hline
    $M_{\nu} = 0$ eV & $w_0 = -1 \, , w_a = 0$ & 0.88& 0.84 \\ 
    
     & $w_0 = -0.9  \, , w_a = -0.3$ & 0.88& 0.84\\ 
    
    & $w_0 = -0.9  \, , w_a = +0.3$ &1.16 &0.92\\ 
    
    & $w_0 = -1.1  \, , w_a = -0.3$ & 0.78 & 0.80\\ 
    
    & $w_0 = -1.1  \, , w_a = +0.3$ &0.88 & 0.83\\ 
    \hline
     $M_{\nu} = 0.16$ eV & $w_0 = -1 \, , w_a = 0$ & 1.05& 0.89\\ 
    
     & $w_0 = -0.9  \, , w_a = -0.3$ & 1.06&0.89\\ 
    
    & $w_0 = -0.9  \, , w_a = +0.3$ & 1.40& 0.97\\ 
    
    & $w_0 = -1.1  \, , w_a = -0.3$ &0.91 &0.84\\ 
    
    & $w_0 = -1.1  \, , w_a = +0.3$ & 1.05 & 0.89\\ 
    \hline
 $M_{\nu} = 0.32$ eV & $w_0 = -1 \, , w_a = 0$ & 1.35 &0.96 \\ 
    
     & $w_0 = -0.9  \, , w_a = -0.3$ & 1.34 & 0.96\\ 
    
    & $w_0 = -0.9  \, , w_a = +0.3$ & 1.77 &1.04\\ 
    
    & $w_0 = -1.1  \, , w_a = -0.3$ & 1.15 & 0.91\\ 
    
    & $w_0 = -1.1  \, , w_a = +0.3$ & 1.33 & 0.96\\ 
    \hline
\end{tabular}
\caption[]{$\alpha$ and $\sigma$ parameters for the skewed Gaussian distribution (Equation~\ref{skewNorm1}) best-fit, for each of the 15 \texttt{DEMNUni} simulations.}\label{pdfTable} 
\end{center}
\end{table}

\begin{figure}[!ht]
    \centering
    \begin{subfigure}[l]{0.6\textwidth}
        \includegraphics[width=\linewidth]{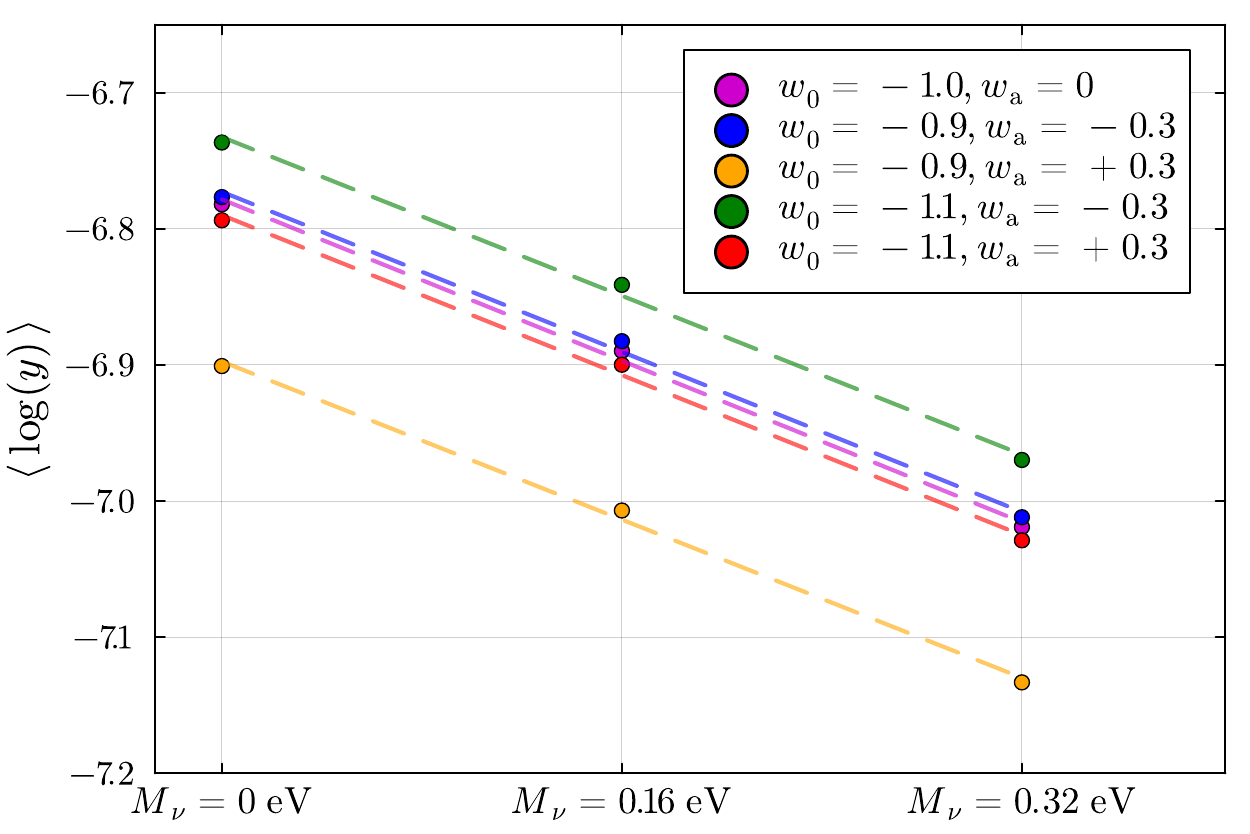}
    \end{subfigure}
    \begin{subfigure}[c]{0.35\textwidth}
        \begin{tabular}{ | c||   c |  c |}
\hline
    \multicolumn{1}{|c||}{$(w_0,w_a)$} & $A_1$ & $A_2$ \\ 
\hline
    $(-1,0)$ & $-16.69$ & $9.91$\\ 
    
    $(-0.9,-0.3)$ & $-16.60$ & $  9.83$\\ 
    
    $(-0.9,+0.3)$ & $-16.61$ & $  9.71$\\ 
    
    $(-1.1,-0.3)$ & $-16.48$ & $  9.75$\\ 
    
    $(-1.1,+0.3)$ & $-16.62$ & $9.83$\\
\hline
\end{tabular}
    \end{subfigure}
    % Main caption for both
    \caption{Mean $\log (y)$ values as a function of $M_\nu$ for all the \texttt{DEMNUni} cosmologies, together with the linear fit in Equation~\eqref{meanyLinFit} for each of the dark energy EoS (the associated parameters are listed in the table on the right panel) the $(w_0 = -0.9,w_a = +0.3)$ models, which are the ones where dark energy significantly slows the accretion of matter (see the right panel in Figure 1 of~\cite{verza_2022b}), the mean $\log (y)$ values are remarkably lower.}
    \label{meanLogyPlot}
\end{figure}

We then look more in detail at the mean $ \log (y)$ (shown in Figure~\ref{meanLogyPlot}) and its trend when varying $M_\nu$. As expected, for a fixed dark energy EoS, an increase in neutrino mass translates to a decrement in the mean logarithmic value. 
What is more interesting is that this trend is mostly linear and so it can be fitted as
\begin{equation}\label{meanyLinFit}
    \langle \, \log (y) \, \rangle = A_1 + A_2 (1-f_\nu) \ ,
\end{equation}
with $f_\nu \equiv \Omega_\nu / \Omega_{\mathrm{m}}$ being the neutrino mass fraction. The best-fit values of $A_1$,$A_2$ for the different dark energy EoS combinations are also presented in Figure~\ref{meanLogyPlot}, and correspond to a standard deviation of order $10^{-2}$ or less. Very interestingly, the observed reduction of approximately $10 f_\nu$ is analogous to the nonlinear damping in the matter power spectrum caused by massive neutrinos~\cite{Carbone_2016}. 

\subsection{Power spectrum}
\begin{figure}
    \centering
    \includegraphics[width=0.7\textwidth]{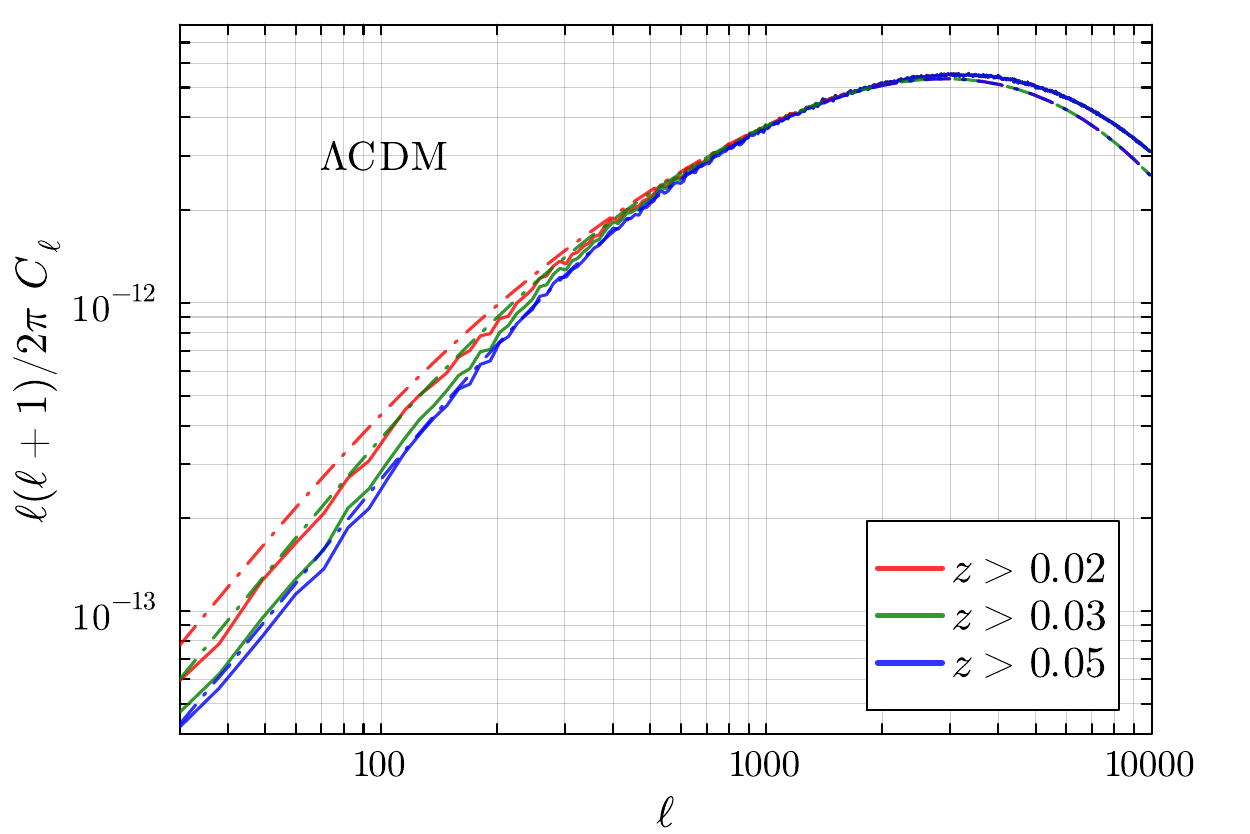}
    \caption{Comparison of the tSZ temperature power spectrum in the $\Lambda$CDM scenario between the \texttt{DEMNUni} measurements (solid lines) and the \texttt{CLASS}$_\mathrm{\texttt{SZ}}$ calculations (dash-dotted lines), with different low-$z$ cuts. We always show the power spectra of the tSZ for $\Delta T / T$ obtained in the low frequency limit where $g(x)\to-2$ (see Eq.~\eqref{deltaToverT}).}
    \label{psThermalComparisonClassSZ}
\end{figure}

We obtain the angular power spectra of Compton-$y$ by directly decomposing the signal over the spherical harmonics basis in the higher resolution ($0.5^\prime$) maps. The computation is affected by numerical error, which we can not smooth out averaging over more realisations because for each simulation only one lightcone is produced. We therefore apply a moving average method over 11 multipoles
\begin{equation}\label{eq::movingAvg}
    \langle C_\ell \rangle = \frac{1}{11} \sum_{\ell-5}^{\ell+5} C_\ell    \ ,
\end{equation}
to produce a smoothed version of all the power spectra to show in the figures. To first test our calculation we compare, in the $\Lambda$CDM model, the \texttt{DEMNUni} results and the halo model expectations from \texttt{CLASS}$_\mathrm{\texttt{SZ}}$~\cite{classSZ,bollietclass}. The latter is computed setting all the cosmological parameters equal to those of the \texttt{DEMNUni} simulations and using the Tinker halo mass function and bias~\cite{tinker2010}. We set the maximum redshift to $2.3$ as in our lightcones, as well as a minimum mass of $3.4\times  10^{13}h^{-1}  \msun$ for the integration in Equations~\eqref{tsz1halo} and~\eqref{tsz2halo}. 
The results of this comparison, for different low-$z$ cuts, are shown in Figure~\ref{psThermalComparisonClassSZ}. This comparison highlights differences between the two approaches both for $\ell\lesssim300$ and $\ell\gtrsim3000$: while for large scales the potential reason of this discrepancy is the single lightcone realisation coupled with the finite simulation volume\footnote{Which limits the power at large scales due to the replication of the same box.}, for the differences at the smallest scales (which at $\ell = 10^4$ consists of $20\%$ more power in our model compared to the halo model from \texttt{CLASS}$_\mathrm{\texttt{SZ}}$) we haven't found a reasonable explanation. Increasing the resolution does not solve the issue, as the effects of the pixelisation appear only for $\ell>10^4$; we find that only by selecting the whole mass sample (without the $3.4\times 10^{13}h^{-1}  \msun$ mass cut) the discrepancy disappears. In fact, both calculations in this scenario have an increased power for $l\gtrsim3000$, more marked for \texttt{CLASS}$_\mathrm{\texttt{SZ}}$ so that it matches the results for the \texttt{DEMNUni} lightcone, which, in contrast, shows only a slight increase. We stress, however, that the modelling of the pressure profiles is appropriate only for massive haloes (see Section~\ref{halocat}): we thus keep our initial halo mass cut.
We also decide to select only haloes at $z>0.05$, so to have a better statistical description of the phenomenon by avoiding local effects, but at the same time significantly reducing power at the larger scales.

\begin{figure}[!ht]
\begin{tabular}{lc} 
\subfloat{\includegraphics[width = 0.5\linewidth]{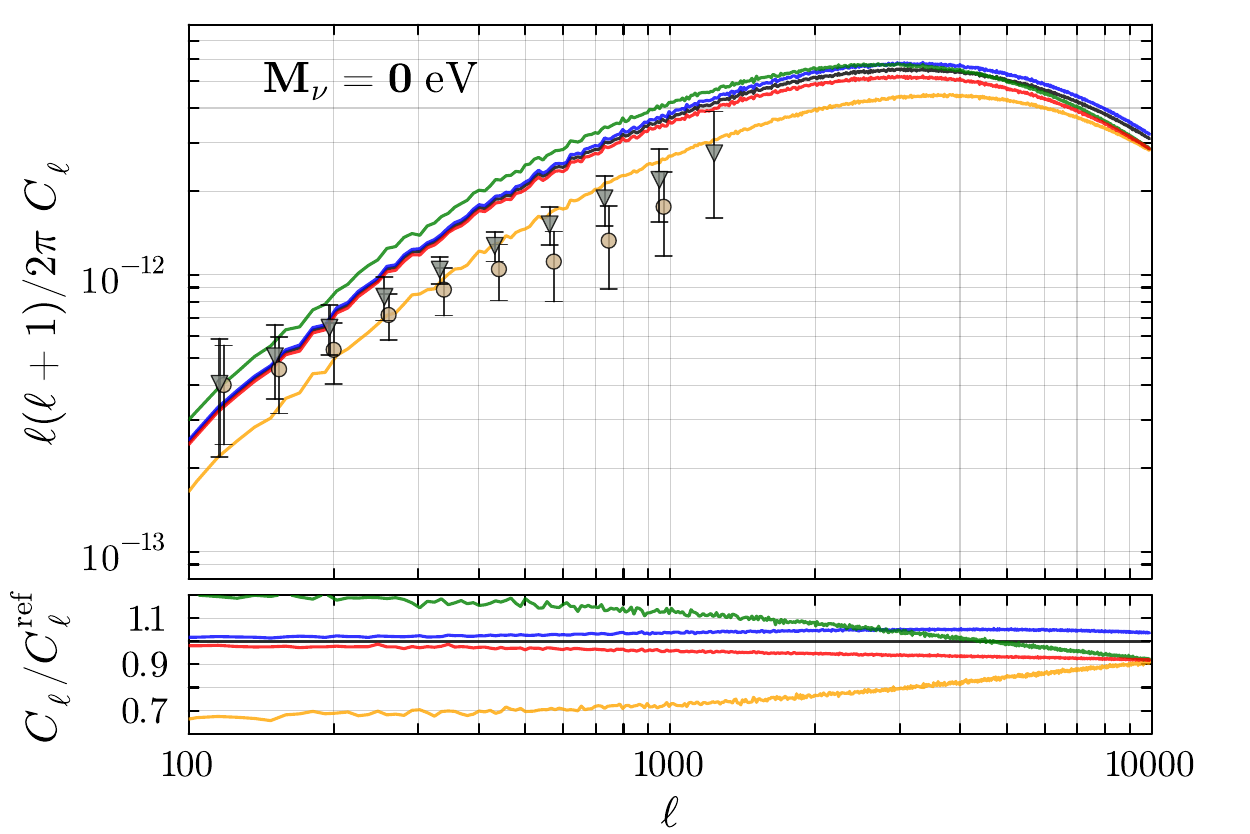}}
\subfloat{\includegraphics[width = 0.5\linewidth]{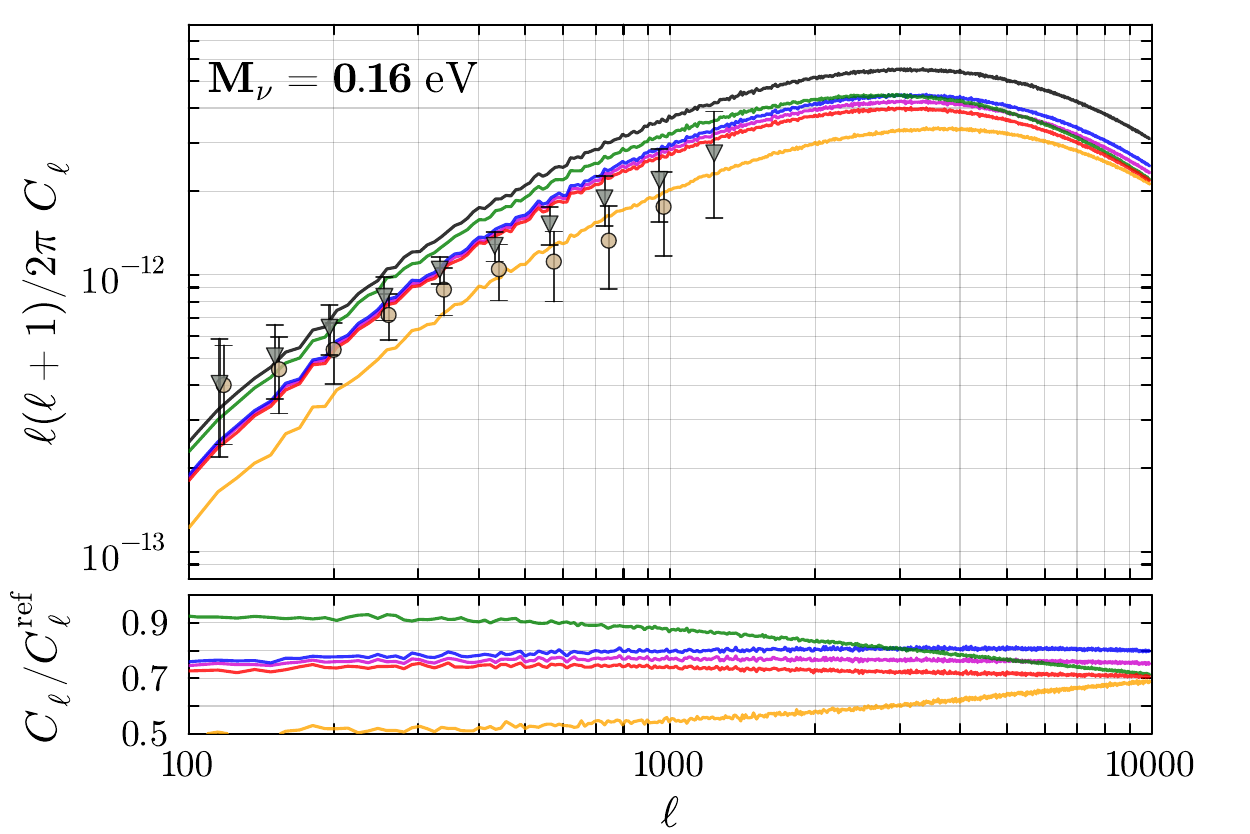}} \\
\subfloat{\includegraphics[width = 0.5\linewidth]{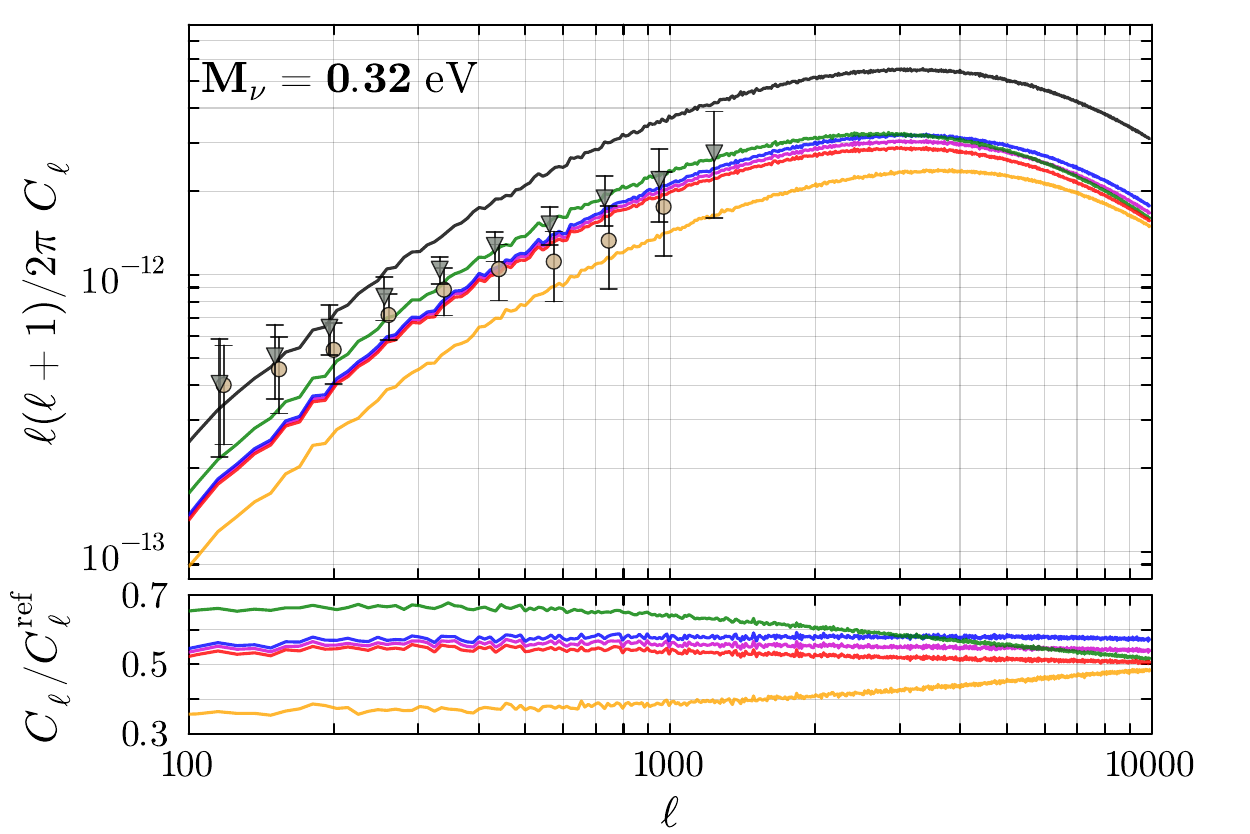}} \hspace{1.02cm}
\subfloat{\raisebox{0.54\height}{\includegraphics[width = 0.27\linewidth]{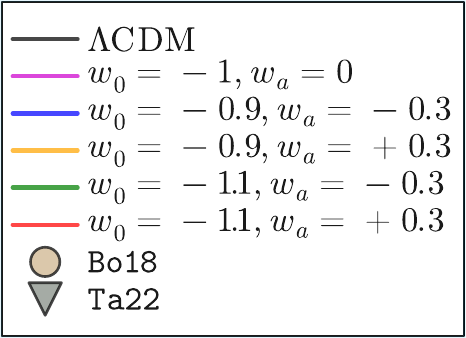}}}
\end{tabular}

\caption{Temperature power spectra for the tSZ effect measured from the 15 \texttt{DEMNUni} simulations for $z>0.05$, compared with the Planck data from~\cite{bolliet2018,tanimura2022} (plotted with a $1\%$ horizontal offset for a better visualisation). The ratio of each power spectrum with respect to the reference $\Lambda$CDM simulation is also present in the sub-panels.}
\label{multiFigPS}
\end{figure}
\begin{table}
\begin{center}

\begin{tabular}{ | c l||  r |r| }
\hline
    \multicolumn{2}{|c||}{\multirow{2}{*}{Simulation}} & \multicolumn{2}{c|}{$\chi^2$}  \\ 
    \cline{3-4}
    \multicolumn{2}{|c||}{} & \texttt{Bo18} & \texttt{Ta22}  \\ 
\hline
    $M_{\nu} = 0$ eV & $w_0 = -1 \, , w_a = 0$ &$10.98$ &$6.93 $ \\ 
    
     & $w_0 = -0.9  \, , w_a = -0.3$ &$12.19$ &$8.03$\\ 
    
    & $w_0 = -0.9  \, , w_a = +0.3$ &$^{\dagger}1.89$ &$^{\dagger}0.28 $\\ 
    
    & $w_0 = -1.1  \, , w_a = -0.3$ & $18.50$ &$14.27$\\ 
    
    & $w_0 = -1.1  \, , w_a = +0.3$ &$9.56$ &$5.69 $\\ 
    \hline
     $M_{\nu} = 0.16$ eV & $w_0 = -1 \, , w_a = 0$ & $3.17$ &$0.76 $\\ 
    
     & $w_0 = -0.9  \, , w_a = -0.3$ & $3.73$ &$1.08 $\\ 
    
    & $w_0 = -0.9  \, , w_a = +0.3$ & $^{\dagger}0.33$ &$2.00 $\\ 
    
    & $w_0 = -1.1  \, , w_a = -0.3$ &$7.09$ &$3.61 $\\ 
    
    & $w_0 = -1.1  \, , w_a = +0.3$ & $2.62$ &$^{\dagger}0.49 $\\ 
    \hline
 $M_{\nu} = 0.32$ eV & $w_0 = -1 \, , w_a = 0$ & $0.33$ &$1.22 $\\ 
    
     & $w_0 = -0.9  \, , w_a = -0.3$ & $0.40$ &$0.98 $\\ 
    
    & $w_0 = -0.9  \, , w_a = +0.3$ & $1.52$ &$7.52 $\\ 
    
    & $w_0 = -1.1  \, , w_a = -0.3$ & $1.12$ &$^{\dagger}0.18 $\\ 
    
    & $w_0 = -1.1  \, , w_a = +0.3$ & $^{\dagger}0.27$ &$1.51 $\\ 
    \hline
\end{tabular}
\caption{Reduced $\chi^2$ for the tSZ temperature power spectra measured from the \texttt{DEMNUni} simulations at $\tilde{\ell}>300$, with data from~\cite{bolliet2018,tanimura2022}. Daggers indicate the best-fitting, lower $\chi^2$ values, for each  dataset and $M_\nu$.}\label{chiSquared}
\end{center}
\end{table}
 The power spectra of all the simulations, computed with the previous specifications, are shown in Figure~\ref{multiFigPS}. 
It is possible to see how both a dynamical dark energy and massive neutrinos greatly impact the outcome. Their effect at first order is similar, almost degenerate, resulting in a rescaling of the tSZ power. While this is true at all scales of interest for massive neutrinos, a dynamical dark energy can impact differently the $\ell\gtrsim2000$ region, thus slightly altering the \textit{shape} of the spectrum. In this region the power spectra (at fixed $M_\nu$) with $(w_0,w_a) = (-0.9, +0.3), (-1.1, -0.3), (-1.1, +0.3)$ ultimately tend to converge at $\ell=10^4$. Indeed, for each neutrino mass the range covered by the power spectra in all the possible dark energy EoS is larger at $\ell=100$  than at $\ell=10^4$ by approximately $60, 63, 56 \%$, respectively for $M_\nu=0,0.16, 0.32$ eV. At fixed $M_\nu$, in the $\ell \lesssim 2000$ range there is little difference between the $(w_0,w_a) = (-1, 0), (-0.9, -0.3), (-1.1, +0.3)$ models, with instead the $(w_0,w_a) = (-0.9, +0.3), (-1.1, -0.3)$ curves remaining respectively lower and higher. Simulations with massless neutrinos mainly overestimate the data from~\cite{bolliet2018} and~\cite{tanimura2022}, denominated respectively \texttt{Bo18} and \texttt{Ta22}, which have been derived analysing Planck Compton-$y$ maps. To better quantify the adherence of each cosmological model with the data we calculate a reduced $\chi^2$ as
\begin{equation}\label{chiSQeq}
    \chi^2 = \frac{1}{N_{\mathrm{points}}}\sum_{\tilde{\ell}} \frac{\left(C_{\tilde{\ell}} - C_{\tilde{\ell}}^{\mathrm{data}}\right)^2}{ \sigma_{\tilde{\ell}} ^2} \ ,
\end{equation}
for $\tilde{\ell}>300$ modes, where $\tilde{\ell}$ indicates the multipole of each data point and $C_{\tilde{\ell}}$ is calculated as in~\eqref{eq::movingAvg}. Selecting only $\tilde{\ell}>300$ ensures we avoid the region affected by our cut at $z=0.05$. We assume the data points to be independent of each other, making use of a Gaussian-only covariance. This assumption leads to an incorrect $\chi^2$ value due to the neglect of the full covariance matrix, but nonetheless proves useful to compare which models are closer to the binned data points.
The results are listed in Table~\ref{chiSquared} and show that, depending on the dataset, the simulations best suited for a description of the data are very different, but in almost all cases are among the ones with massive neutrinos.

\begin{figure}[!ht]
\begin{tabular}{lc} 
\subfloat{\includegraphics[width = 0.5\linewidth]{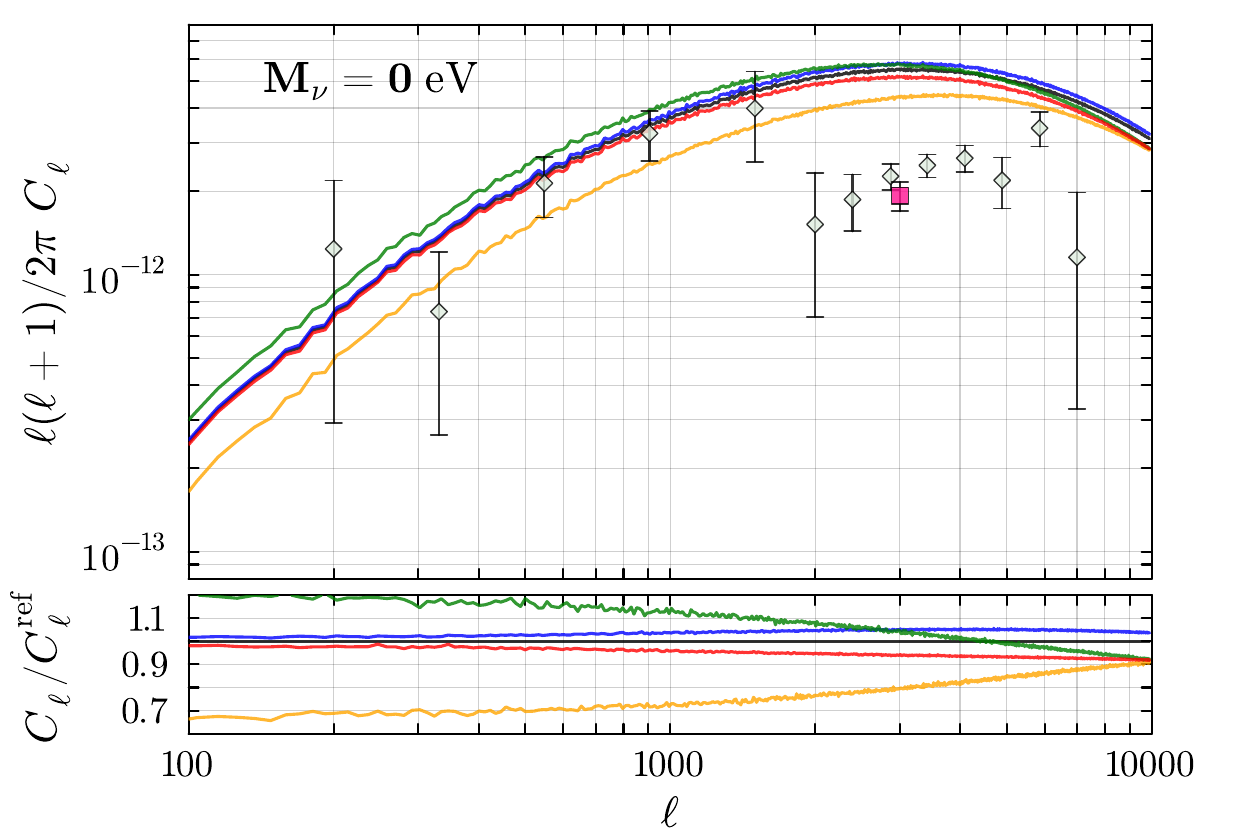}}
\subfloat{\includegraphics[width = 0.5\linewidth]{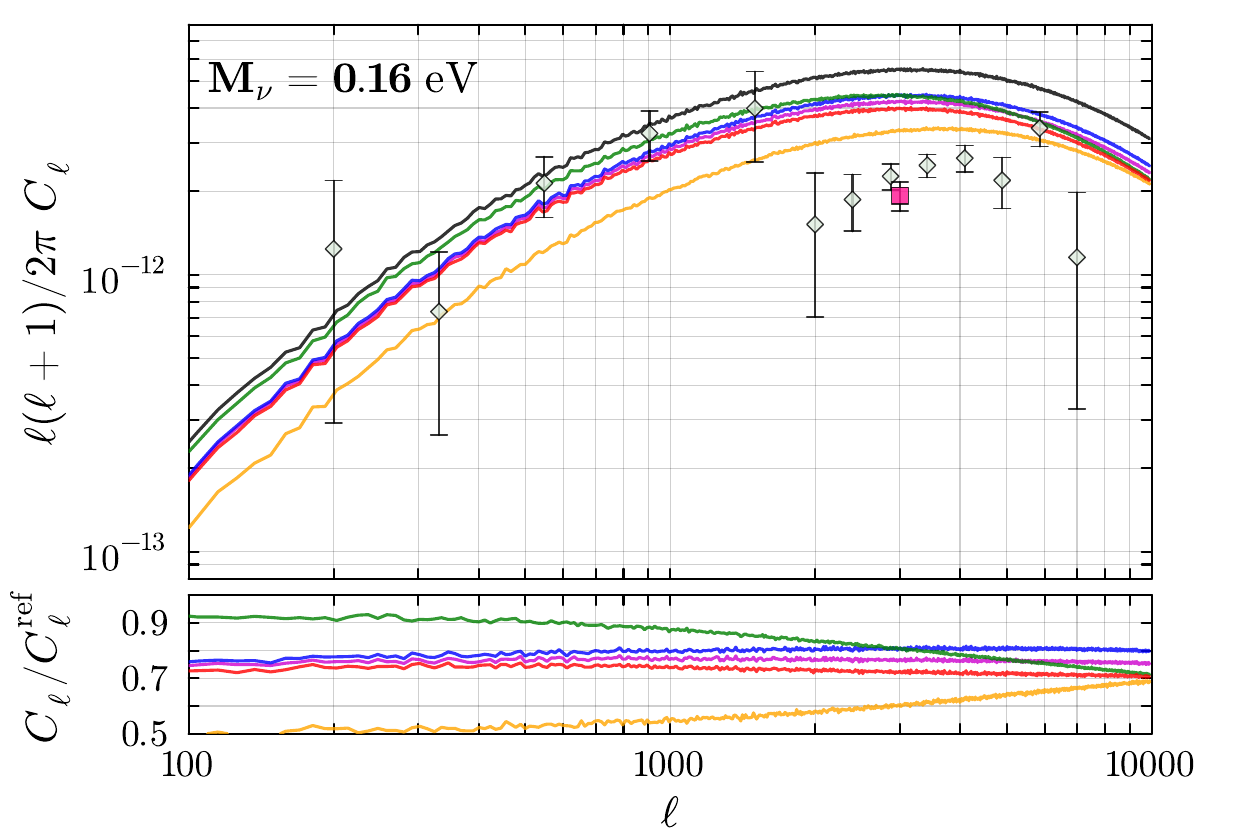}} \\
\subfloat{\includegraphics[width = 0.5\linewidth]{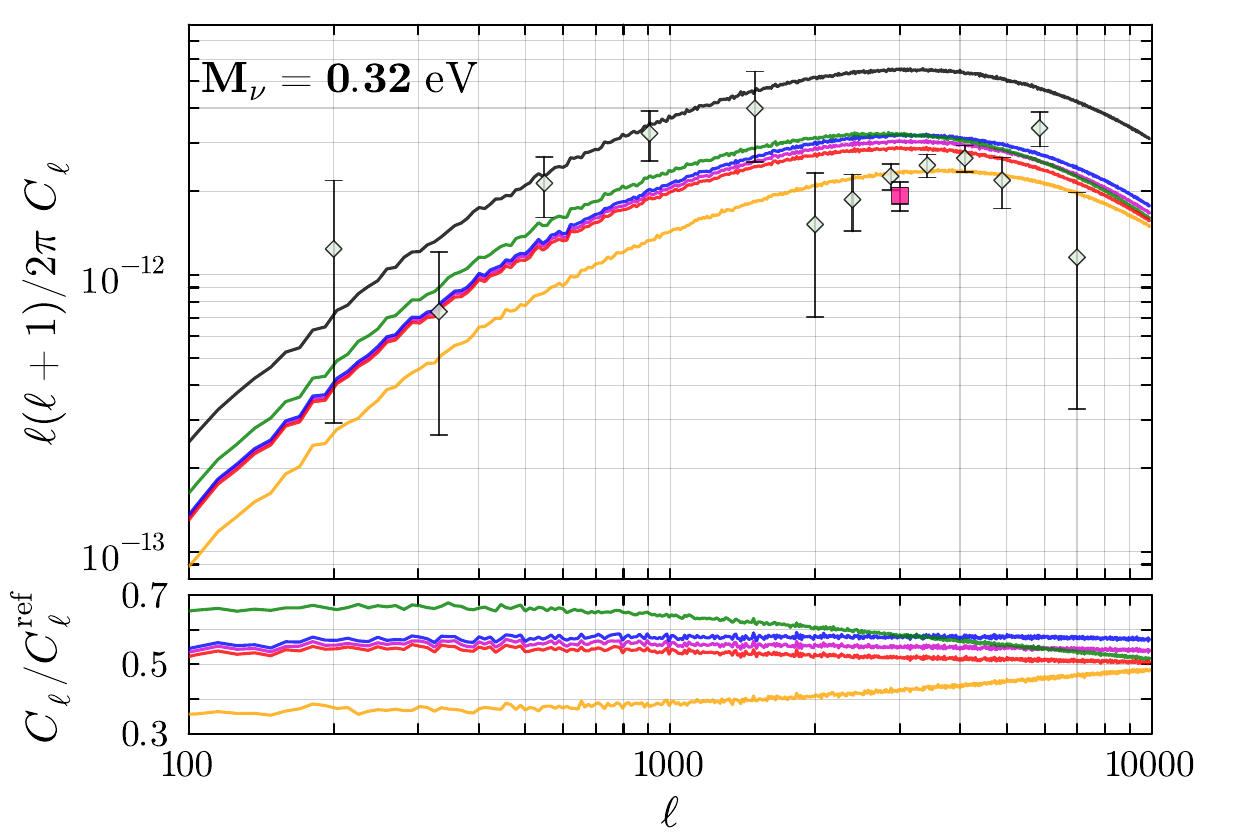}} \hspace{1.02cm}
\subfloat{\raisebox{0.54\height}{\includegraphics[width = 0.27\linewidth]{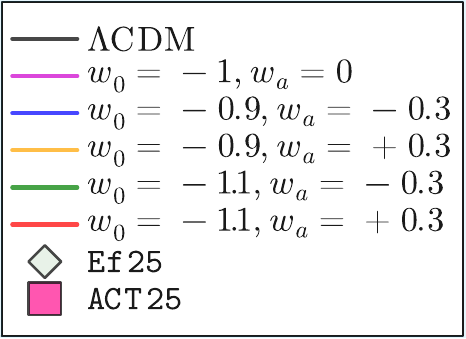}}}
\end{tabular}

\caption{Same as Figure~\ref{multiFigPS}, but here compared to Refs.~\cite{eftsathiou2025,act2025ps}. For \texttt{ACT25} we show only the single amplitude inferred, at $\ell=3000$.}
\label{multiFigPS_newdata}
\end{figure}

Additional data of the tSZ effect were recently published in the independent analyses of~\cite{eftsathiou2025} and~\cite{act2025ps}, which we denote respectively as \texttt{Ef25} and \texttt{ACT25}. In \texttt{Ef25} the authors made use of temperature maps from Planck, ACT and SPT at 100 GHz, and here we report the results of their template-free analysis. Instead, in \texttt{ACT25} the tSZ power was constrained as a part of the multi-frequency CMB inference, assuming a parametrisation where a baseline template is allowed to vary in amplitude and shape. In Figure~\ref{multiFigPS_newdata} we compare \texttt{Ef25} and \texttt{ACT25} with the results from the \texttt{DEMNUni} simulations. Both the datasets are characterised by a flatter shape than usual, with a significant increase in power at large scales, making it difficult to describe them with any of the \texttt{DEMNUni} maps. Additionally, the discontinuity found in \texttt{Ef25} around $\ell =2000$ is hard to recover with any physically motivated model.
In both these works the results for the tSZ power spectrum are quoted to be consistent with astrophysical models with an enhanced baryonic feedback.
Motivated by these more recent results, and by the differential cosmic history imprinted by a dynamical dark energy which leads to an altered HMF, we investigate quantitatively the shape variation of the tSZ spectrum in the \texttt{DEMNUni} set, for $\ell>300$. We thus fit each of the 14 non-$\Lambda$CDM models with a functional form analogous to that used in \texttt{ACT25}:
\begin{equation}\label{eq::ACT-like_parametrisation}
    C_\ell = C_\ell^{\rm{ref}} A \, \left( \frac{\ell}{\ell_0}\right)^B \, ,
\end{equation}
where $C_\ell^{\rm{ref}}$ is our $\Lambda$CDM result, $A$ and $B$ are the parameters which alter the amplitude and shape of the power spectrum, while $\ell_0 = 3000$.
In Figure~\ref{multiFigPS_newdata} we show the residuals between the fit and the outputs from the simulated maps, which indicate that the functional form of Equation~\eqref{eq::ACT-like_parametrisation} constitutes a fairly good approximation for the \texttt{DEMNUni} models, being able to describe the different spectra with a discrepancy of less than $10\%$ for $\ell > 300$ ($5\%$ for $\ell>500$). We find that the models with $(w_0 ,w_a) = (-0.9,+0.3), (-1.1,-0.3)$ are the ones with a shape that differs the most from $\Lambda$CDM. Indeed, regardless of $M_\nu$, the fit for these cosmologies presents respectively $B\approx+0.1$ and $B\approx-0.1$; the latter highlights a non-negligible increase in the share of power at large scales, that is nonetheless far from being compatible with the value of $B=-0.6$ found in \texttt{ACT25}. Our analysis shows that, when considering current constraints on the values of $w_0$ and $w_a$, dynamical dark energy can not justify the observed shape of the auto spectrum of the tSZ effect.

\begin{figure}[!ht]
    \centering
    \includegraphics[width=0.7\textwidth]{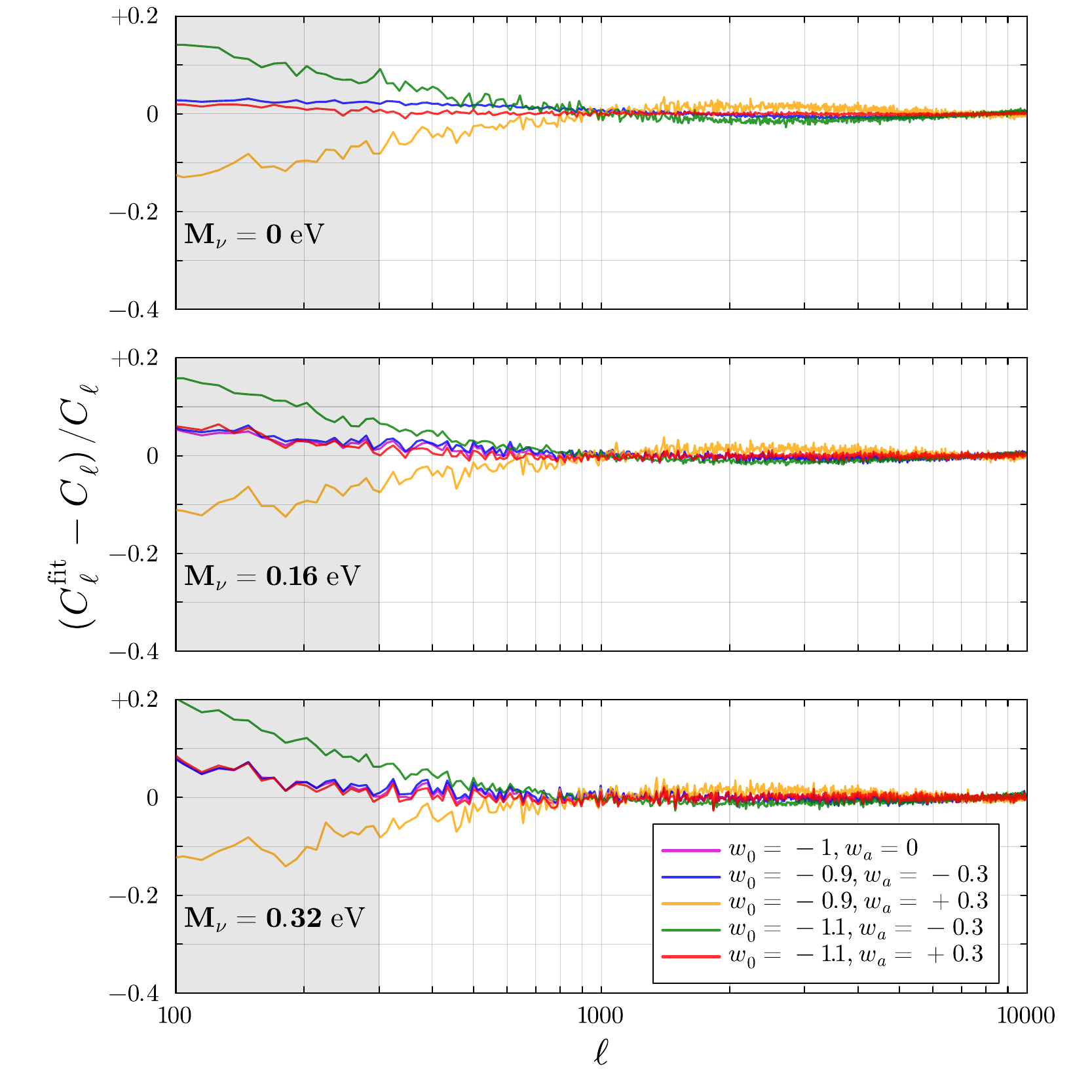}
    \caption{Residuals of the comparison between the best-fit curves of Equation~\eqref{eq::ACT-like_parametrisation} and the direct measurements from the \texttt{DEMNUni} synthetic maps. The grey area represents the region where the fit is extrapolated, as it was performed considering $\ell>300$. }
    \label{fig::residuals_ACTlaw}
\end{figure}

\begin{figure}[!ht]
    \centering
    \subfloat{\includegraphics[width = 0.5\linewidth]{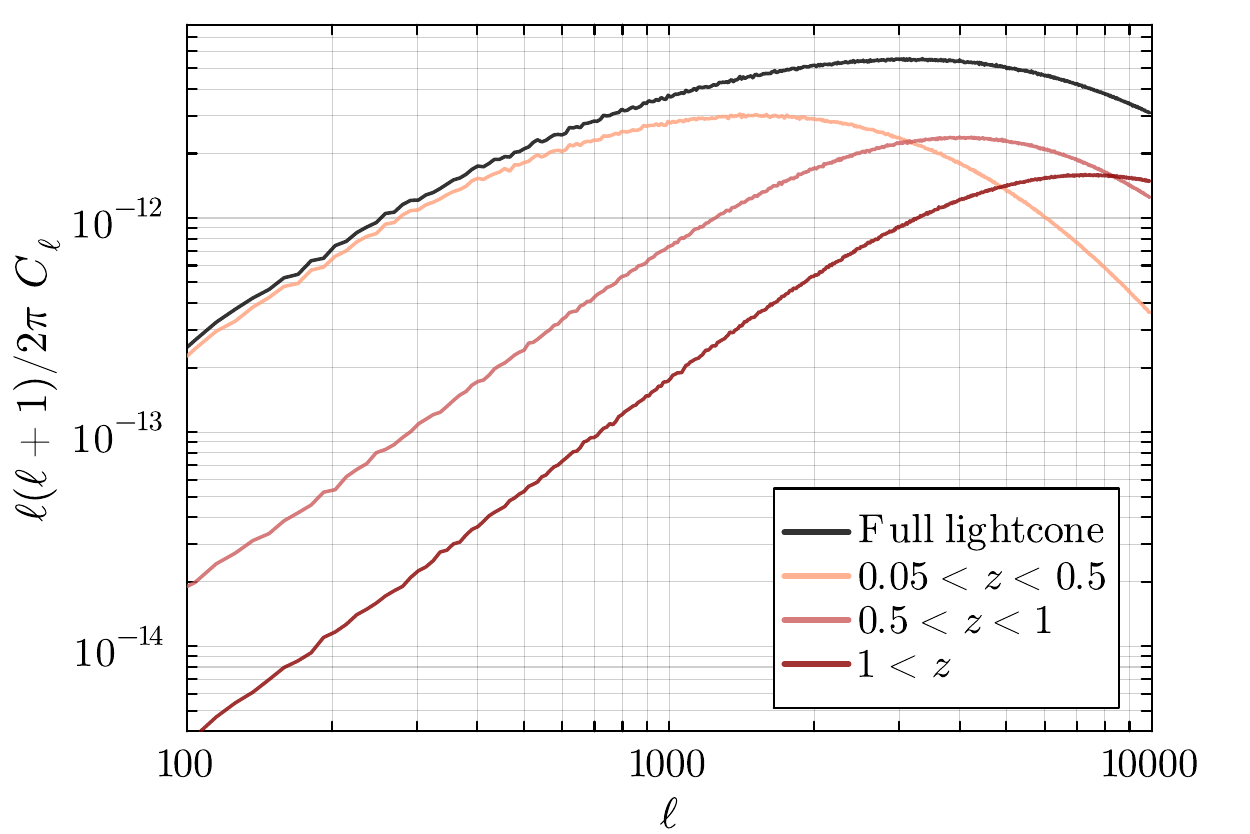}}
    \subfloat{\includegraphics[width = 0.5\linewidth]{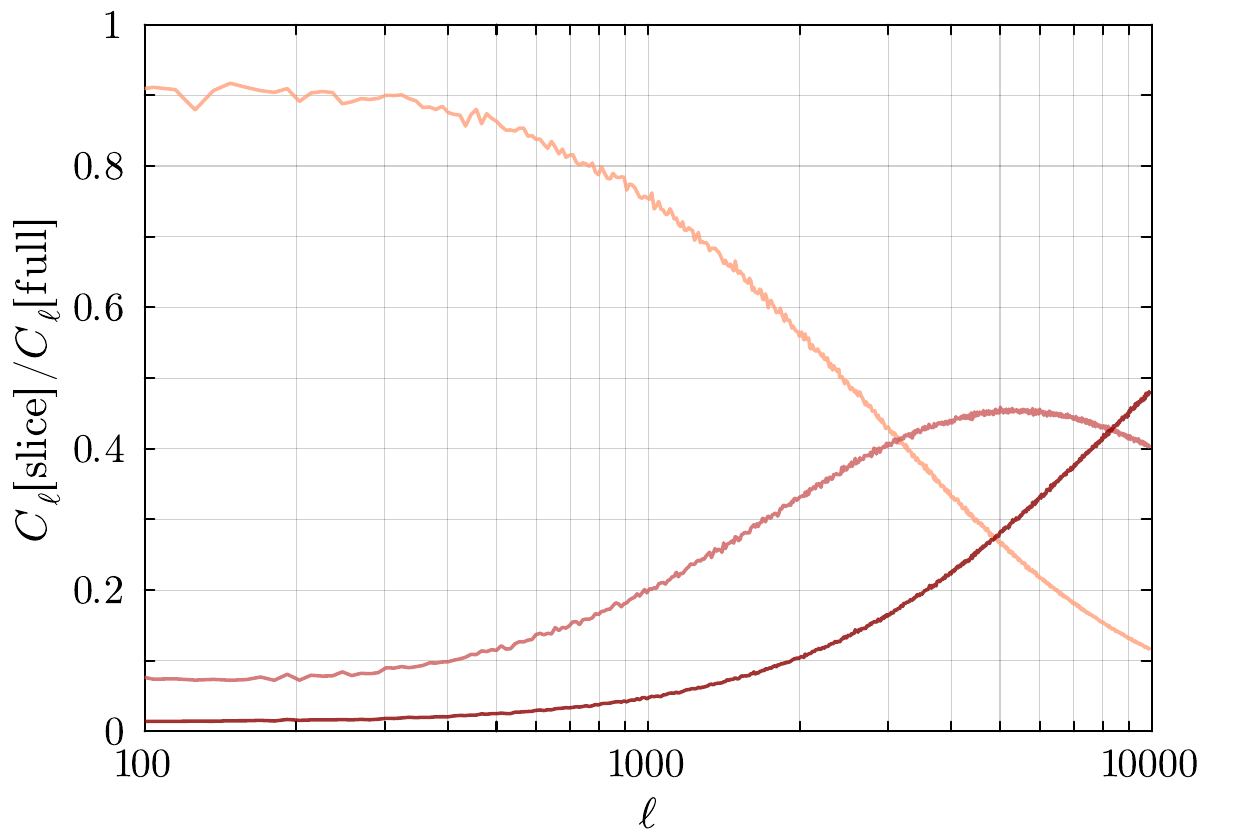}}\\ 
    \caption{tSZ temperature power spectra for three different redshift slices, compared to the full lightcone measurements, in the \texttt{DEMNUni} $\Lambda$CDM simulation. The behaviour in the other 14 simulations is analogous to the $\Lambda$CDM one, with no major differences.}
    \label{lcdmRedshiftShare}
\end{figure}

 Furthermore, we explore the share of power for the \texttt{DEMNUni} outputs in different redshift intervals, which we choose to be $0.05<z_1<0.5$, $0.5<z_2<1$ and $z_3>1$. This is done to investigate a potential different redshift dependence between the 15 scenarios, mainly due to the variability in the EoS for dark energy, which is \textit{not} observed. In fact, all the models show the same trend: up to $\ell=3000$ the dominant slice is $z_1$, while at $\ell\approx 10^4$ $z_2$ and $z_3$ each amount to $40$--$50$\% of the power of the whole redshift sample (see Figure~\ref{lcdmRedshiftShare}).

 We also investigate the scaling of the power spectrum, in the interval $300<\ell<1000$, with respect to $\sigma_8 (z=0)$, calculated via \texttt{CAMB}~\cite{cambpaper,cambsoft}; typical values reported in the literature are $C_\ell \propto \sigma_8^{7\text{--}9}$~\cite{KomatsuSeljScaling,ShawScaling,TracScaling,bolliet2020,planck2015}. The authors of~\cite{bolliet2020} stressed the importance of the  ‘cb' (Cold DM and Baryons) prescription when in presence of massive neutrinos~\cite{cbcastorina,cbLoverde}, that is to say using the same parametrisation of the HMF~\cite{tinker2008} as in massless neutrino cosmological models
\begin{equation}\label{eq::hmfPar}
    n(M,z) = \frac{\rho_\mathrm{m}}{M} f (\sigma,z) \frac{\mathrm{d}\ln{\sigma^{-1}}}{\mathrm{d}\ln{M}} \ ,
\end{equation}
but with the density and the matter power spectrum variance evaluated only for baryons and CDM, so $\rho_{\mathrm{cb}} = \rho_{\mathrm{m}} - \rho_{\nu} $ and $\sigma^{\mathrm{cb}}_M(z)$. This prescription reflects the effects of neutrino free-streaming and should allow precise estimates of $n(M,z)$ by calibrating $f(\sigma,z)$ purely in massless neutrinos simulations.
We accordingly focus on the ‘cb' quantities. Fixing the neutrino total mass and allowing $w_0, w_a$ to vary, we find that $\sigma_8^{\mathrm{cb}}$ is not capable of tracing the observed variation in the power spectrum. As an example, for better clarity: the $(w_0=-0.9, w_a=-0.3)$ simulations show more power when compared to the cosmological constant scenarios (see again Figure~\ref{multiFigPS}) but exhibit smaller values of $\sigma_8^{\mathrm{cb}}$. Conversely, we can instead study the scaling of the power spectrum with this parameter
\begin{equation}\label{scalingSigma8cb}
    C_\ell \propto (\sigma_8^{\mathrm{cb}})^q \ 
\end{equation}
in a fixed dark energy scenario, allowing the neutrino mass to vary\footnote{This parametrization does not consider the dependence on $\Omega_{\mathrm{cb}}$, however this effect is negligible since $\Omega_{\mathrm{cb}}$ varies by at most $3\%$ in our simulation set and the tSZ power spectrum is expected to scale linearly on this parameter~\cite{sabyr2025_minkowski}.}.
 
This results in 5 values for $q$ presented in Table~\ref{scalingtSZtable}. Even though our simulation-based scaling is obtained by fitting only a limited number of curves (and also neglecting the small $\Omega_{\mathrm{cb}}$ contribution), the $q$ values we find differ from those in~\cite{bolliet2018,bolliet2020}. We therefore suggest that future work should take care in assuming a fixed scaling relation with the cosmological parameters, having shown in the case of $\sigma_8^{\mathrm{cb}}$ how the relation itself is dependent on the pressure profile choice and on the nature of dark energy.

\begin{table}
\begin{center}

\begin{tabular}{ | c||   c |  }
\hline
    $(w_0,w_a)$ & $q$ \\
\hline
    $(-1,0)$ & $7.5$ \\ 
    
    $(-0.9,-0.3)$ & $7.5$\\ 
    
    $(-0.9,+0.3)$ & $8.1$\\ 
    
    $(-1.1,-0.3)$ & $7.3$ \\ 
    
    $(-1.1,+0.3)$ & $7.5$\\
\hline
\end{tabular}
\caption{Power-law exponents for the scaling of the tSZ temperature power spectrum with respect to $\sigma_8^{\mathrm{cb}}$ alone, obtained keeping fixed dark energy EoS while varying $M_\nu$.} \label{scalingtSZtable}
\end{center}
\end{table}

\subsection{Forecasts of the Signal-To-Noise ratio}\label{subsec::forecasts}
\begin{figure}[!ht]
    \centering
    \includegraphics[width=0.6\textwidth]{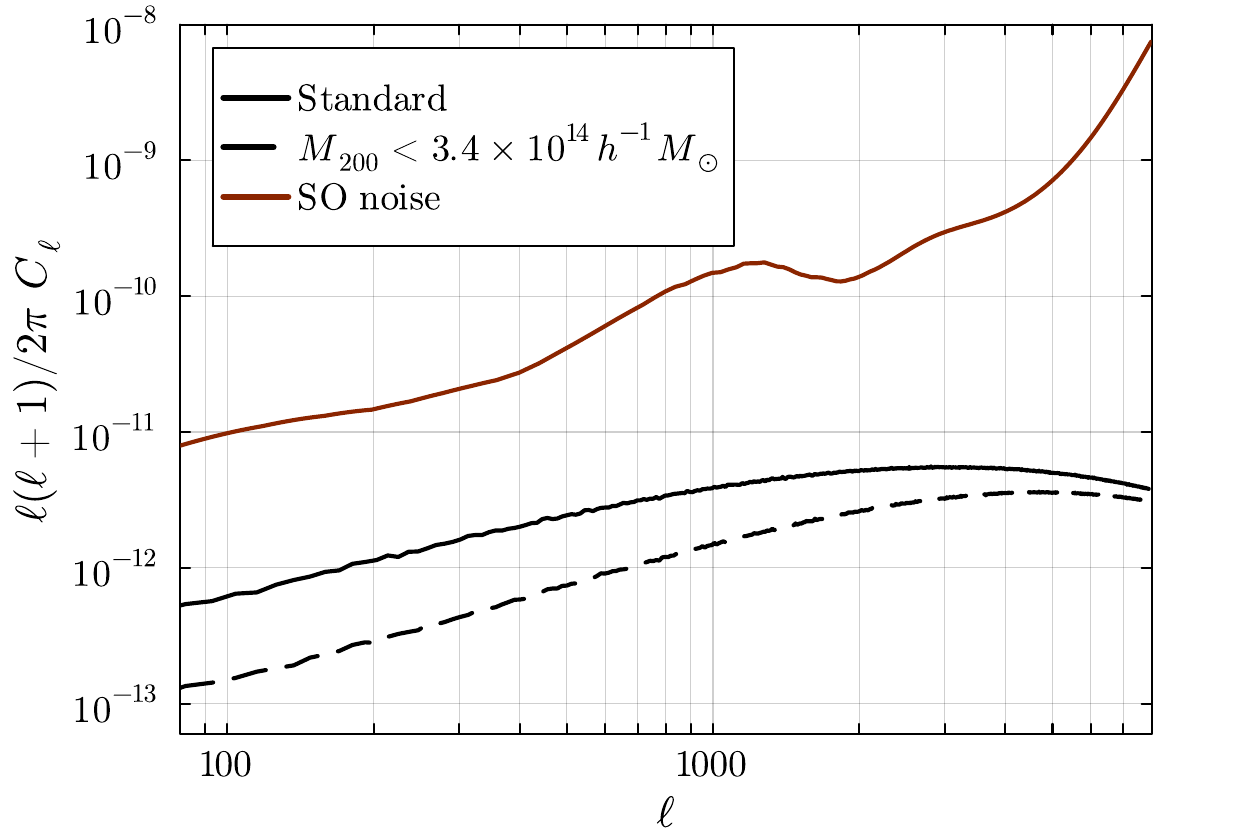}
    \caption{Simons Observatory LAT noise model and \texttt{DEMNUni} $\Lambda$CDM tSZ power spectra compared. For the latter there are both our standard calculation ($M_{200}>3.4\times10^{13}h^{-1}\msun$) and the one adopted in Section~\ref{subsec::forecasts} ($3.4\times10^{13}h^{-1}\msun<M_{200}<3.4\times 10^{14} h^{-1}\msun$). The noise is the same as the dotted orange line in Figure 36 of~\cite{simonsObs}, here shown in the low frequency limit.}
    \label{noiseVScl}
\end{figure}

\begin{figure}[!ht]
    \centering
    \begin{tabular}{ll}
    \subfloat{\includegraphics[width = 0.45\linewidth]{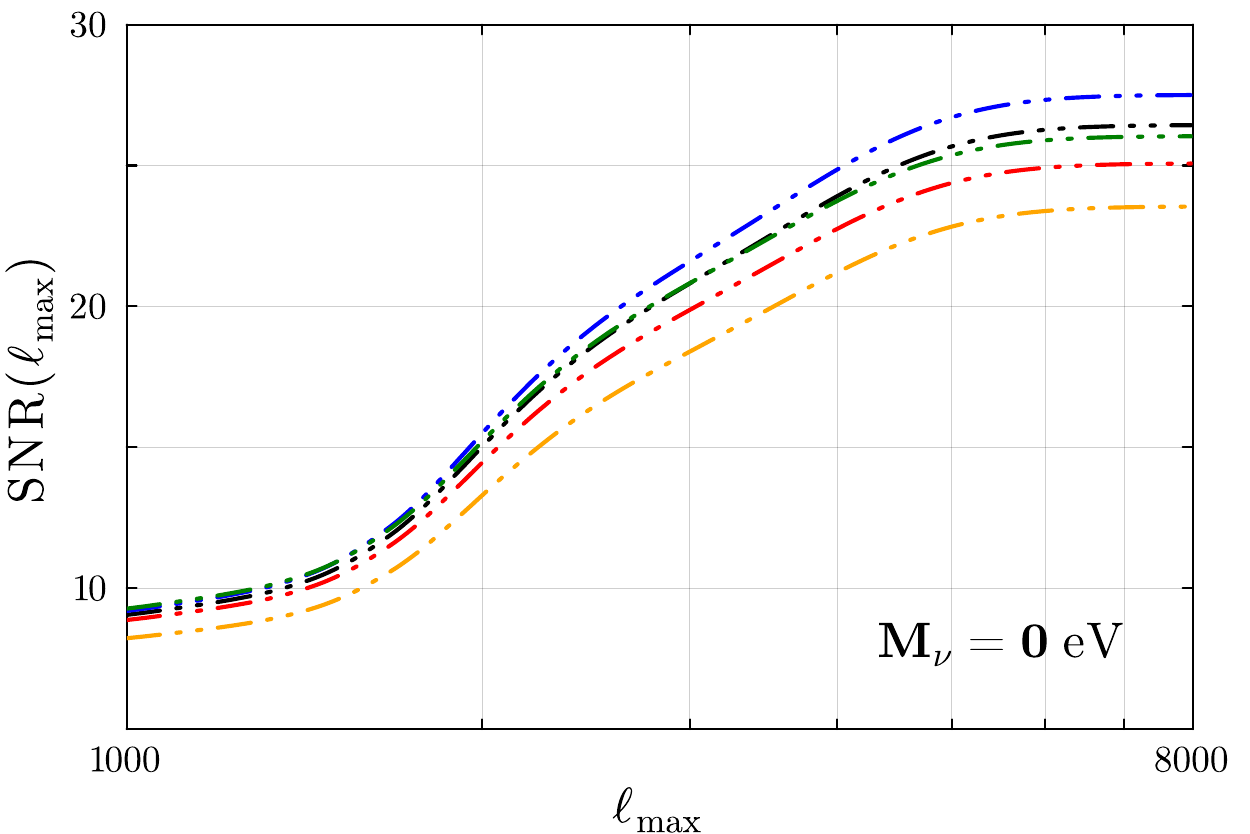}}
    \subfloat{\includegraphics[width = 0.45\linewidth]{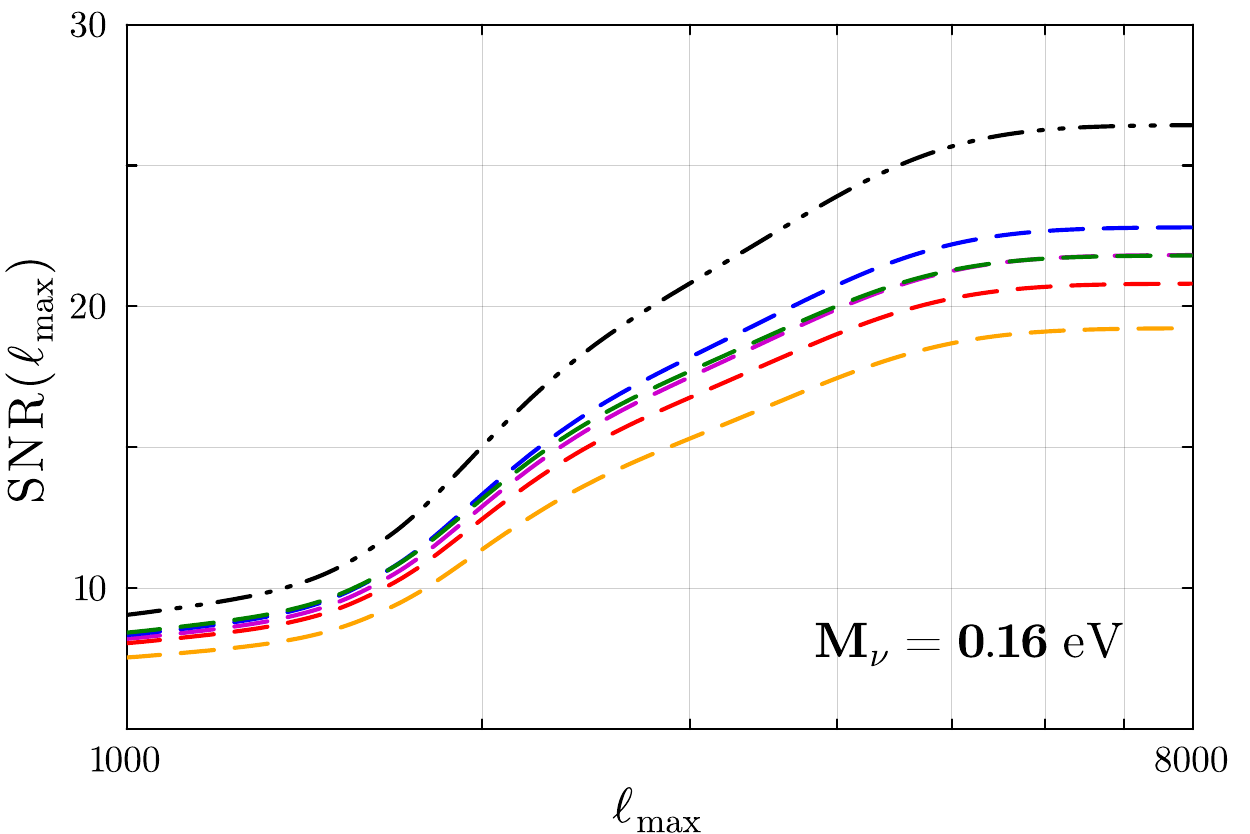}} \\
    \subfloat{\includegraphics[width = 0.45\linewidth]{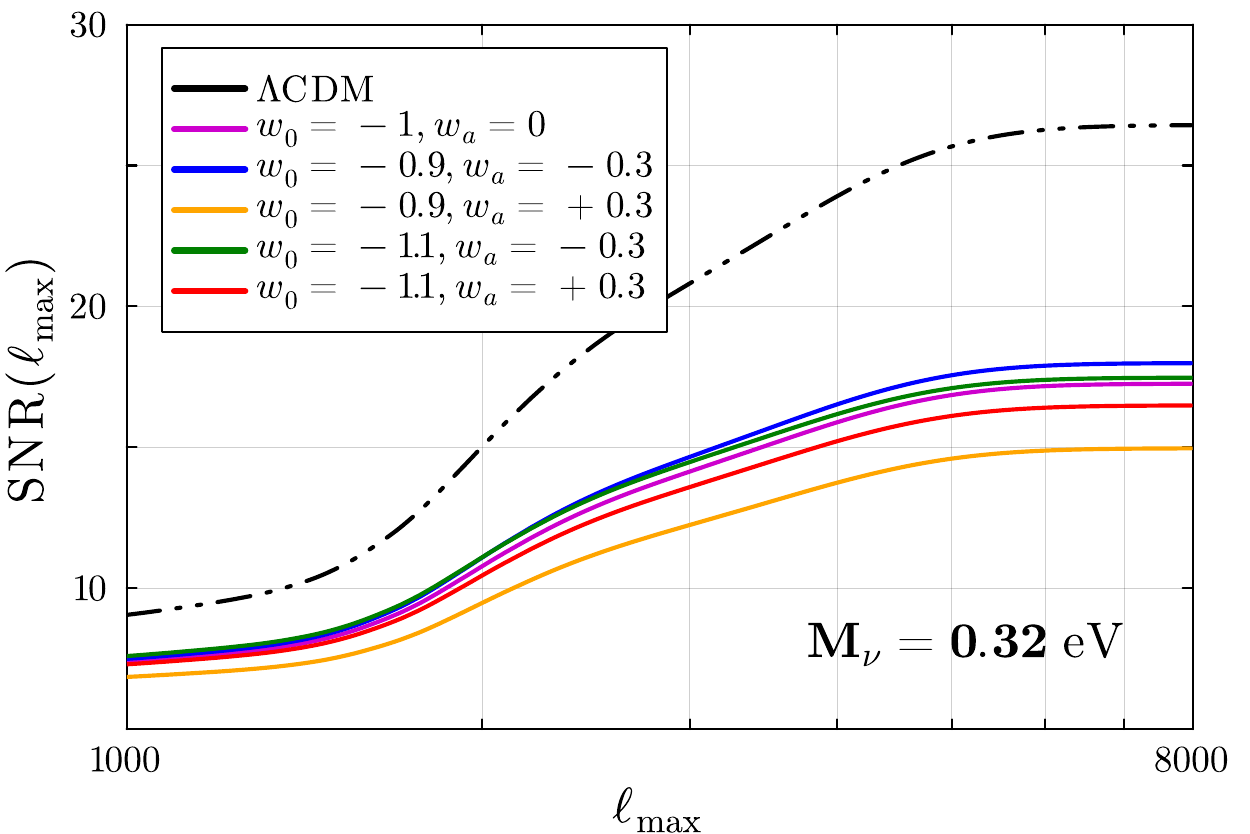}}
    \subfloat{\includegraphics[width = 0.45\linewidth]{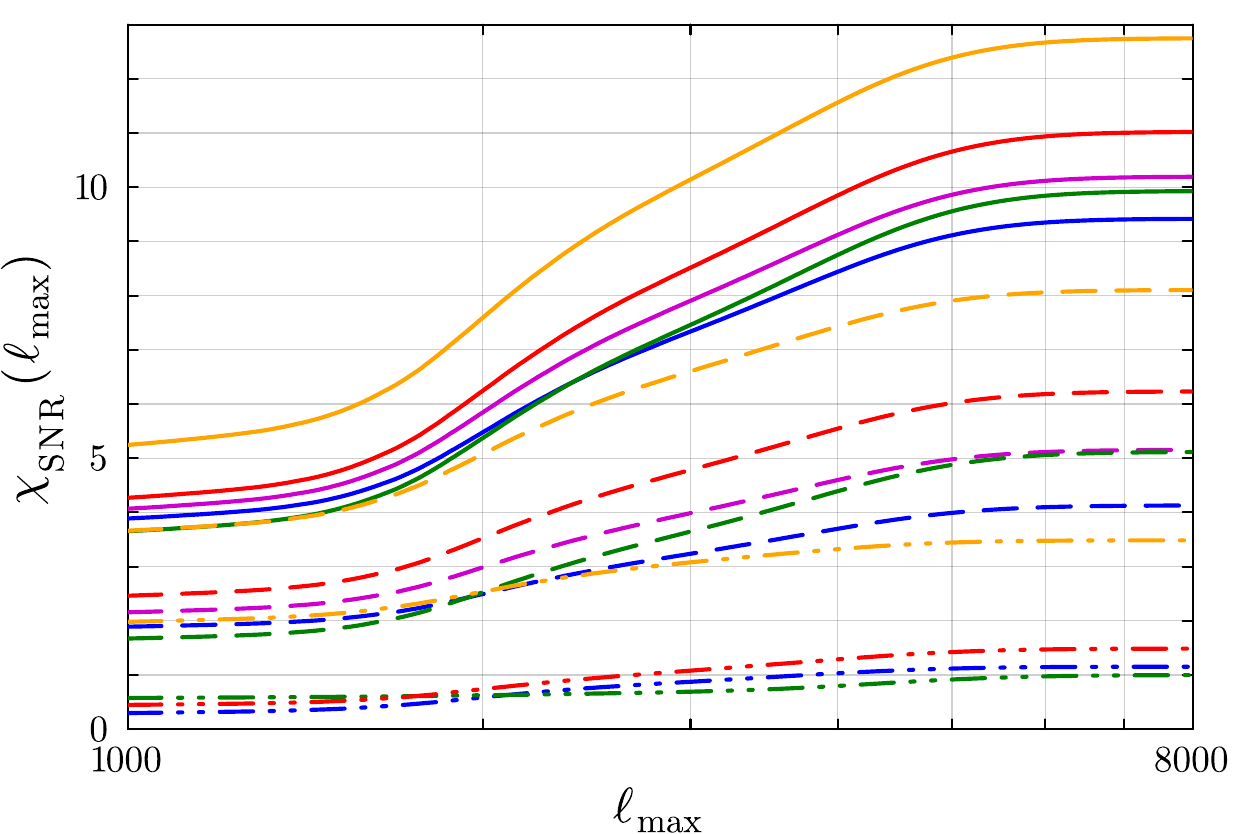}}
    \end{tabular}
    \caption{Top-left, top-right and bottom-left panels show the cumulative signal-to-noise ratio for observations with the SO LAT instrument for the tSZ effect, as calculated from the different \texttt{DEMNUni} simulations. The bottom-right plot shows the $\chi^2$, in Eq.~\eqref{chisqForecast}, with respect to the reference $\Lambda$CDM simulation: dash-double dotted, dashed and solid lines refer to $M_\nu = 0, 0.16, 0.32$ eV, respectively.}
    \label{cumSNRfig}
\end{figure}

Here we provide the forecast for the signal-to-noise ratio (SNR) for the detection of the tSZ effect, specifically regarding the observations with the LAT instrument of Simons Observatory (SO)~\cite{simonsObs}.
The SNR computation starts by evaluating the covariance matrix
\begin{equation}\label{covariance}
    \mathcal{M}_{\ell \ell'} = \frac{1}{4\pi f_{\mathrm{sky}}} \left( 4\pi \delta_{\ell \ell'} \frac{2 (C_\ell + N_\ell)^2}{2\ell+1} + T_{\ell \ell'}\right) \ ,
\end{equation}
where $N_\ell$ is the noise power spectrum and $T_{\ell \ell'}$ is the tri-spectrum for the tSZ~\cite{hillPaj2013,SNRsilk}, while $f_{\mathrm{sky}}$ is the fraction of the sky covered (equal to $0.4$ for SO). The tri-spectrum gives a non negligible non-Gaussian contribution up to moderate $\ell$ (see e.g.~\cite{makiyaTrisp}), but it can be suppressed by eliminating the most massive clusters~\cite{bolliet2020,hillPaj2013}, as their large angular size in the observed sky inevitably couples different scales. As the calculation of the tri-spectrum is computationally expensive, we decide to apply an additional cut in mass, selecting now only the haloes in our sample with $M_{200}<3.4\times 10^{14} h^{-1}\msun$. This choice strongly reduces the observed power, as the very hot ICM in the largest clusters is the main source of the tSZ signal, but provides a purely Gaussian covariance
\begin{equation}
    \mathcal{M}_\ell = \frac{2}{f_{\mathrm{sky}}(2\ell + 1)} (C_\ell + N_\ell)^2 \ ,
\end{equation}
and a more accurate signal-to-noise estimate. Observationally, this would correspond to the application of a sky mask on the regions where clusters above the same mass threshold are identified.
For what concerns the noise, we opt for the one obtained by SO collaboration by constraining both CMB and CIB, which constitutes the least optimistic but more realistic option, ranging from $\ell=80$ to $\ell=7979$ (see Figure~\ref{noiseVScl}). Once the covariance matrix has been estimated with $N_\ell$, the corresponding \textit{cumulative} signal-to-noise ratio is given by
\begin{equation}\label{cumSNR}
    \mathrm{SNR} (\ell_{\mathrm{max}}) = \sqrt{\sum_{\ell,\ell'}^{\ell_{\mathrm{max}}} \frac{C_\ell C_{\ell'}}{\mathcal{M}_{\ell \ell'}}} \overset{\mathrm{Gauss.}}{=} \sqrt{\sum_{\ell}^{\ell_{\mathrm{max}}} \frac{C_\ell^{2}}{\mathcal{M}_\ell}} \ .
\end{equation}
The cumulative SNR results are plotted in Figure~\ref{cumSNRfig}. If the masking of the more massive haloes can be obtained, Simons Observatory should reach values up to $15 \lesssim \mathrm{SNR} \lesssim 27.5$ for observations of the tSZ effect. Due to the specific shapes of both the signal and the noise power spectrum the crucial interval for obtaining a significant signal-to-noise is approximately $(1500,6000)$, after which the SNR values stabilise, following the smoothing effect of the arcminute-size beams of the LAT instrument.

 Besides, with the same formalism described above, it is possible to quantify the detectability of deviations from a certain reference cosmological model~\cite{chisqEstimator}. This is done by substituting the $C_\ell$'s at the numerator in Equation~\eqref{cumSNR} with 
\begin{equation}\label{deltaCl}
    \Delta C_\ell = C_\ell^{\mathrm{ref}} - C_\ell \ ,
\end{equation}
i.~e. the difference between the power spectra calculated respectively in the reference model and in the model investigated; then we calculate the signal-to-noise $\chi^2$ with
\begin{equation}\label{chisqForecast}
    \chi^2_{\mathrm{SNR}} = \sum_{\ell}^{\ell_{\mathrm{max}}}\frac{{\Delta C_\ell}^2}{\mathcal{M}_\ell^{\mathrm{ref}}} \ ,
\end{equation}
where the covariance is meant to be evaluated for the reference scenario.
We use this estimator to probe whether the different cosmological models characterising the \texttt{DEMNUni} set could be potentially distinguished from the reference $\Lambda$CDM with statistical significance with data of the tSZ effect from Simons Observatory, that we recall range from $\ell=80$ to $\ell=7979$.
In Figure~\ref{cumSNRfig} we show the results obtained for $\sqrt{\chi ^2}$, in analogy to the cumulative SNR. Assuming that the detection is only possible at more than $5\sigma$'s, we conclude that SO will be able to discriminate between $\Lambda$CDM and $w_0w_a\nu$CDM cosmologies with $M_\nu = 0.32$ eV, as well as most of those with $M_\nu = 0.16$ eV (aside from the $(w_0,w_a)=(-0.9,-0.3)$ one), but not between the different dark energy EoS of Table~\ref{tab:demnuni_sets_table} in a massless neutrino scenario.

\section{Conclusions}\label{sec::Conclusions}
In this work we study how extensions of the standard $\Lambda$CDM cosmology, namely massive neutrinos and dynamical dark energy, affect the properties of the tSZ effect from the large-scale structure of the Universe. We use the 15 \texttt{DEMNUni} cosmological $N$-body simulations~\cite{Carbone_2016, parimbelli_2022}, that describe the evolution of the cosmic structures assuming different neutrino masses and dark energy EoS (summarised in Table~\ref{tab:demnuni_sets_table}), and build up a model to describe the pressure of the baryonic component, modelling the tSZ signal arising from galaxy clusters/groups via scaling relations.

Starting from the simulation outputs, we identify the dark matter haloes via the \texttt{FoF} and \texttt{Subfind} algorithms, and build mock lightcones using their catalogues. We then use the prescription by~\cite{battaglia1} to model the pressure profile for each halo with $M_{200} > 3.4 \times 10^{13} \, h^{-1} M_\odot$ and construct a set of full-sky maps of the Compton-$y$ parameter in order to analyse the properties of the thermal SZ effect in the $w_0w_a\nu$CDM cosmologies of the \texttt{DEMNUni} simulations. Our main results can be summarised as follows.

\begin{itemize}
\item We observe that the mean logarithmic Compton-$y$ parameter reduces approximately as $10f_\nu$, regardless of the dark energy EoS. This is analogous to the nonlinear damping caused by massive neutrinos in the matter power spectrum~\cite{bird2012}.
\item For all the cosmologies considered in Table~\ref{tab:demnuni_sets_table}, the distribution of the Compton-$y$ in logarithmic scale is well described by a skewed Gaussian. At fixed dark energy EoS, an increase in $M_\nu$ translates into an increase of both the skewness and the variance of the distribution.
\item We observe that, at fixed dark energy EoS, the tSZ power spectrum scales as a power-law with respect to the parameter $\sigma_8^{\mathrm{cb}}$, with an exponent varying from $7.3$ to $8.1$.
This strong dependence reflects the importance of the high-mass end of the HMF for sourcing the tSZ effect.
Conversely, we find that $\sigma_8^{\mathrm{cb}}$ can not trace the variation of the power spectrum when varying the dark energy EoS at fixed $M_\nu$.
\item The models best suited for a description of data derived from Planck Compton-$y$ maps vary depending on the exact choice of the dataset; overall a better agreement is found for cosmologies with massive neutrinos.
\item We produce cumulative SNR forecasts for observations of the tSZ effect with the Simons Observatory LAT instrument, based on its most realistic noise model, with values ranging from $\mathrm{SNR}=15$ ($M_\nu=0.32$ eV with $(w_0,w_a)=(-0.9,+0.3)$) to $\mathrm{SNR}=27$ ($M_\nu=0$ with $(w_0,w_a)=(-0.9,-0.3)$). 
Also, we estimate the detectability of the different cosmologies from the reference massless neutrino $\Lambda$CDM model via signal-to-noise ratios $\chi^2$; these show that most likely SO will be able to discriminate between $\Lambda$CDM and $w_0w_a\nu$CDM cosmologies with $M_\nu = 0.32$ eV, as well as those with $M_\nu = 0.16$ eV (aside from the $(w_0,w_a)=(-0.9,-0.3)$ one), but not between different dark energy EoS in the $M_\nu = 0$ scenario.
\end{itemize}
Our work shows that the tSZ effect is a powerful probe of cosmological models beyond the standard $\Lambda$CDM, due to its strong dependence on the matter clustering and the abundance of virialised structures. On the one hand, the impact of a non-zero neutrino mass, also in the presence of a dynamical dark energy, is particularly significant on the tSZ temperature power spectra and will most likely be detected by future surveys. On the other hand, for the massless neutrino case, the dark energy equations of state considered in the \texttt{DEMNUni} set might not lead to detectable differences yet, especially for those leading to tSZ power spectra very close to the cosmological constant scenario. Further investigations will be needed to gain more precise insights on this phenomenon, in particular combining the data from the upcoming CMB surveys with tailored hydrodynamical or baryonified N-body simulations, taking into account both massive neutrinos and dynamical dark energy.

\acknowledgments
For this work DL was partially supported by the Italian national inter-university PhD programme in Space Science and Technology (CUP 40400445), with the project SPACE-IT-UP, by the Italian Space Agency and Ministry of University and Research, Contract Number 2024-5-E.0. The \texttt{DEMNUni} simulations were carried out in the framework of ``The Dark Energy and Massive-Neutrino Universe" project, using the Tier-0 IBM BG/Q Fermi machine and the Tier-0 Intel OmniPath Cluster Marconi-A1 of the Centro Interuniversitario del Nord-Est per il Calcolo Elettronico (CINECA). We acknowledge a generous CPU and storage allocation by the Italian Super-Computing Resource Allocation (ISCRA) as well as from the coordination of the ``Accordo Quadro MoU per lo svolgimento di attività congiunta di ricerca Nuove frontiere in Astrofisica: HPC e Data Exploration di nuova generazione'', together with storage from INFN-CNAF and INAF-IA2. MC is partially supported by the 2023/24 and 2024/25 ``Research and Education'' grant from Fondazione CRT. The OAVdA is managed by the Fondazione Cl\'ement Fillietroz-ONLUS, which is supported by the Regional Government of the Aosta Valley, the Town Municipality of Nus and the ``Unit\'e des Communes vald\^otaines Mont-\'Emilius''.
%\clearpage
%\newpage
\bibliography{biblio}

\providecommand{\href}[2]{#2}\begingroup\raggedright\begin{thebibliography}{10}

\bibitem{wmap1}
D.N.~{Spergel}, L.~{Verde}, H.V.~{Peiris}, E.~{Komatsu}, M.R.~{Nolta}, C.L.~{Bennett} et~al., \emph{{First-Year Wilkinson Microwave Anisotropy Probe (WMAP) Observations: Determination of Cosmological Parameters}}, \href{https://doi.org/10.1086/377226}{\emph{The Astrophysical Journal Supplement Series} {\bfseries 148} (2003) 175} [\href{https://arxiv.org/abs/astro-ph/0302209}{{\ttfamily astro-ph/0302209}}].

\bibitem{wmap7}
E.~{Komatsu}, K.M.~{Smith}, J.~{Dunkley}, C.L.~{Bennett}, B.~{Gold}, G.~{Hinshaw} et~al., \emph{{Seven-year Wilkinson Microwave Anisotropy Probe (WMAP) Observations: Cosmological Interpretation}}, \href{https://doi.org/10.1088/0067-0049/192/2/18}{\emph{\apjs} {\bfseries 192} (2011) 18} [\href{https://arxiv.org/abs/1001.4538}{{\ttfamily 1001.4538}}].

\bibitem{planck2013Cosmo}
{Planck Collaboration}, P.A.R.~{Ade}, N.~{Aghanim}, C.~{Armitage-Caplan}, M.~{Arnaud}, M.~{Ashdown} et~al., \emph{{Planck 2013 results. XVI. Cosmological parameters}}, \href{https://doi.org/10.1051/0004-6361/201321591}{\emph{\aap} {\bfseries 571} (2014) A16} [\href{https://arxiv.org/abs/1303.5076}{{\ttfamily 1303.5076}}].

\bibitem{planck2015cosmo}
{Planck Collaboration}, P.A.R.~{Ade}, N.~{Aghanim}, M.~{Arnaud}, M.~{Ashdown}, J.~{Aumont} et~al., \emph{{Planck 2015 results. XIII. Cosmological parameters}}, \href{https://doi.org/10.1051/0004-6361/201525830}{\emph{\aap} {\bfseries 594} (2016) A13} [\href{https://arxiv.org/abs/1502.01589}{{\ttfamily 1502.01589}}].

\bibitem{planck2018}
{Planck Collaboration}, N.~{Aghanim}, Y.~{Akrami}, M.~{Ashdown}, J.~{Aumont}, C.~{Baccigalupi} et~al., \emph{{Planck 2018 results. VI. Cosmological parameters}}, \href{https://doi.org/10.1051/0004-6361/201833910}{\emph{\aap} {\bfseries 641} (2020) A6} [\href{https://arxiv.org/abs/1807.06209}{{\ttfamily 1807.06209}}].

\bibitem{act2025ps}
T.~{Louis}, A.~{La Posta}, Z.~{Atkins}, H.T.~{Jense}, I.~{Abril-Cabezas}, G.E.~{Addison} et~al., \emph{{The Atacama Cosmology Telescope: DR6 power spectra, likelihoods and {\ensuremath{\Lambda}}CDM parameters}}, \href{https://doi.org/10.1088/1475-7516/2025/11/062}{\emph{\jcap} {\bfseries 2025} (2025) 062} [\href{https://arxiv.org/abs/2503.14452}{{\ttfamily 2503.14452}}].

\bibitem{act2025extended}
E.~{Calabrese}, J.C.~{Hill}, H.T.~{Jense}, A.~{La Posta}, I.~{Abril-Cabezas}, G.E.~{Addison} et~al., \emph{{The Atacama Cosmology Telescope: DR6 constraints on extended cosmological models}}, \href{https://doi.org/10.1088/1475-7516/2025/11/063}{\emph{\jcap} {\bfseries 2025} (2025) 063} [\href{https://arxiv.org/abs/2503.14454}{{\ttfamily 2503.14454}}].

\bibitem{desiBao2025}
{DESI Collaboration}, M.~{Abdul-Karim}, J.~{Aguilar}, S.~{Ahlen}, S.~{Alam}, L.~{Allen} et~al., \emph{{DESI DR2 Results II: Measurements of Baryon Acoustic Oscillations and Cosmological Constraints}}, \href{https://doi.org/10.48550/arXiv.2503.14738}{\emph{arXiv e-prints} (2025) arXiv:2503.14738} [\href{https://arxiv.org/abs/2503.14738}{{\ttfamily 2503.14738}}].

\bibitem{Carbone_2016}
C.~{Carbone}, M.~{Petkova} and K.~{Dolag}, \emph{{DEMNUni: ISW, Rees-Sciama, and weak-lensing in the presence of massive neutrinos}}, \href{https://doi.org/10.1088/1475-7516/2016/07/034}{\emph{\jcap} {\bfseries 2016} (2016) 034} [\href{https://arxiv.org/abs/1605.02024}{{\ttfamily 1605.02024}}].

\bibitem{parimbelli_2022}
G.~{Parimbelli}, C.~{Carbone}, J.~{Bel}, B.~{Bose}, M.~{Calabrese}, E.~{Carella} et~al., \emph{{DEMNUni: comparing nonlinear power spectra prescriptions in the presence of massive neutrinos and dynamical dark energy}}, \href{https://doi.org/10.1088/1475-7516/2022/11/041}{\emph{\jcap} {\bfseries 2022} (2022) 041} [\href{https://arxiv.org/abs/2207.13677}{{\ttfamily 2207.13677}}].

\bibitem{flamingoClusters}
R.~{Kugel}, J.~{Schaye}, M.~{Schaller}, I.G.~{McCarthy}, J.~{Braspenning}, J.C.~{Helly} et~al., \emph{{The FLAMINGO project: a comparison of galaxy cluster samples selected on mass, X-ray luminosity, Compton-Y parameter, or galaxy richness}}, \href{https://doi.org/10.1093/mnras/stae2218}{\emph{\mnras} {\bfseries 534} (2024) 2378} [\href{https://arxiv.org/abs/2406.03180}{{\ttfamily 2406.03180}}].

\bibitem{Xray+tSZ}
A.~{La Posta}, D.~{Alonso}, N.E.~{Chisari}, T.~{Ferreira} and C.~{Garc{\'\i}a-Garc{\'\i}a}, \emph{{Insights on gas thermodynamics from the combination of x-ray and thermal Sunyaev-Zel'dovich data cross correlated with cosmic shear}}, \href{https://doi.org/10.1103/m77z-w7pl}{\emph{\prd} {\bfseries 112} (2025) 043525} [\href{https://arxiv.org/abs/2412.12081}{{\ttfamily 2412.12081}}].

\bibitem{wl+kSZBigwood}
L.~{Bigwood}, A.~{Amon}, A.~{Schneider}, J.~{Salcido}, I.G.~{McCarthy}, C.~{Preston} et~al., \emph{{Weak lensing combined with the kinetic Sunyaev-Zel'dovich effect: a study of baryonic feedback}}, \href{https://doi.org/10.1093/mnras/stae2100}{\emph{\mnras} {\bfseries 534} (2024) 655} [\href{https://arxiv.org/abs/2404.06098}{{\ttfamily 2404.06098}}].

\bibitem{tSZACT+DESI}
R.H.~{Liu}, S.~{Ferraro}, E.~{Schaan}, R.~{Zhou}, J.N.~{Aguilar}, S.~{Ahlen} et~al., \emph{{Measurements of the thermal Sunyaev-Zel'dovich effect with ACT and DESI luminous red galaxies}}, \href{https://doi.org/10.1103/jqn8-19gx}{\emph{\prd} {\bfseries 112} (2025) 083561} [\href{https://arxiv.org/abs/2502.08850}{{\ttfamily 2502.08850}}].

\bibitem{pairwise_kSZ_cmbS4}
E.~{Schiappucci}, S.~{Raghunathan}, C.~{To}, F.~{Bianchini}, C.L.~{Reichardt}, N.~{Battaglia} et~al., \emph{{Constraining cosmological parameters using the pairwise kinematic Sunyaev-Zel'dovich effect with CMB-S4 and future galaxy cluster surveys}}, \href{https://doi.org/10.1103/PhysRevD.111.063541}{\emph{\prd} {\bfseries 111} (2025) 063541} [\href{https://arxiv.org/abs/2409.18368}{{\ttfamily 2409.18368}}].

\bibitem{pairwisekSZ_soergel}
B.~{Soergel}, A.~{Saro}, T.~{Giannantonio}, G.~{Efstathiou} and K.~{Dolag}, \emph{{Cosmology with the pairwise kinematic SZ effect: calibration and validation using hydrodynamical simulations}}, \href{https://doi.org/10.1093/mnras/sty1324}{\emph{\mnras} {\bfseries 478} (2018) 5320} [\href{https://arxiv.org/abs/1712.05714}{{\ttfamily 1712.05714}}].

\bibitem{chiang2020}
Y.-K.~{Chiang}, R.~{Makiya}, B.~{M{\'e}nard} and E.~{Komatsu}, \emph{{The Cosmic Thermal History Probed by Sunyaev-Zeldovich Effect Tomography}}, \href{https://doi.org/10.3847/1538-4357/abb403}{\emph{\apj} {\bfseries 902} (2020) 56} [\href{https://arxiv.org/abs/2006.14650}{{\ttfamily 2006.14650}}].

\bibitem{superK}
Y.~{Fukuda}, T.~{Hayakawa}, E.~{Ichihara}, K.~{Inoue}, K.~{Ishihara}, H.~{Ishino} et~al., \emph{{Evidence for Oscillation of Atmospheric Neutrinos}}, \href{https://doi.org/10.1103/PhysRevLett.81.1562}{\emph{\prl} {\bfseries 81} (1998) 1562} [\href{https://arxiv.org/abs/hep-ex/9807003}{{\ttfamily hep-ex/9807003}}].

\bibitem{Roncarelli_2015}
M.~{Roncarelli}, C.~{Carbone} and L.~{Moscardini}, \emph{{The effect of massive neutrinos on the Sunyaev-Zel'dovich and X-ray observables of galaxy clusters}}, \href{https://doi.org/10.1093/mnras/stu2546}{\emph{\mnras} {\bfseries 447} (2015) 1761} [\href{https://arxiv.org/abs/1409.4285}{{\ttfamily 1409.4285}}].

\bibitem{mauro2017}
M.~{Roncarelli}, F.~{Villaescusa-Navarro} and M.~{Baldi}, \emph{{The kinematic Sunyaev-Zel'dovich effect of the large-scale structure (I): dependence on neutrino mass}}, \href{https://doi.org/10.1093/mnras/stx170}{\emph{\mnras} {\bfseries 467} (2017) 985} [\href{https://arxiv.org/abs/1702.00676}{{\ttfamily 1702.00676}}].

\bibitem{sdssBaoClustering2017}
S.~{Alam}, M.~{Ata}, S.~{Bailey}, F.~{Beutler}, D.~{Bizyaev}, J.A.~{Blazek} et~al., \emph{{The clustering of galaxies in the completed SDSS-III Baryon Oscillation Spectroscopic Survey: cosmological analysis of the DR12 galaxy sample}}, \href{https://doi.org/10.1093/mnras/stx721}{\emph{\mnras} {\bfseries 470} (2017) 2617} [\href{https://arxiv.org/abs/1607.03155}{{\ttfamily 1607.03155}}].

\bibitem{desY3ext}
T.M.C.~{Abbott}, M.~{Aguena}, A.~{Alarcon}, O.~{Alves}, A.~{Amon}, F.~{Andrade-Oliveira} et~al., \emph{{Dark Energy Survey Year 3 results: Constraints on extensions to {\ensuremath{\Lambda}} CDM with weak lensing and galaxy clustering}}, \href{https://doi.org/10.1103/PhysRevD.107.083504}{\emph{\prd} {\bfseries 107} (2023) 083504} [\href{https://arxiv.org/abs/2207.05766}{{\ttfamily 2207.05766}}].

\bibitem{kSZtomography}
A.J.~{Tishue}, S.C.~{Hotinli}, P.~{Adshead}, E.D.~{Kovetz} and M.S.~{Madhavacheril}, \emph{{Neutrino Mass Constraints from kSZ Tomography}}, \href{https://doi.org/10.48550/arXiv.2502.05260}{\emph{arXiv e-prints} (2025) arXiv:2502.05260} [\href{https://arxiv.org/abs/2502.05260}{{\ttfamily 2502.05260}}].

\bibitem{lesgourgues2006}
J.~{Lesgourgues} and S.~{Pastor}, \emph{{Massive neutrinos and cosmology}}, \href{https://doi.org/10.1016/j.physrep.2006.04.001}{\emph{\physrep} {\bfseries 429} (2006) 307} [\href{https://arxiv.org/abs/astro-ph/0603494}{{\ttfamily astro-ph/0603494}}].

\bibitem{lesgNeutrinoBook}
J.~{Lesgourgues}, G.~{Mangano}, G.~{Miele} and S.~{Pastor}, \emph{{Neutrino Cosmology}}, Cambridge University Press (2013).

\bibitem{pdg2022}
R.L.~{Workman}, V.D.~{Burkert}, V.~{Crede}, E.~{Klempt}, U.~{Thoma}, L.~{Tiator} et~al., \emph{{Review of Particle Physics}}, \href{https://doi.org/10.1093/ptep/ptac097}{\emph{Progress of Theoretical and Experimental Physics} {\bfseries 2022} (2022) 083C01}.

\bibitem{chevPolanki}
M.~{Chevallier} and D.~{Polarski}, \emph{{Accelerating Universes with Scaling Dark Matter}}, \href{https://doi.org/10.1142/S0218271801000822}{\emph{International Journal of Modern Physics D} {\bfseries 10} (2001) 213} [\href{https://arxiv.org/abs/gr-qc/0009008}{{\ttfamily gr-qc/0009008}}].

\bibitem{linder}
E.V.~{Linder}, \emph{{Exploring the Expansion History of the Universe}}, \href{https://doi.org/10.1103/PhysRevLett.90.091301}{\emph{\prl} {\bfseries 90} (2003) 091301} [\href{https://arxiv.org/abs/astro-ph/0208512}{{\ttfamily astro-ph/0208512}}].

\bibitem{Hernandez_2024a}
B.~{Hern{\'a}ndez-Molinero}, C.~{Carbone}, R.~{Jimenez} and C.~{Pe{\~n}a Garay}, \emph{{Cosmic background neutrinos deflected by gravity: DEMNUni simulation analysis}}, \href{https://doi.org/10.1088/1475-7516/2024/01/006}{\emph{\jcap} {\bfseries 2024} (2024) 006} [\href{https://arxiv.org/abs/2301.12430}{{\ttfamily 2301.12430}}].

\bibitem{Hernandez_2024b}
B.~{Hern{\'a}ndez-Molinero}, C.~{Carbone}, R.~{Jimenez} and C.P.~{Garay}, \emph{{Neutrino halo profiles: HR-DEMNUni simulation analysis}}, \href{https://doi.org/10.1088/1475-7516/2024/09/033}{\emph{\jcap} {\bfseries 2024} (2024) 033} [\href{https://arxiv.org/abs/2407.12694}{{\ttfamily 2407.12694}}].

\bibitem{Castorina_2015}
E.~{Castorina}, C.~{Carbone}, J.~{Bel}, E.~{Sefusatti} and K.~{Dolag}, \emph{{DEMNUni: the clustering of large-scale structures in the presence of massive neutrinos}}, \href{https://doi.org/10.1088/1475-7516/2015/07/043}{\emph{\jcap} {\bfseries 7} (2015) 043} [\href{https://arxiv.org/abs/1505.07148}{{\ttfamily 1505.07148}}].

\bibitem{Moresco_2017}
M.~{Moresco}, F.~{Marulli}, L.~{Moscardini}, E.~{Branchini}, A.~{Cappi}, I.~{Davidzon} et~al., \emph{{The VIMOS Public Extragalactic Redshift Survey (VIPERS) . Exploring the dependence of the three-point correlation function on stellar mass and luminosity at 0.5 <z < 1.1}}, \href{https://doi.org/10.1051/0004-6361/201628589}{\emph{\aap} {\bfseries 604} (2017) A133} [\href{https://arxiv.org/abs/1603.08924}{{\ttfamily 1603.08924}}].

\bibitem{Zennaro_2017}
M.~{Zennaro}, J.~{Bel}, F.~{Villaescusa-Navarro}, C.~{Carbone}, E.~{Sefusatti} and L.~{Guzzo}, \emph{{Initial conditions for accurate N-body simulations of massive neutrino cosmologies}}, \href{https://doi.org/10.1093/mnras/stw3340}{\emph{\mnras} {\bfseries 466} (2017) 3244} [\href{https://arxiv.org/abs/1605.05283}{{\ttfamily 1605.05283}}].

\bibitem{ruggeri_2018}
R.~{Ruggeri}, E.~{Castorina}, C.~{Carbone} and E.~{Sefusatti}, \emph{{DEMNUni: massive neutrinos and the bispectrum of large scale structures}}, \href{https://doi.org/10.1088/1475-7516/2018/03/003}{\emph{\jcap} {\bfseries 2018} (2018) 003} [\href{https://arxiv.org/abs/1712.02334}{{\ttfamily 1712.02334}}].

\bibitem{bel_2019}
J.~{Bel}, A.~{Pezzotta}, C.~{Carbone}, E.~{Sefusatti} and L.~{Guzzo}, \emph{{Accurate fitting functions for peculiar velocity spectra in standard and massive-neutrino cosmologies}}, \href{https://doi.org/10.1051/0004-6361/201834513}{\emph{\aap} {\bfseries 622} (2019) A109} [\href{https://arxiv.org/abs/1809.09338}{{\ttfamily 1809.09338}}].

\bibitem{parimbelli_2021}
G.~{Parimbelli}, S.~{Anselmi}, M.~{Viel}, C.~{Carbone}, F.~{Villaescusa-Navarro}, P.S.~{Corasaniti} et~al., \emph{{The effects of massive neutrinos on the linear point of the correlation function}}, \href{https://doi.org/10.1088/1475-7516/2021/01/009}{\emph{\jcap} {\bfseries 2021} (2021) 009} [\href{https://arxiv.org/abs/2007.10345}{{\ttfamily 2007.10345}}].

\bibitem{Guidi_2022}
M.~{Guidi}, A.~{Veropalumbo}, E.~{Branchini}, A.~{Eggemeier} and C.~{Carbone}, \emph{{Modelling the next-to-leading order matter three-point correlation function using FFTLog}}, \href{https://doi.org/10.1088/1475-7516/2023/08/066}{\emph{\jcap} {\bfseries 2023} (2023) 066} [\href{https://arxiv.org/abs/2212.07382}{{\ttfamily 2212.07382}}].

\bibitem{Baratta_2022}
P.~{Baratta}, J.~{Bel}, S.~{Gouyou Beauchamps} and C.~{Carbone}, \emph{{COVMOS: A new Monte Carlo approach for galaxy clustering analysis}}, \href{https://doi.org/10.1051/0004-6361/202245683}{\emph{\aap} {\bfseries 673} (2023) A1} [\href{https://arxiv.org/abs/2211.13590}{{\ttfamily 2211.13590}}].

\bibitem{Gouyou_Beauchamps_2023}
S.~{Gouyou Beauchamps}, P.~{Baratta}, S.~{Escoffier}, W.~{Gillard}, J.~{Bel}, J.~{Bautista} et~al., \emph{{Cosmological inference including massive neutrinos from the matter power spectrum: Biases induced by uncertainties in the covariance matrix}}, \href{https://doi.org/10.1051/0004-6361/202347164}{\emph{\aap} {\bfseries 693} (2025) A226} [\href{https://arxiv.org/abs/2306.05988}{{\ttfamily 2306.05988}}].

\bibitem{Verdiani_2025}
F.~{Verdiani}, E.~{Bellini}, C.~{Moretti}, E.~{Sefusatti}, C.~{Carbone} and M.~{Viel}, \emph{{Redshift-Space Distortions in Massive Neutrinos Cosmologies}}, \href{https://doi.org/10.48550/arXiv.2503.06655}{\emph{arXiv e-prints} (2025) arXiv:2503.06655} [\href{https://arxiv.org/abs/2503.06655}{{\ttfamily 2503.06655}}].

\bibitem{SHAM-Carella_in_prep}
E.~{Carella}, C.~{Carbone}, M.~{Zennaro}, G.~{Girelli}, M.~{Bolzonella}, F.~{Marulli} et~al., \emph{{DEMNUni: The galaxy-halo connection in the presence of dynamical dark energy and massive neutrinos}}, {\emph{In prep} }.

\bibitem{Ingoglia_2024}
{Euclid Collaboration}, L.~{Ingoglia}, M.~{Sereno}, S.~{Farrens}, C.~{Giocoli}, L.~{Baumont} et~al., \emph{{Euclid preparation: LXV. Determining the weak lensing mass accuracy and precision for galaxy clusters}}, \href{https://doi.org/10.1051/0004-6361/202452122}{\emph{\aap} {\bfseries 695} (2025) A280} [\href{https://arxiv.org/abs/2409.02783}{{\ttfamily 2409.02783}}].

\bibitem{Fabbian_2018}
G.~{Fabbian}, M.~{Calabrese} and C.~{Carbone}, \emph{{CMB weak-lensing beyond the Born approximation: a numerical approach}}, \href{https://doi.org/10.1088/1475-7516/2018/02/050}{\emph{\jcap} {\bfseries 2018} (2018) 050} [\href{https://arxiv.org/abs/1702.03317}{{\ttfamily 1702.03317}}].

\bibitem{kreisch_2019}
C.D.~{Kreisch}, A.~{Pisani}, C.~{Carbone}, J.~{Liu}, A.J.~{Hawken}, E.~{Massara} et~al., \emph{{Massive neutrinos leave fingerprints on cosmic voids}}, \href{https://doi.org/10.1093/mnras/stz1944}{\emph{\mnras} {\bfseries 488} (2019) 4413} [\href{https://arxiv.org/abs/1808.07464}{{\ttfamily 1808.07464}}].

\bibitem{schuster_2019}
N.~{Schuster}, N.~{Hamaus}, A.~{Pisani}, C.~{Carbone}, C.D.~{Kreisch}, G.~{Pollina} et~al., \emph{{The bias of cosmic voids in the presence of massive neutrinos}}, \href{https://doi.org/10.1088/1475-7516/2019/12/055}{\emph{\jcap} {\bfseries 2019} (2019) 055} [\href{https://arxiv.org/abs/1905.00436}{{\ttfamily 1905.00436}}].

\bibitem{verza_2019}
G.~{Verza}, A.~{Pisani}, C.~{Carbone}, N.~{Hamaus} and L.~{Guzzo}, \emph{{The void size function in dynamical dark energy cosmologies}}, \href{https://doi.org/10.1088/1475-7516/2019/12/040}{\emph{\jcap} {\bfseries 2019} (2019) 040} [\href{https://arxiv.org/abs/1906.00409}{{\ttfamily 1906.00409}}].

\bibitem{verza_2022a}
G.~{Verza}, C.~{Carbone} and A.~{Renzi}, \emph{{The Halo Bias inside Cosmic Voids}}, \href{https://doi.org/10.3847/2041-8213/ac9d98}{\emph{\apjl} {\bfseries 940} (2022) L16} [\href{https://arxiv.org/abs/2207.04039}{{\ttfamily 2207.04039}}].

\bibitem{verza_2022b}
G.~{Verza}, C.~{Carbone}, A.~{Pisani} and A.~{Renzi}, \emph{{DEMNUni: disentangling dark energy from massive neutrinos with the void size function}}, \href{https://doi.org/10.1088/1475-7516/2023/12/044}{\emph{\jcap} {\bfseries 2023} (2023) 044} [\href{https://arxiv.org/abs/2212.09740}{{\ttfamily 2212.09740}}].

\bibitem{Vielzeuf_2022}
P.~{Vielzeuf}, M.~{Calabrese}, C.~{Carbone}, G.~{Fabbian} and C.~{Baccigalupi}, \emph{{DEMNUni: the imprint of massive neutrinos on the cross-correlation between cosmic voids and CMB lensing}}, \href{https://doi.org/10.1088/1475-7516/2023/08/010}{\emph{\jcap} {\bfseries 2023} (2023) 010} [\href{https://arxiv.org/abs/2303.10048}{{\ttfamily 2303.10048}}].

\bibitem{Cuozzo2022}
V.~{Cuozzo}, C.~{Carbone}, M.~{Calabrese}, E.~{Carella} and M.~{Migliaccio}, \emph{{DEMNUni: cross-correlating the nonlinear ISWRS effect with CMB-lensing and galaxies in the presence of massive neutrinos}}, \href{https://doi.org/10.1088/1475-7516/2024/04/073}{\emph{\jcap} {\bfseries 2024} (2024) 073} [\href{https://arxiv.org/abs/2307.15711}{{\ttfamily 2307.15711}}].

\bibitem{Springel_2005}
V.~Springel, \emph{The cosmological simulation code gadget-2}, \href{https://doi.org/10.1111/j.1365-2966.2005.09655.x}{\emph{\mnras} {\bfseries 364} (2005) 1105–1134}.

\bibitem{Viel_2010}
M.~Viel, M.G.~Haehnelt and V.~Springel, \emph{The effect of neutrinos on the matter distribution as probed by the intergalactic medium}, \href{https://doi.org/10.1088/1475-7516/2010/06/015}{\emph{\jcap} {\bfseries 2010} (2010) 015–015}.

\bibitem{springel01}
V.~{Springel}, S.D.M.~{White}, G.~{Tormen} and G.~{Kauffmann}, \emph{{Populating a cluster of galaxies - I. Results at z=0}}, \href{https://doi.org/10.1046/j.1365-8711.2001.04912.x}{\emph{\mnras} {\bfseries 328} (2001) 726} [\href{https://arxiv.org/abs/astro-ph/0012055}{{\ttfamily astro-ph/0012055}}].

\bibitem{dolag09}
K.~{Dolag}, S.~{Borgani}, G.~{Murante} and V.~{Springel}, \emph{{Substructures in hydrodynamical cluster simulations}}, \href{https://doi.org/10.1111/j.1365-2966.2009.15034.x}{\emph{\mnras} {\bfseries 399} (2009) 497} [\href{https://arxiv.org/abs/0808.3401}{{\ttfamily 0808.3401}}].

\bibitem{melitaLensing}
C.~{Carbone}, V.~{Springel}, C.~{Baccigalupi}, M.~{Bartelmann} and S.~{Matarrese}, \emph{{Full-sky maps for gravitational lensing of the cosmic microwave background}}, \href{https://doi.org/10.1111/j.1365-2966.2008.13544.x}{\emph{\mnras} {\bfseries 388} (2008) 1618} [\href{https://arxiv.org/abs/0711.2655}{{\ttfamily 0711.2655}}].

\bibitem{calabrese2015}
M.~{Calabrese}, C.~{Carbone}, G.~{Fabbian}, M.~{Baldi} and C.~{Baccigalupi}, \emph{{Multiple lensing of the cosmic microwave background anisotropies}}, \href{https://doi.org/10.1088/1475-7516/2015/03/049}{\emph{\jcap} {\bfseries 3} (2015) 049} [\href{https://arxiv.org/abs/1409.7680}{{\ttfamily 1409.7680}}].

\bibitem{Hilbert_2020}
S.~{Hilbert}, A.~{Barreira}, G.~{Fabbian}, P.~{Fosalba}, C.~{Giocoli}, S.~{Bose} et~al., \emph{{The accuracy of weak lensing simulations}}, \href{https://doi.org/10.1093/mnras/staa281}{\emph{\mnras} {\bfseries 493} (2020) 305} [\href{https://arxiv.org/abs/1910.10625}{{\ttfamily 1910.10625}}].

\bibitem{euclidOverview}
{Euclid Collaboration}, Y.~{Mellier}, {Abdurro'uf}, J.A.~{Acevedo Barroso}, A.~{Ach{\'u}carro}, J.~{Adamek} et~al., \emph{{Euclid: I. Overview of the Euclid mission}}, \href{https://doi.org/10.1051/0004-6361/202450810}{\emph{\aap} {\bfseries 697} (2025) A1} [\href{https://arxiv.org/abs/2405.13491}{{\ttfamily 2405.13491}}].

\bibitem{kompaneets}
A.S.~{Kompaneets}, \emph{{The Establishment of Thermal Equilibrium between Quanta and Electrons}}, {\emph{Soviet Journal of Experimental and Theoretical Physics} {\bfseries 4} (1957) 730}.

\bibitem{sz}
Y.B.~{Zel'dovich} and R.A.~{Sunyaev}, \emph{{The Interaction of Matter and Radiation in a Hot-Model Universe}}, \href{https://doi.org/10.1007/BF00661821}{\emph{\apss} {\bfseries 4} (1969) 301}.

\bibitem{szKinematic2}
R.A.~{Sunyaev} and Y.B.~{Zel'dovich}, \emph{{Small-Scale Fluctuations of Relic Radiation}}, \href{https://doi.org/10.1007/BF00653471}{\emph{\apss} {\bfseries 7} (1970) 3}.

\bibitem{sunyaevKinematic}
R.A.~{Sunyaev}, \emph{{Fluctuations in microwave background radiation due to secondary ionization of the intergalactic gas in the universe.}}, {\emph{Pisma v Astronomicheskii Zhurnal} {\bfseries 3} (1977) 491}.

\bibitem{mauro2018}
M.~{Roncarelli}, M.~{Baldi} and F.~{Villaescusa-Navarro}, \emph{{The kinematic Sunyaev-Zel'dovich effect of the large-scale structure (II): the effect of modified gravity}}, \href{https://doi.org/10.1093/mnras/sty2225}{\emph{\mnras} {\bfseries 481} (2018) 2497} [\href{https://arxiv.org/abs/1805.11607}{{\ttfamily 1805.11607}}].

\bibitem{battaglia1}
N.~{Battaglia}, J.R.~{Bond}, C.~{Pfrommer} and J.L.~{Sievers}, \emph{{On the Cluster Physics of Sunyaev-Zel'dovich and X-Ray Surveys. II. Deconstructing the Thermal SZ Power Spectrum}}, \href{https://doi.org/10.1088/0004-637X/758/2/75}{\emph{\apj} {\bfseries 758} (2012) 75} [\href{https://arxiv.org/abs/1109.3711}{{\ttfamily 1109.3711}}].

\bibitem{cooray2005}
A.~{Cooray}, D.~{Baumann} and K.~{Sigurdson}, \emph{{Statistical imprints of SZ effects in the cosmic microwave background}},  in \emph{Background Microwave Radiation and Intracluster Cosmology}, F.~{Melchiorri} and Y.~{Rephaeli}, eds., p.~309, Jan., 2005, \href{https://doi.org/10.48550/arXiv.astro-ph/0410006}{DOI} [\href{https://arxiv.org/abs/astro-ph/0410006}{{\ttfamily astro-ph/0410006}}].

\bibitem{hillPaj2013}
J.C.~{Hill} and E.~{Pajer}, \emph{{Cosmology from the thermal Sunyaev-Zel'dovich power spectrum: Primordial non-Gaussianity and massive neutrinos}}, \href{https://doi.org/10.1103/PhysRevD.88.063526}{\emph{\prd} {\bfseries 88} (2013) 063526} [\href{https://arxiv.org/abs/1303.4726}{{\ttfamily 1303.4726}}].

\bibitem{carWCS}
M.R.~{Calabretta} and E.W.~{Greisen}, \emph{{Representations of celestial coordinates in FITS}}, \href{https://doi.org/10.1051/0004-6361:20021327}{\emph{\aap} {\bfseries 395} (2002) 1077} [\href{https://arxiv.org/abs/astro-ph/0207413}{{\ttfamily astro-ph/0207413}}].

\bibitem{azzaliniSkew}
A.~{Azzalini}, \emph{A class of distributions which includes the normal ones}, {\emph{Scandinavian Journal of Statistics} {\bfseries 12} (1985) 171}.

\bibitem{classSZ}
B.~{Bolliet}, A.~{Kusiak}, F.~{McCarthy}, A.~{Sabyr}, K.~{Surrao}, J.C.~{Hill} et~al., \emph{{class\_sz I: Overview}}, \href{https://doi.org/10.48550/arXiv.2310.18482}{\emph{arXiv e-prints} (2023) arXiv:2310.18482} [\href{https://arxiv.org/abs/2310.18482}{{\ttfamily 2310.18482}}].

\bibitem{bollietclass}
B.~{Bolliet}, J.~{Colin Hill}, S.~{Ferraro}, A.~{Kusiak} and A.~{Krolewski}, \emph{{Projected-field kinetic Sunyaev-Zel'dovich Cross-correlations: halo model and forecasts}}, \href{https://doi.org/10.1088/1475-7516/2023/03/039}{\emph{\jcap} {\bfseries 2023} (2023) 039} [\href{https://arxiv.org/abs/2208.07847}{{\ttfamily 2208.07847}}].

\bibitem{tinker2010}
J.L.~{Tinker}, B.E.~{Robertson}, A.V.~{Kravtsov}, A.~{Klypin}, M.S.~{Warren}, G.~{Yepes} et~al., \emph{{The Large-scale Bias of Dark Matter Halos: Numerical Calibration and Model Tests}}, \href{https://doi.org/10.1088/0004-637X/724/2/878}{\emph{The Astrophysical Journal} {\bfseries 724} (2010) 878} [\href{https://arxiv.org/abs/1001.3162}{{\ttfamily 1001.3162}}].

\bibitem{bolliet2018}
B.~{Bolliet}, B.~{Comis}, E.~{Komatsu} and J.F.~{Mac{\'\i}as-P{\'e}rez}, \emph{{Dark energy constraints from the thermal Sunyaev-Zeldovich power spectrum}}, \href{https://doi.org/10.1093/mnras/sty823}{\emph{\mnras} {\bfseries 477} (2018) 4957} [\href{https://arxiv.org/abs/1712.00788}{{\ttfamily 1712.00788}}].

\bibitem{tanimura2022}
H.~{Tanimura}, M.~{Douspis}, N.~{Aghanim} and L.~{Salvati}, \emph{{Constraining cosmology with a new all-sky Compton parameter map from the Planck PR4 data}}, \href{https://doi.org/10.1093/mnras/stab2956}{\emph{\mnras} {\bfseries 509} (2022) 300} [\href{https://arxiv.org/abs/2110.08880}{{\ttfamily 2110.08880}}].

\bibitem{eftsathiou2025}
G.~{Efstathiou} and F.~{McCarthy}, \emph{{The power spectrum of the thermal Sunyaev{\textendash}Zeldovich effect}}, \href{https://doi.org/10.1093/mnras/staf709}{\emph{\mnras} {\bfseries 540} (2025) 1055} [\href{https://arxiv.org/abs/2502.10232}{{\ttfamily 2502.10232}}].

\bibitem{cambpaper}
A.~{Lewis}, A.~{Challinor} and A.~{Lasenby}, \emph{{Efficient Computation of Cosmic Microwave Background Anisotropies in Closed Friedmann-Robertson-Walker Models}}, \href{https://doi.org/10.1086/309179}{\emph{\apj} {\bfseries 538} (2000) 473} [\href{https://arxiv.org/abs/astro-ph/9911177}{{\ttfamily astro-ph/9911177}}].

\bibitem{cambsoft}
A.~{Lewis} and A.~{Challinor}, ``{CAMB: Code for Anisotropies in the Microwave Background}.'' Astrophysics Source Code Library, Feb, 2011.

\bibitem{KomatsuSeljScaling}
E.~{Komatsu} and U.~{Seljak}, \emph{{The Sunyaev-Zel'dovich angular power spectrum as a probe of cosmological parameters}}, \href{https://doi.org/10.1046/j.1365-8711.2002.05889.x}{\emph{\mnras} {\bfseries 336} (2002) 1256} [\href{https://arxiv.org/abs/astro-ph/0205468}{{\ttfamily astro-ph/0205468}}].

\bibitem{ShawScaling}
L.D.~{Shaw}, D.~{Nagai}, S.~{Bhattacharya} and E.T.~{Lau}, \emph{{Impact of Cluster Physics on the Sunyaev-Zel'dovich Power Spectrum}}, \href{https://doi.org/10.1088/0004-637X/725/2/1452}{\emph{\apj} {\bfseries 725} (2010) 1452} [\href{https://arxiv.org/abs/1006.1945}{{\ttfamily 1006.1945}}].

\bibitem{TracScaling}
H.~{Trac}, P.~{Bode} and J.P.~{Ostriker}, \emph{{Templates for the Sunyaev-Zel'dovich Angular Power Spectrum}}, \href{https://doi.org/10.1088/0004-637X/727/2/94}{\emph{\apj} {\bfseries 727} (2011) 94} [\href{https://arxiv.org/abs/1006.2828}{{\ttfamily 1006.2828}}].

\bibitem{bolliet2020}
B.~{Bolliet}, T.~{Brinckmann}, J.~{Chluba} and J.~{Lesgourgues}, \emph{{Including massive neutrinos in thermal Sunyaev Zeldovich power spectrum and cluster counts analyses}}, \href{https://doi.org/10.1093/mnras/staa1835}{\emph{\mnras} {\bfseries 497} (2020) 1332} [\href{https://arxiv.org/abs/1906.10359}{{\ttfamily 1906.10359}}].

\bibitem{planck2015}
{Planck Collaboration}, N.~{Aghanim}, M.~{Arnaud}, M.~{Ashdown}, J.~{Aumont}, C.~{Baccigalupi} et~al., \emph{{Planck 2015 results. XXII. A map of the thermal Sunyaev-Zeldovich effect}}, \href{https://doi.org/10.1051/0004-6361/201525826}{\emph{\aap} {\bfseries 594} (2016) A22} [\href{https://arxiv.org/abs/1502.01596}{{\ttfamily 1502.01596}}].

\bibitem{cbcastorina}
E.~{Castorina}, E.~{Sefusatti}, R.K.~{Sheth}, F.~{Villaescusa-Navarro} and M.~{Viel}, \emph{{Cosmology with massive neutrinos II: on the universality of the halo mass function and bias}}, \href{https://doi.org/10.1088/1475-7516/2014/02/049}{\emph{\jcap} {\bfseries 2014} (2014) 049} [\href{https://arxiv.org/abs/1311.1212}{{\ttfamily 1311.1212}}].

\bibitem{cbLoverde}
M.~{LoVerde}, \emph{{Spherical collapse in {\ensuremath{\nu}}{\ensuremath{\Lambda}}CDM}}, \href{https://doi.org/10.1103/PhysRevD.90.083518}{\emph{\prd} {\bfseries 90} (2014) 083518} [\href{https://arxiv.org/abs/1405.4858}{{\ttfamily 1405.4858}}].

\bibitem{tinker2008}
J.~{Tinker}, A.V.~{Kravtsov}, A.~{Klypin}, K.~{Abazajian}, M.~{Warren}, G.~{Yepes} et~al., \emph{{Toward a Halo Mass Function for Precision Cosmology: The Limits of Universality}}, \href{https://doi.org/10.1086/591439}{\emph{\apj} {\bfseries 688} (2008) 709} [\href{https://arxiv.org/abs/0803.2706}{{\ttfamily 0803.2706}}].

\bibitem{sabyr2025_minkowski}
A.~{Sabyr}, J.C.~{Hill} and Z.~{Haiman}, \emph{{Constraining cosmology with thermal Sunyaev-Zel'dovich maps: Minkowski functionals, peaks, minima, and moments}}, \href{https://doi.org/10.1103/PhysRevD.111.103536}{\emph{\prd} {\bfseries 111} (2025) 103536} [\href{https://arxiv.org/abs/2410.21247}{{\ttfamily 2410.21247}}].

\bibitem{simonsObs}
P.~{Ade}, J.~{Aguirre}, Z.~{Ahmed}, S.~{Aiola}, A.~{Ali}, D.~{Alonso} et~al., \emph{{The Simons Observatory: science goals and forecasts}}, \href{https://doi.org/10.1088/1475-7516/2019/02/056}{\emph{\jcap} {\bfseries 2019} (2019) 056} [\href{https://arxiv.org/abs/1808.07445}{{\ttfamily 1808.07445}}].

\bibitem{SNRsilk}
P.~{Singh}, B.B.~{Nath}, S.~{Majumdar} and J.~{Silk}, \emph{{CMB distortion from circumgalactic gas}}, \href{https://doi.org/10.1093/mnras/stv155}{\emph{\mnras} {\bfseries 448} (2015) 2384} [\href{https://arxiv.org/abs/1408.4896}{{\ttfamily 1408.4896}}].

\bibitem{makiyaTrisp}
R.~{Makiya}, S.~{Ando} and E.~{Komatsu}, \emph{{Joint analysis of the thermal Sunyaev-Zeldovich effect and 2MASS galaxies: probing gas physics in the local Universe and beyond}}, \href{https://doi.org/10.1093/mnras/sty2031}{\emph{\mnras} {\bfseries 480} (2018) 3928} [\href{https://arxiv.org/abs/1804.05008}{{\ttfamily 1804.05008}}].

\bibitem{chisqEstimator}
A.R.~{Pullen}, S.~{Alam} and S.~{Ho}, \emph{{Probing gravity at large scales through CMB lensing}}, \href{https://doi.org/10.1093/mnras/stv554}{\emph{\mnras} {\bfseries 449} (2015) 4326} [\href{https://arxiv.org/abs/1412.4454}{{\ttfamily 1412.4454}}].

\bibitem{bird2012}
S.~{Bird}, M.~{Viel} and M.G.~{Haehnelt}, \emph{{Massive neutrinos and the non-linear matter power spectrum}}, \href{https://doi.org/10.1111/j.1365-2966.2011.20222.x}{\emph{\mnras} {\bfseries 420} (2012) 2551} [\href{https://arxiv.org/abs/1109.4416}{{\ttfamily 1109.4416}}].

\bibitem{Mauro_2007}
M.~{Roncarelli}, L.~{Moscardini}, S.~{Borgani} and K.~{Dolag}, \emph{{The Sunyaev-Zel'dovich effects from a cosmological hydrodynamical simulation: large-scale properties and correlation with the soft X-ray signal}}, \href{https://doi.org/10.1111/j.1365-2966.2007.11914.x}{\emph{\mnras} {\bfseries 378} (2007) 1259} [\href{https://arxiv.org/abs/astro-ph/0701680}{{\ttfamily astro-ph/0701680}}].

\bibitem{shawksz}
L.D.~{Shaw}, D.H.~{Rudd} and D.~{Nagai}, \emph{{Deconstructing the Kinetic SZ Power Spectrum}}, \href{https://doi.org/10.1088/0004-637X/756/1/15}{\emph{\apj} {\bfseries 756} (2012) 15} [\href{https://arxiv.org/abs/1109.0553}{{\ttfamily 1109.0553}}].

\bibitem{flamingo_2025_kSZfeedback}
I.G.~{McCarthy}, A.~{Amon}, J.~{Schaye}, E.~{Schaan}, R.E.~{Angulo}, J.~{Salcido} et~al., \emph{{FLAMINGO: combining kinetic SZ effect and galaxy{\textendash}galaxy lensing measurements to gauge the impact of feedback on large-scale structure}}, \href{https://doi.org/10.1093/mnras/staf731}{\emph{\mnras} {\bfseries 540} (2025) 143} [\href{https://arxiv.org/abs/2410.19905}{{\ttfamily 2410.19905}}].

\bibitem{nfw}
J.F.~{Navarro}, C.S.~{Frenk} and S.D.M.~{White}, \emph{{The Structure of Cold Dark Matter Halos}}, \href{https://doi.org/10.1086/177173}{\emph{\apj} {\bfseries 462} (1996) 563} [\href{https://arxiv.org/abs/astro-ph/9508025}{{\ttfamily astro-ph/9508025}}].

\bibitem{chiu2016_baryons}
I.~{Chiu}, J.~{Mohr}, M.~{McDonald}, S.~{Bocquet}, M.L.N.~{Ashby}, M.~{Bayliss} et~al., \emph{{Baryon content of massive galaxy clusters at 0.57 < z < 1.33}}, \href{https://doi.org/10.1093/mnras/stv2303}{\emph{\mnras} {\bfseries 455} (2016) 258} [\href{https://arxiv.org/abs/1412.7823}{{\ttfamily 1412.7823}}].

\bibitem{bahamas_2017}
I.G.~{McCarthy}, J.~{Schaye}, S.~{Bird} and A.M.C.~{Le Brun}, \emph{{The BAHAMAS project: calibrated hydrodynamical simulations for large-scale structure cosmology}}, \href{https://doi.org/10.1093/mnras/stw2792}{\emph{\mnras} {\bfseries 465} (2017) 2936} [\href{https://arxiv.org/abs/1603.02702}{{\ttfamily 1603.02702}}].

\bibitem{the300_2018}
W.~{Cui}, A.~{Knebe}, G.~{Yepes}, F.~{Pearce}, C.~{Power}, R.~{Dave} et~al., \emph{{The Three Hundred project: a large catalogue of theoretically modelled galaxy clusters for cosmological and astrophysical applications}}, \href{https://doi.org/10.1093/mnras/sty2111}{\emph{\mnras} {\bfseries 480} (2018) 2898} [\href{https://arxiv.org/abs/1809.04622}{{\ttfamily 1809.04622}}].

\bibitem{flamingo_2023}
J.~{Schaye}, R.~{Kugel}, M.~{Schaller}, J.C.~{Helly}, J.~{Braspenning}, W.~{Elbers} et~al., \emph{{The FLAMINGO project: cosmological hydrodynamical simulations for large-scale structure and galaxy cluster surveys}}, \href{https://doi.org/10.1093/mnras/stad2419}{\emph{\mnras} {\bfseries 526} (2023) 4978} [\href{https://arxiv.org/abs/2306.04024}{{\ttfamily 2306.04024}}].

\bibitem{duffy2008}
A.R.~{Duffy}, J.~{Schaye}, S.T.~{Kay} and C.~{Dalla Vecchia}, \emph{{Dark matter halo concentrations in the Wilkinson Microwave Anisotropy Probe year 5 cosmology}}, \href{https://doi.org/10.1111/j.1745-3933.2008.00537.x}{\emph{\mnras} {\bfseries 390} (2008) L64} [\href{https://arxiv.org/abs/0804.2486}{{\ttfamily 0804.2486}}].

\end{thebibliography}\endgroup
\bibliographystyle{JHEP}

%\newpage
\appendix

\section{The kSZ from the \texttt{DEMNUni} clusters}\label{sec::kszResults}
We present here our description of the kSZ effect, carried out with the same \texttt{DEMNUni} halo sample as in the main analysis. It is well known that the kSZ effect receives a substantial contribution from low mass halos and from the diffuse material of the LSS, i.e. gas from filaments and sheets~\cite{Mauro_2007,shawksz,mauro2018,mauro2017}. Moreover, recent works show that baryon feedback may affect substantially the amplitude of the kSZ power spectrum~\cite{flamingo_2025_kSZfeedback}. Since the \texttt{DEMNUni} simulations are gravity only, a precise description of the kSZ signal is clearly beyond the scope of this work. We thus decided to apply an halo-based approach, extending our tSZ modelling to the kSZ effect, acknowledging that the description presented here is partial and meant to describe only the kSZ component associated to the bulk motion of galaxy clusters. We provide this as an interesting addition, to be potentially investigated in the future with measurements based on cluster samples.
\subsection{Modelling of the electron density}
\label{ksz}
We model the density of baryons inside the haloes with an NFW~\cite{nfw} profile
\begin{equation}\label{nfwBaryons}
    \rho_{\mathrm{b}} (r) = (1-f_*) f_{\mathrm{b}} \, \rho (r) =  (1-f_*) f_{\mathrm{b}} \frac{\rho_0}{(r/r_{\mathrm{s}})(1 + r/r_{\mathrm{s}})^2} \ ,
\end{equation}
where $f_*$ is the stellar mass fraction, held fixed at a representative value of $0.1$~\cite{chiu2016_baryons,bahamas_2017}. Indeed, the large scatter observed in the relation between this parameter and both halo mass and redshift~\cite{the300_2018,flamingo_2023} makes it challenging to implement a physically reasonable analytical model. We then account for the dependencies of $r_\mathrm{s}$ and $\rho_0$ on $M_{200}$ and $z$ through the Duffy concentration parameter
\begin{equation}\label{duffyConc}
    R_{200} / r_{\mathrm{s}} \equiv c_{200} = 5.71 \left ( \frac{M_{200}}{2 \times 10^{12} \, h^{-1} \msun}\right)^{-0.084} (1+z)^{-0.47} \ ,
\end{equation}
derived from N-body simulations~\cite{duffy2008}, retrieving $\rho_0$ by normalising $\rho(r)$ to $M_{200}$ at $R_{200}$. The resulting density profiles are shown in Figure~\ref{multiFigProfiles}.
The electron density can then be computed by rescaling the baryon density as 
\begin{equation}\label{electronDens}
    n_\mathrm{e} = \frac{1}{\mu m_\mathrm{p}} \rho_{\mathrm{b}} \ ,
\end{equation}
where $\mu = \left[ f_\mathrm{H} + \frac{1}{2}(1-f_\mathrm{H}) \right] ^{-1} \simeq 1.14$ is the mean molecular weight per electron of a fully ionised cosmological gas in units of the proton mass $m_\mathrm{p}$. Finally, the kSZ induced CMB temperature variation is derived for each halo with Equation~\eqref{kszeq}, considering the density profile as in Eq.~\eqref{electronDens} and $v_{\mathrm{los}}$ taken for each halo from their individual peculiar velocities, as extracted from the catalogues (Section~\ref{halocat}), thus neglecting internal motions. Since we are considering objects at relatively low redshift ($z<2.5$) we can safely assume $\exp(-\tau) \simeq 1$.
We then produce $\Delta T/T$ maps with the same pixelisation scheme and resolution as for the tSZ effect (see Section~\ref{mapmaking}); examples can be found in Figure~\ref{multiFigMaps}.

\subsection{One-point statistics}
\begin{figure}
    \centering
    \includegraphics[width=0.6\textwidth]{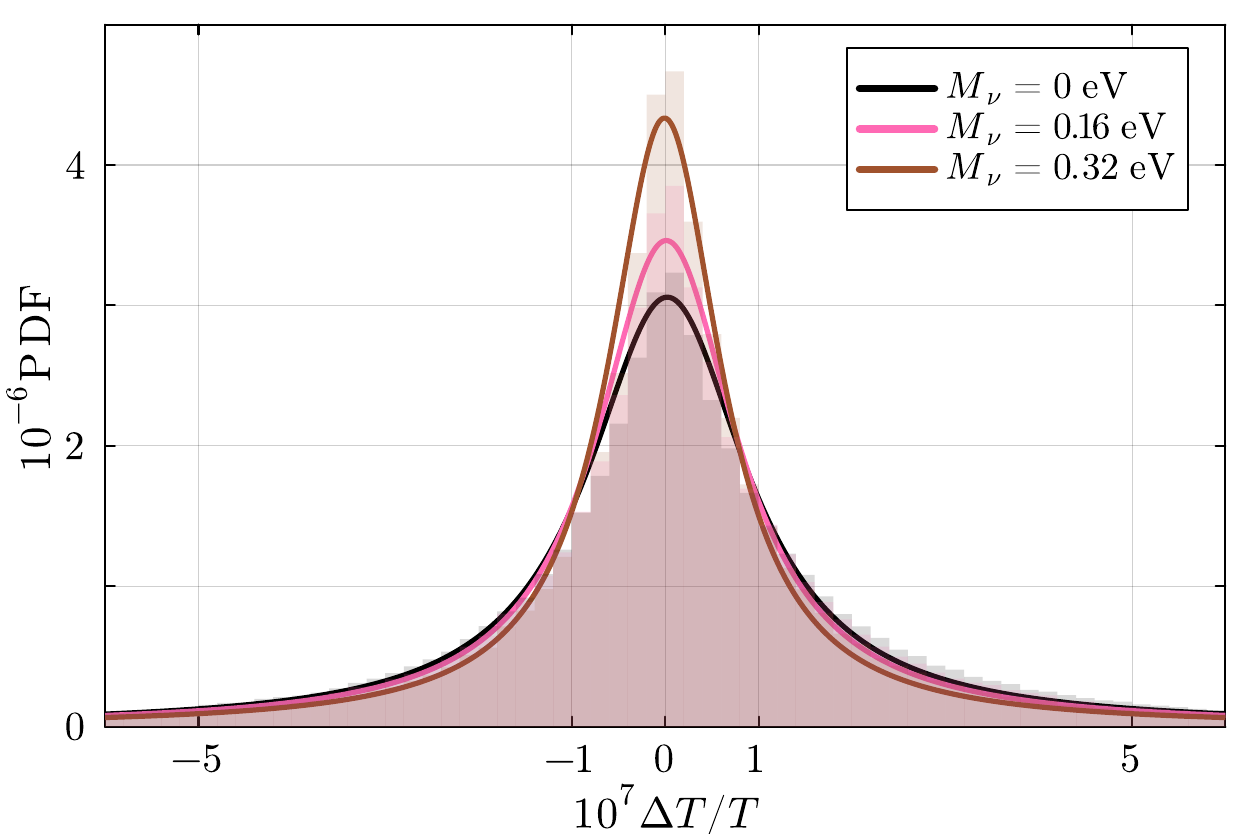}
    \caption{Best-fit Cauchy distribution (solid lines) for the kSZ $\Delta T / T$ maps from the $\nu\Lambda$CDM cosmologies in Table~\ref{tab:demnuni_sets_table}, compared to the corresponding normalised distributions of the data.}
    \label{pdfKinetic}
\end{figure}

Looking first at the right column of Figure~\ref{multiFigMaps}, it is possible to appreciate the symmetric nature of the kSZ signal, even on a small fraction of the sky, which is due to the nature of the peculiar velocity field. Moreover, as already pointed out, we can also observe how the increase in $M_\nu$ seems to impact the kinematic SZ less than the thermal one.
We now study in detail the pixel distribution of this kSZ $\Delta T /T$ signal in the lower resolution ($0.5^{\circ}$) maps, from which we remove the null pixels, as for the previous case. We find the best-fit with a Cauchy distribution 
\begin{equation}\label{Cauchy}
    P(x) = \frac{1}{\pi C}\frac{1}{1+\left( \frac{x-\mu}{C}\right)^2} \, 
\end{equation}
with mean $\mu$ and scale parameter $C$. 
The fitting values for the two parameters are listed in Table~\ref{pdfTablekSZ}: as expected, the mean is in all simulations comparable to zero as it is at least one order of magnitude smaller with respect to $C$. When increasing neutrino mass, at fixed dark energy EoS, the scale parameter decreases leading to a more peaked distribution.
Again it is possible to see the effect of incrementing the mass budget for neutrinos: due to the reduced number of haloes in the high-mass end the values closer to zero are favoured, as they derive from smaller structures with an accordingly smaller optical depth. Examples of the distributions are shown in Figure~\ref{pdfKinetic}. Our finding differs substantially from those of~\cite{mauro2017,mauro2018}, obtained through hydrodynamical simulations, where the resulting distribution is a Gaussian curve rather than a Cauchy one. This confirms how our cluster-only analysis gives a partial picture of kSZ, missing an important part of signal with lower intensity, arising in diffuse gas. 

\begin{table}
\begin{center}
\begin{tabular}{ | c l||   c |  c| }
\hline
    \multicolumn{2}{|c||}{Simulation} & $10^8 \mu$ & $10^8 C$ \\ 
\hline
    $M_{\nu} = 0$ eV & $w_0 = -1 \, , w_a = 0$ & 0.22 &10.41\\ 
    
     & $w_0 = -0.9  \, , w_a = -0.3$ & 0.24 &10.77\\ 
    
    & $w_0 = -0.9  \, , w_a = +0.3$ & 0.02 &8.94\\ 
    
    & $w_0 = -1.1  \, , w_a = -0.3$ & 0.28 & 11.36\\ 
    
    & $w_0 = -1.1  \, , w_a = +0.3$ &0.20 & 10.33 \\ 
    \hline
     $M_{\nu} = 0.16$ eV & $w_0 = -1 \, , w_a = 0$ & 0.11 & 9.20\\ 
    
     & $w_0 = -0.9  \, , w_a = -0.3$ & 0.05 & 9.31\\ 
    
    & $w_0 = -0.9  \, , w_a = +0.3$ & -0.07 & 7.33\\ 
    
    & $w_0 = -1.1  \, , w_a = -0.3$ &0.12 & 9.70 \\ 
    
    & $w_0 = -1.1  \, , w_a = +0.3$ & 0.04 & 8.87\\ 
    \hline
 $M_{\nu} = 0.32$ eV & $w_0 = -1 \, , w_a = 0$ & -0.08 & 7.34 \\ 
    
     & $w_0 = -0.9  \, , w_a = -0.3$ & -0.01 & 7.22\\ 
    
    & $w_0 = -0.9  \, , w_a = +0.3$ & -0.11 & 5.72\\ 
    
    & $w_0 = -1.1  \, , w_a = -0.3$ & -0.02 & 7.90 \\ 
    
    & $w_0 = -1.1  \, , w_a = +0.3$ & -0.07 & 7.06 \\ 
    \hline
\end{tabular}
\caption[]{Best-fit parameters $C$ and $\mu$ for the Cauchy distribution in Equation~\eqref{Cauchy} for all the \texttt{DEMNUni} simulations.}\label{pdfTablekSZ} 
\end{center}
\end{table}

\subsection{Power spectrum}
\begin{figure}[!ht]
    \centering
    \begin{tabular}{lc}
    \subfloat{\includegraphics[width = 0.5\linewidth]{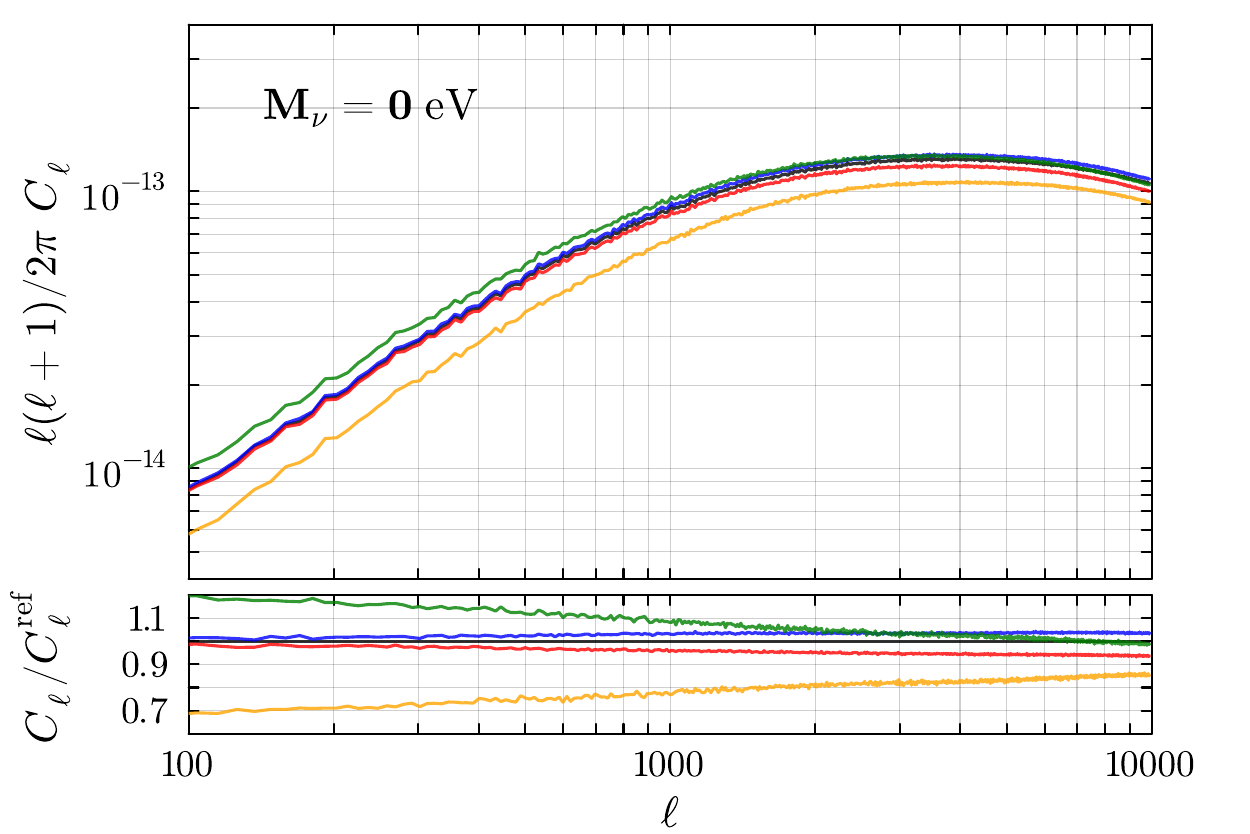}}
    \subfloat{\includegraphics[width = 0.5\linewidth]{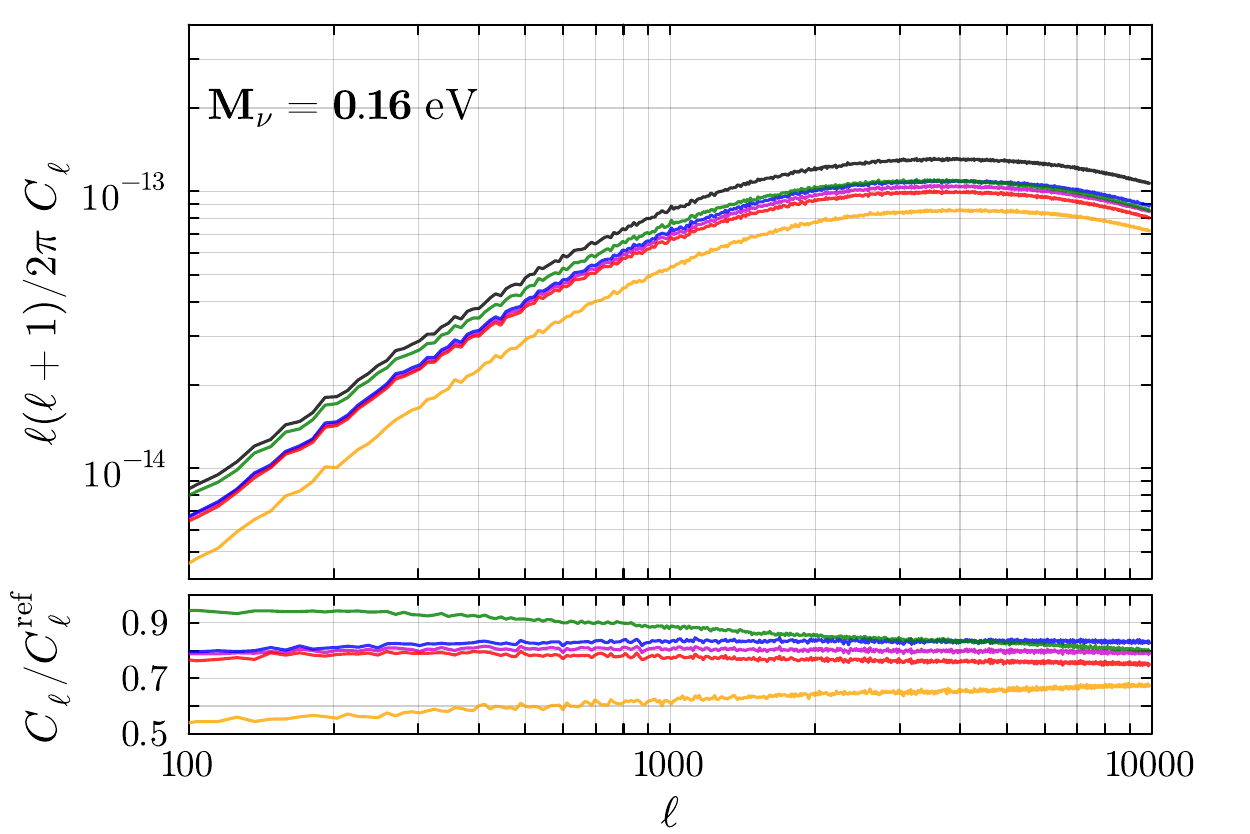}}\\ 
    \subfloat{\includegraphics[width = 0.5\linewidth]{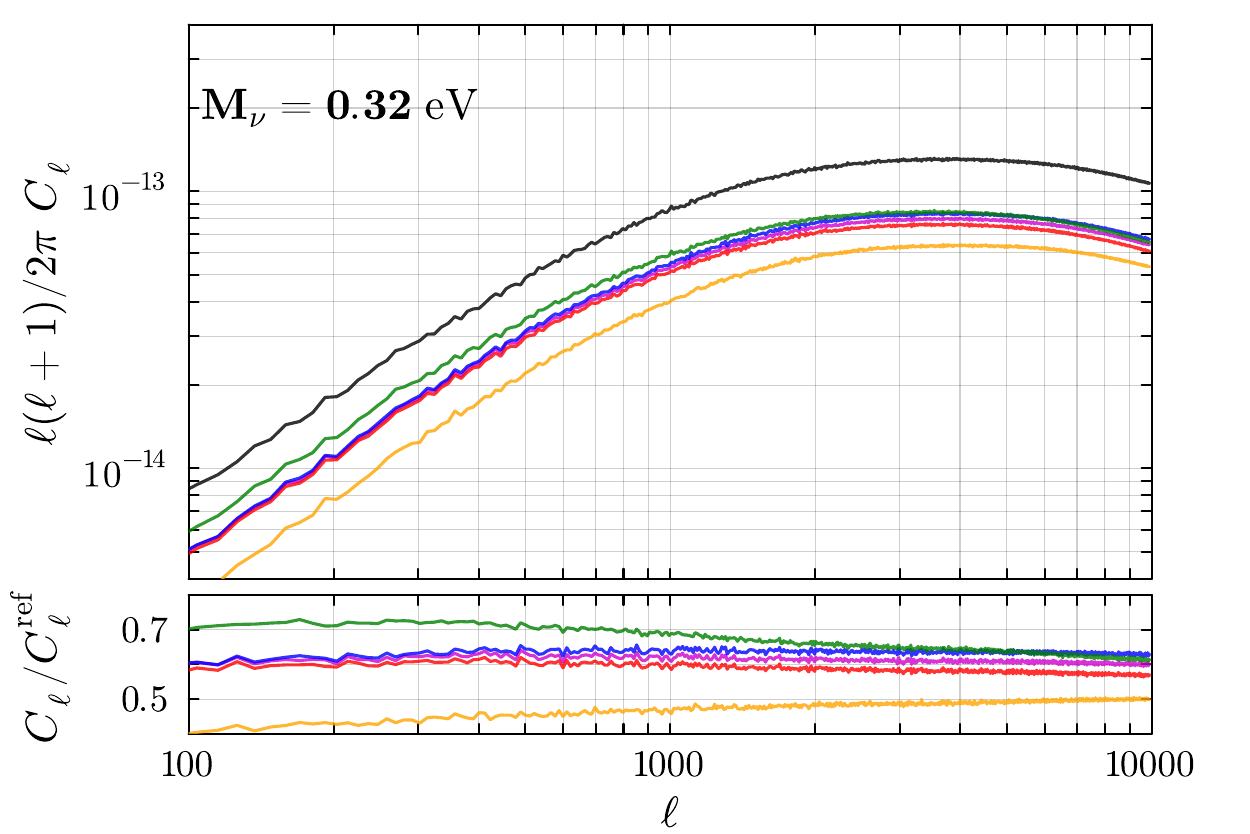}} \hspace{1.02cm}
    \subfloat{\raisebox{0.6\height}{\includegraphics[width = 0.3\linewidth]{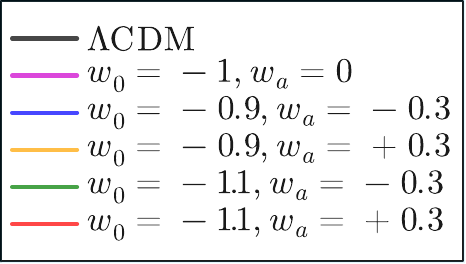}}}
    \end{tabular}
    \caption{Temperature power spectra for the kSZ effect measured at $z>0.05$ from all the 15 \texttt{DEMNUni} simulations. Notice how in our cluster-only model the shape is similar to that of the tSZ effect (see Figure~\ref{multiFigPS}). The sub-panels also show the ratio of the different spectra with respect to the $\Lambda$CDM simulation.}
    \label{kineticPS}
\end{figure}

\begin{figure}[!ht]
    \centering
    \begin{tabular}{lc}
    \subfloat{\includegraphics[width = 0.5\linewidth]{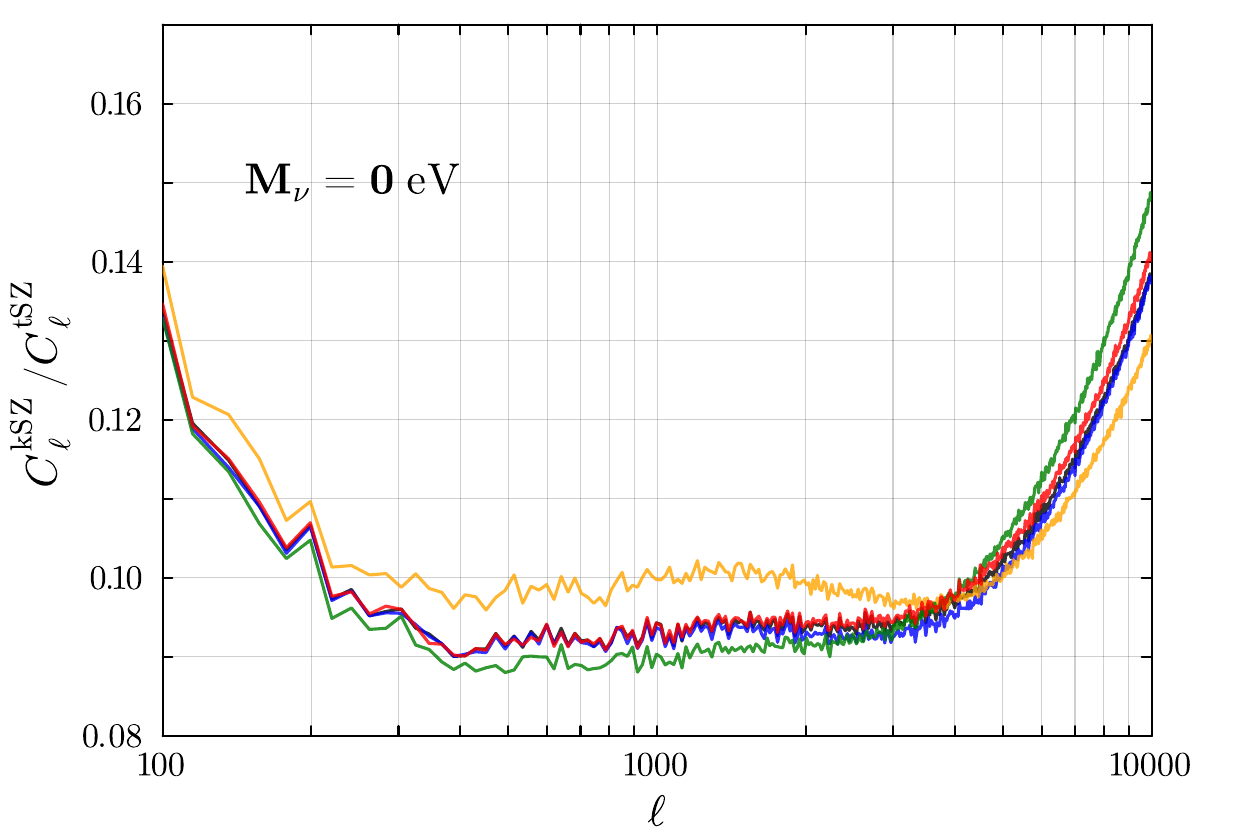}}
    \subfloat{\includegraphics[width = 0.5\linewidth]{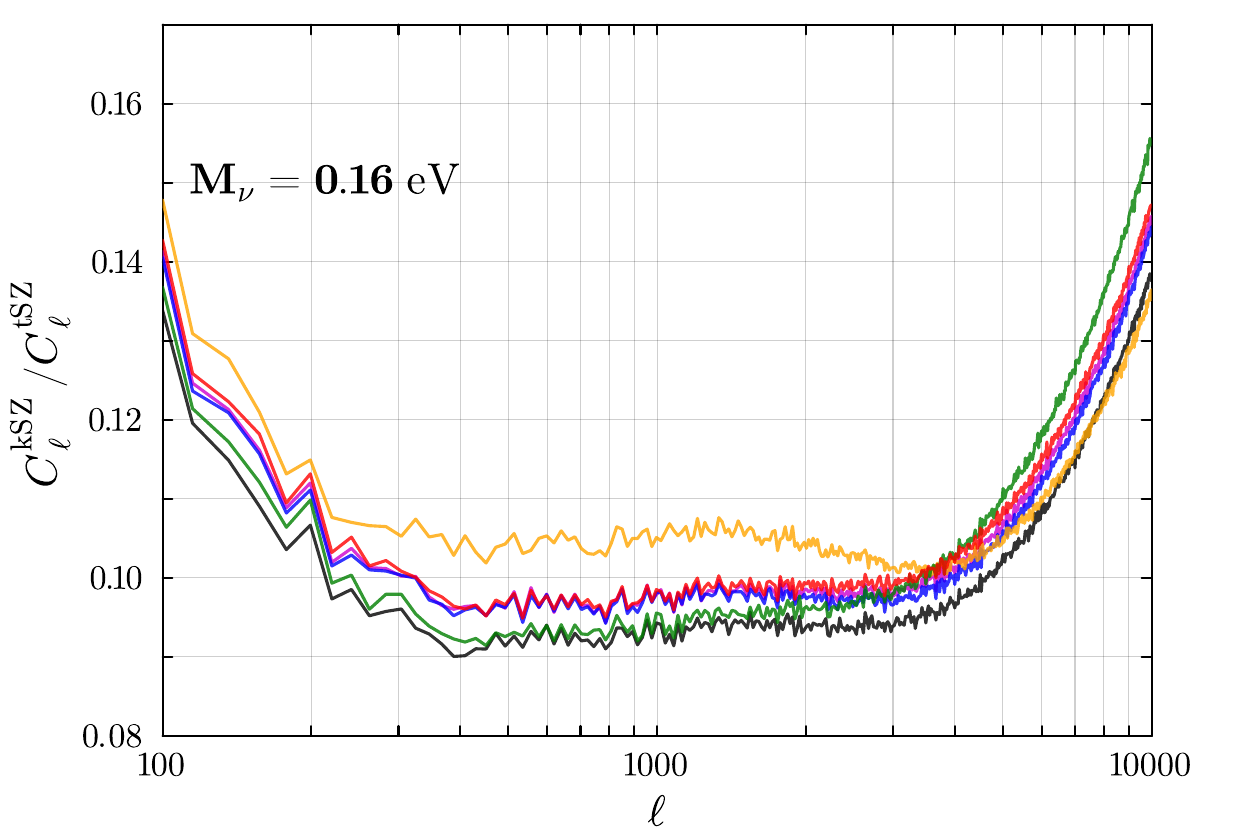}}\\ 
    \subfloat{\includegraphics[width = 0.5\linewidth]{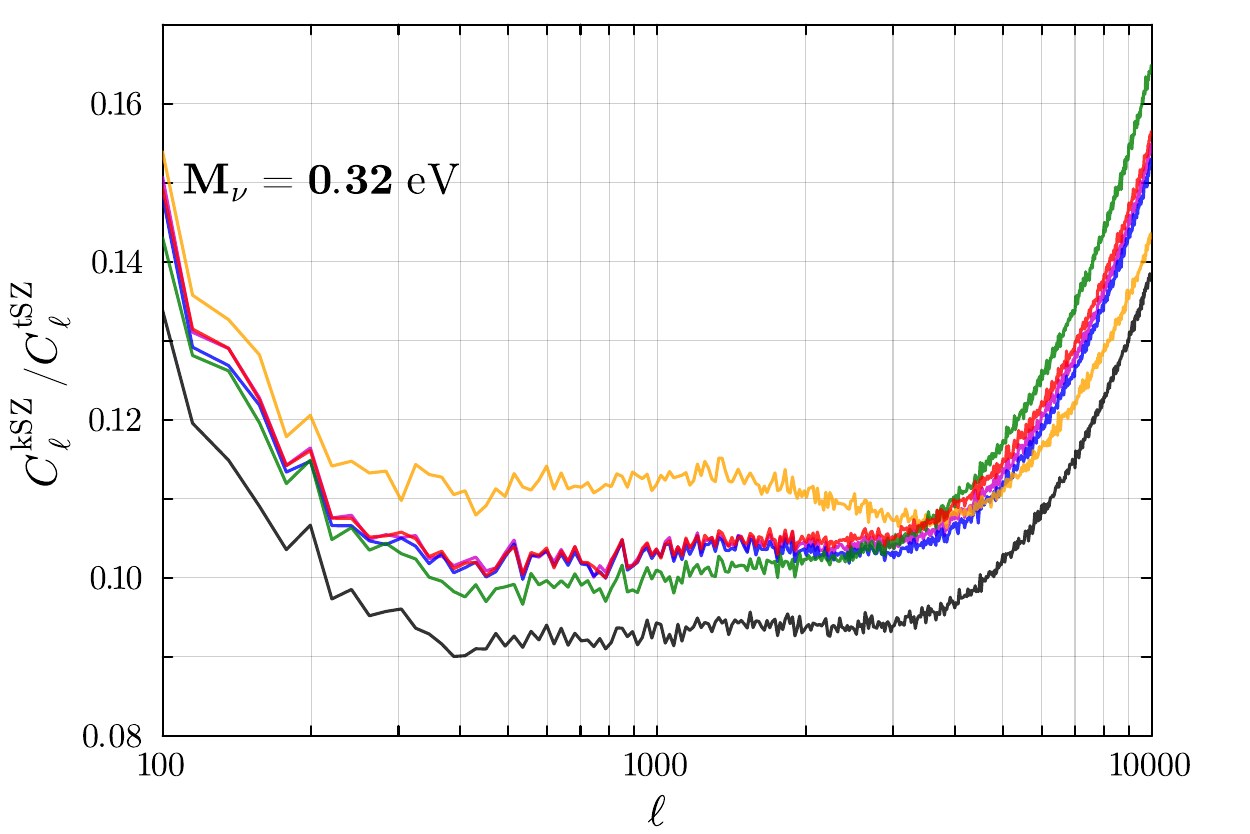}} \hspace{0.89cm}
    \subfloat{\raisebox{0.6\height}{\includegraphics[width = 0.3\linewidth]{Images/legendOnly_kinetic_cropped.pdf}}}
    \end{tabular}
    \caption{Temperature power spectrum ratios, $C_\ell^{\mathrm{kSZ}}/C_\ell^{\mathrm{tSZ}}$, measured from the \texttt{DEMNUni} simulated maps, in the case of a channel frequency of $280$ GHz, for which $g(x)=1$. This ratio slightly increases as $M_\nu$ increases, consistently with the reduced dependence of the kinematic effect, if compared to the tSZ, on the abundance of massive clusters which is largely suppressed by $M_\nu$.}
    \label{ratioAll}
\end{figure}

We compute the power spectrum of the cluster component of the kinematic SZ effect in the same way as for the tSZ, and obtain the results shown in Figure~\ref{kineticPS}.
Due to the partial description that our model provides the kSZ lacks power if compared to a more precise modelling (again see e.g.~\cite{mauro2017,mauro2018}). Nonetheless, a similar behaviour to that of the thermal effect can be observed, where both a different dark energy EoS and a different $M_\nu$ have a significant, and almost degenerate, impact on the power spectrum.
With respect to the tSZ, for the kSZ we observe a less prominent dependence on the mass of the neutrino component: this is due to the increased contribution to the total power from the smaller haloes, which have an abundance that is less influenced by a change in $M_\nu$. 
Also, at fixed $M_\nu$, there is less tendency to converge at the smallest scales with respect to the tSZ, as for $\ell = 100$ the range in power covered by the possible EoS is greater than at $\ell = 10^4$ by $44,43$ and $39$\% for the $M_\nu=0,0.16$ and $0.32$ eV models, respectively.
The differences in the two effects become even more clear when explicitly looking at the ratio kSZ/tSZ, shown in Figure~\ref{ratioAll}, where a growth both towards smaller scales and larger scales is present, respectively for $\ell \gtrsim 3000$ and $\ell \lesssim 300$. 
The former is given by the non-negligible contribution that the kSZ receives from haloes of lower mass, differently from the tSZ, as these haloes are both smaller in size and  with a significant abundance also at higher redshifts.
The latter, instead, is due to the peculiar velocity field, which has significant variability on very large scales (up to $400$ Mpc) as it is mainly driven by linear, large scale evolution, while the density field varies on scales which are smaller.


 We then study the dependence of the power spectrum on $\sigma_8^{\mathrm{cb}}$ also for the case of the kSZ. In the literature there are reported values for the scaling with different parameters~\cite{shawksz}, but always referred to the total kinematic effect, where diffuse matter is also present. Assuming that it is possible to neglect the small variations in $\Omega_{\mathrm{cb}}$, like we did for the tSZ, we fit the power-law scaling of the power spectrum as in Eq.~\eqref{scalingSigma8cb}, and find the results listed in Table~\ref{scalingkSZtable}.
As expected, the dependencies are weaker with respect to those of the thermal effect in our investigation, as the number of smaller haloes is less affected by $\sigma_8^{\mathrm{cb}}$. Conversely, these values are larger than the ones from the literature regarding the total kSZ, as the lack of a diffuse component in our analysis makes the kinematic power spectrum more sensitive to clustering. 
\begin{table}[!ht]
\begin{center}

\begin{tabular}{ | c||   c |  }
\hline
    $(w_0,w_a)$ & $q$ \\ 
\hline
    $(-1,0)$ & $6.1$ \\ 
    
    $(-0.9,-0.3)$ & $6.1$\\ 
    
    $(-0.9,+0.3)$ & $6.5$\\ 
    
    $(-1.1,-0.3)$ & $5.9$ \\ 
    
    $(-1.1,+0.3)$ & $6.1$\\
\hline
\end{tabular}
\caption{Power-law exponents for the scaling of the temperature power spectrum for the kSZ effect with respect to $\sigma_8^{\mathrm{cb}}$ alone. The calculation is analogous to the one for tSZ, i.e. done keeping fixed the dark energy EoS and varying $M_\nu$.} \label{scalingkSZtable}
\end{center}
\end{table}

\end{document}